\documentclass[a4paper]{article}
\pdfoutput=1
\usepackage{jheppub}

\allowdisplaybreaks

\usepackage[utf8]{inputenc}

\usepackage[table ]{ xcolor}

\usepackage{multirow}

\usepackage{amsmath}

\usepackage{tikz}
\usetikzlibrary{shapes,arrows}
\tikzstyle{startstop} = [rectangle,rounded corners, minimum width=3cm,minimum height=1cm,text centered, draw=black,fill=red!30]
\tikzstyle{io} = [trapezium, trapezium left angle = 70,trapezium right angle=110,minimum width=3cm,minimum height=1cm,text centered,draw=black,fill=blue!30]
\tikzstyle{process} = [rectangle,minimum width=3cm,minimum height=1cm,text centered,text width =3cm,draw=black,fill=orange!30]
\tikzstyle{decision} = [diamond,minimum width=3cm,minimum height=1cm,shape aspect=3,inner sep = 0.4pt,text centered,draw=black,fill=green!30]
\tikzstyle{arrow} = [thick,->,>=stealth]
\tikzstyle{shadow}=[preaction={fill=black,opacity=.5,transform canvas={xshift=0.5mm,yshift=-0.5mm},shading=radial,shading angle=20},fill=red]

\tikzstyle{ellipse}=[draw, rectangle, minimum width=2.8em, rounded corners=6pt,line width=0.5pt]
\tikzstyle{pxsbx}=[trapezium, trapezium left angle=75, trapezium right angle=105, minimum width=3em, text centered, draw = black, fill=white,line width=0.5pt] 
\tikzstyle{lingxing}=[draw,diamond,shape aspect=3,inner sep = 0.4pt,thick,font=\itshape,line width=0.5pt]

\usepackage{subcaption}

\usepackage{amssymb}
\usepackage{graphicx,color}

\usepackage{amsmath,amsthm}

\usepackage{mathrsfs}

\usepackage{bm}

\def\beq{\begin{equation}}
\def\eeq{\end{equation}}
\newcommand{\bea}{\begin{eqnarray}}
\newcommand{\eea}{\end{eqnarray}}
\def\bi{\begin{itemize}}
\def\ei{\end{itemize}}
\def\ba{\begin{array}}
\def\ea{\end{array}}
\def\bfig{\begin{figure}}
\def\efig{\end{figure}}

\def\R{\mathbb{R}}

\def\sgn{\text{sgn}}

\newcommand{\Su}{\mathrm{SU}(2)}

\def\be{\begin{eqnarray}}
\def\ee{\end{eqnarray}}

\newcommand{\cc}{\mathcal C}

\newcommand{\cf}{\mathcal F}

\newcommand{\ck}{\mathcal K}

\newcommand{\cs}{\mathcal S}

\newcommand{\cw}{\mathcal W}
\newcommand{\cx}{\mathcal X}
\newcommand{\cy}{\mathcal Y}

\newcommand{\sm}{\mathscr{M}}

\newcommand{\fa}{\mathfrak{a}}

\renewcommand{\a}{\alpha}
\renewcommand{\b}{\beta}
\newcommand{\g}{\gamma}
\newcommand{\G}{\Gamma}

\newcommand{\sig}{\sigma}
\newcommand{\Sig}{\Sigma}
\renewcommand{\l}{\lambda}
\renewcommand{\L }{\Lambda}
\renewcommand{\o}{\omega}
\renewcommand{\O}{\Omega}
\renewcommand{\t}{\tau}

\newcommand{\dd}{\mathrm d}
\newcommand{\rmd}{\mathrm d}

\newcommand{\lt}{\left}
\newcommand{\rt}{\right}

\title{Covariant $\bm{\bar{\mu}}$-scheme effective dynamics, mimetic gravity, and non-singular black holes: Applications to spherical symmetric quantum gravity and CGHS model}

\author[1,2]{Muxin Han}  

\author[1]{\ Hongguang Liu}

\affiliation[1]{Institut f\"ur Quantengravitation, Friedrich-Alexander Universit\"at Erlangen-N\"urnberg, Staudtstr. 7/B2, 91058 Erlangen, Germany}

\affiliation[2]{Department of Physics, Florida Atlantic University, 777 Glades Road, Boca Raton, FL 33431-0991, USA}

\emailAdd{hanm(At)fau.edu}
\emailAdd{hongguang.liu(AT)gravity.fau.de}

\abstract{We propose a new $\bar{\mu}$-scheme Hamiltonian effective dynamics in the spherical symmetric sector of Loop Quantum Gravity (LQG). The effective dynamics is generally covariant as derived from a covariant Lagrangian. The Lagrangian belongs to the class of extended mimetic gravity Lagrangians in 4 dimensions. We apply the effective dynamics to both cosmology and black hole. The effective dynamics reproduces the non-singular Loop-Quantum-Cosmology (LQC) effective dynamics. From the effective dynamics, we obtain the non-singular black hole solution, which has a killing symmetry in addition to the spherical symmetry and reduces to the Schwarzschild geometry asymptotically near the infinity. The black hole spacetime resolves the classical singularity and approaches asymptotically the Nariai geometry $\mathrm{dS}_2\times S^2$ at the future infinity in the interior of the black hole. The resulting black hole spacetime has the complete future null infinity $\mathscr{I}^+$. Thanks to the general covariance, the effective dynamics can be reformulated in the light-cone gauge. We generalize the covariant $\bar{\mu}$-scheme effective dynamics to the Callan-Giddings-Harvey-Strominger (CGHS) model and apply the light-cone formulation to the CGHS black hole solution with the null-shell collapse. We focus on the effective dynamics projected along the null shell. The result shows that both the 2d scalar curvature and the derivative of dilaton field are finite, in contrast to the divergence in the CGHS model.}


\begin{document}

\maketitle

\section{Introduction}

The effective dynamics of quantum gravity is an interesting approach to extracting physical predictions of quantum gravity without involving noncommutativity of quantum-geometry operators. The effective dynamics is described by $c$-number gravity and matter fields satisfying certain differential equations modifying the Einstein equation, while the quantum gravity effects are incorporated by the modification. Some remarkable progress has been made by the effective dynamics for the symmetry reduced models in Loop Quantum Gravity (LQG), such as Loop Quantum Cosmology (LQC) and quantum black holes, which both the big-bang and black-hole singularities are shown to be resolved, see e.g. \cite{Bojowald:2001xe,Ashtekar:2006wn,Agullo:2016tjh,Bojowald:2019ckm,Ashtekar:2020ifw,Ashtekar:2005qt,Modesto:2005zm,Bohmer:2007wi,Dadhich:2015ora,Ashtekar:2010qz,Chiou:2012pg,Gambini:2013hna,Bianchi:2018mml,DAmbrosio:2020mut,Brunnemann:2005in,Olmedo:2017lvt,Ashtekar:2018cay,Bojowald:2018xxu,Bodendorfer:2019cyv,Alesci:2019pbs,Assanioussi:2019twp,Han:2020uhb,Kelly:2020lec,Gambini:2020nsf,Giesel:2021dug,Lewandowski:2022zce}. The effective dynamics of quantum gravity closely relates to the program of modified gravity (see e.g. \cite{Clifton:2011jh,Chamseddine:2013kea,Sebastiani:2016ras}). The modified gravity theories define the Lagrangian that modifies the Einstein-Hilbert Lagrangian by adding the higher derivative corrections. The equations of motion from the modified gravity Lagrangian gives the modified Einstein equation, which connects to the effective dynamics of quantum gravity when we relate the higher derivative corrections to the quantum gravity effect (see e.g. \cite{BenAchour:2018khr,Langlois:2017hdf,BenAchour:2017ivq,Bodendorfer:2017bjt,Chamseddine:2016ktu,Frolov:2021afd}).

In the effective dynamics of LQG, the modification of the Einstein equation is given by the so-called holonomy-correction. Namely, the basic variables of the effective dynamics include the holonomy of the Ashtekar-Barbero connection, which is responsible for the correction to the classical Einstein equation. The classical Einstein equation is recovered only by linearizing the holonomy. The holonomy-correction is of the higher derivative type, because it contains the higher orders in the connection, which is the derivative of the metric. The effective dynamics of LQG is mostly formulated in the canonical formulation based on a $3+1$ decomposition. It is often not manifest whether the effective dynamics is covariant or relying on the special foliation, and whether the dynamics is free of the Lorentz violation. Indeed, there is the long-standing debate in the LQG community about the covariance of the effective dynamics \cite{Bojowald:2022zog,Gambini:2022dec,Bojowald:2020dkb,Bojowald:2015zha,Tibrewala:2013kba}. The effective dynamics of LQC has been shown to be covariant, because it can be derived from a covariant scalar tensor Lagrangian belonging to the extended mimetic gravity family \cite{Langlois:2017hdf,Bodendorfer:2017bjt,Chamseddine:2016uef} (see also \cite{Olmo:2008nf} for a different approach). The recent debate largely focuses on the effective black hole models in the spherical symmetric LQG.

The effective dynamics of quantum gravity is covariant if it can be derived from a manifestly covariant Lagrangian. The lesson of LQC suggests that the mimetic gravity should be a useful tool for constructing the covariant Lagrangian for the effective dynamics. Mimetic gravity is a theory of modified gravity which belongs to the family of scalar-tensor theories. The field content of the mimetic gravity contains the gravitational field $g_{\mu\nu}$ and a scalar field $\phi$, as well as a Lagrangian multiplier $\l$. The variation of the mimetic gravity Lagrangian with respect to $\l$ results in the mimetic constraint $\nabla_\mu\phi\nabla^\mu\phi =-1$, which implies that the constant $\phi$ slices are always spacelike. The (extended) mimetic gravity theory belongs to the family of Degenerate Higher-Order
Scalar-Tensor (DHOST) theories \cite{Langlois:2015skt,BenAchour:2016fzp}, which propagates (up to) only three
degrees of freedom: one scalar and two gravity tensorial modes. It is possible to use $\phi$ as the clock field, whose value defines the internal time for the effective dynamics. The mimetic gravity contains higher derivative couplings between $g_{\mu\nu}$ and $\phi$, which is the source of the higher derivative modification to the Einstein gravity. The initial physical motivation of the mimetic gravity has been to propose an alternative to cold dark matter in the universe \cite{Chamseddine:2013kea}. But here the mimetic gravity is viewed as an effective theory of quantum gravity.

One of the purpose of this work is to construct the covariant effective dynamics of the spherical symmetric LQG by using the mimetic gravity Lagrangian. As a result, the effective Hamiltonian $H$ is obtained for generating the effective dynamics in the spherical symmetric sector of LQG. This Hamiltonian is of the $\bar{\mu}$-type because it is based on the $\bar{\mu}$-scheme holonomies depending on both the connection and triad. The $\bar{\mu}$-scheme holonomies are along curves with fixed Planckian length measured by the triad. Importantly, the effective Hamiltonian dynamics can be derived from the covariant mimetic gravity Lagrangian with certain prescribed higher derivative coupling. Therefore the effective dynamics is covariant thus is called the covariant $\bar{\mu}$-scheme effective dynamics. Here the Hamiltonian $H$ is not a linear combination of constraints but a physical Hamiltonian, which relates to the internal time defined by the mimetic scalar $\phi$. Indeed the constant-$\phi$ slices define the foliation for the Hamiltonian effective dynamics. This foliation is a gauge fixing which make the covariance not manifest at the Hamiltonian level. But the covariance is manifest at the level of Lagrangian. The covariant mimetic gravity Lagrangian is formulated in 4 dimensions, and its spherical symmetry reduction results in the covariant $\bar{\mu}$-scheme effective dynamics. The covariant $\bar{\mu}$-scheme effective dynamics is a 2-dimensional field theory relating to the mimetic extension of dilaton-gravity models.

This work may be seen as a continuation from the early attempt \cite{BenAchour:2017ivq}, as well as the non-singlar black hole model from the limiting curvature hypothesis \cite{Chamseddine:2016ktu}. The covariant $\bar{\mu}$-scheme effective dynamics proposed here has the following advantage comparing to the earlier models: The Hamiltonian of spherical symmetric gravity depends on 2 components $A_1,A_2$ of the Ashtekar-Barbero connection, which gives 2 different $\bar{\mu}$-scheme holonomies. A true $\bar{\mu}$-scheme Hamiltonian of LQG should depend on both holonomies, rather than $A_1$ or $A_2$ itself. This requirement is satisfied by our covariant $\bar{\mu}$-scheme Hamiltonian $H$ but is not satisfied by the earlier models\footnote{The earlier models depend on one of $A_1, A_2$ instead of its holonomy. }.

We apply the covariant $\bar{\mu}$-scheme effective dynamics to both the homogeneous-isotropic cosmology and spherical symmetric black holes. The homogeneous and isotropic symmetry recovers the effective dynamics to the $\bar{\mu}$-scheme effective dynamics in the $K$-quantization LQC, and thus the covariant $\bar{\mu}$-scheme effective dynamics includes the LQC effective dynamics as a subsector. The effective dynamics resolves the big-bang singularity with a non-singular bounce. In the spherical symmetric effective dynamics, we impose an additional killing symmetry and the boundary condition that the spacetime from the effective dynamics should reduces to the Schwarzschild geometry at infinity. An advantage of our approach is that both the black hole exterior and interior are treated uniformly with a single set of effective Hamiltonian equations, relating to the fact that our spherical symmetric effective dynamics is a 1+1 dimensional field theory. As a result, the solution of the effective equations gives a non-singular black hole: The solution reduces to the Schwarzschild geometry in the low curvature regime and replaces the classical singularity by the non-singular Planckian curvature regime. Due to the singularity resolution, the effective dynamics extends the spacetime in the Planckian curvature regime. The spacetime approaches asymptotically to the Nariai geometry $\mathrm{dS}_2\times S^2$ at the future infinity in the interior of the black hole (this asymptotic geometry is similar to the earlier results in \cite{Han:2020uhb,Bohmer:2007wi}). The entire spacetime from the covariant $\bar{\mu}$-scheme effective dynamics is non-singular and has the complete future null infinity $\mathscr{I}^+$, which contains a spacelike part corresponding to the $\mathscr{I}^+$ of $\mathrm{dS}_2$ and a null part corresponding to the $\mathscr{I}^+$ of the Schwarzschild geometry.

The covariant $\bar{\mu}$-scheme effective dynamics is important conceptually because guarantees the general covariance of the effective theory. Moreover, the covariant $\bar{\mu}$-scheme is also important technically, because the formulation does not rely on the 3+1 decomposition and can adapt to any coordinate system. In particular, the effective dynamics can be formulated in the light-cone gauge, which is useful in the black hole model with null-shell collapse. Here we consider a 1-parameter family of 1+1 dimensional dilaton-gravity models coupled to the mimetic scalar with higher derivative interactions. These models contains both the spherical symmetry reduction of 4d gravity and the Callan-Giddings-Harvey-Strominger (CGHS) model living in 1+1 dimensions. The covariant $\bar{\mu}$-scheme effective dynamics is generalized to the family of dilaton-gravity models, and the results give a family of mimetic-dilaton-gravity Lagrangians that are covariant in 1+1 dimensions. We focus on the mimetic-CGHS model and formulate the effective dynamics in the light-cone gauge to study 1+1 dimensional black hole with null-shell collapse. In this work, we only consider the effective dynamics along the null shell and leave the full study of 1+1 dimensions to the future research (see e.g. \cite{Laddha:2006fr,Corichi:2016nkp,Eyheralde:2018htf,Bojowald:2016vlj} for some earlier results on the quantum dynamics of dilaton-gravity models). We impose the boundary condition that the spacetime reduces to the classical CGHS black hole solution at $\mathscr{I}^-$. The effective dynamics shows that both the 2d scalar curvature and the derivative of dilaton field are finite along the null shell, in contrast to the the divergence in the CGHS model \cite{Strominger:1994tn,Russo:1992ht,Ashtekar:2010qz}. In addition, in the neighborhood of the null shell, The spacetime extends to the infinity in the future null direction and finds an asymptotically flat regime there. 

The structure of this paper is summarized as the following: Section \ref{effective dynamics of spherical symmetric quantum gravity} reviews the spherical symmetry reduction of LQG and the idea of $\bar{\mu}$-scheme effective dynamics. An example of the covariant $\bar{\mu}$-scheme effective Hamiltonian is introduced in this section. Section \ref{Mimetic gravity in four dimensions} reviews the mimetic gravity Lagrangian and equations of motion in 4 dimensions. Section \ref{Spherical symmetry reduction and 2d mimetic-dilaton-gravity models} discusses the spherical symmetry reduction of mimetic gravity and introduce a family of 2d mimetic-dilaton-gravity models. We also discuss the gauge fixing that leads to the foliation with constant-$\phi$ slices. Section \ref{Hamiltonian dynamics} studies the Hamiltonian from the mimetic gravity and/or the mimetic-dilaton-gravity models. We propose the higher derivative interactions that lead to the covariant $\bar{\mu}$-scheme effective Hamiltonian. Section \ref{Application I} shows that the LQC effective dynamics can be reproduced as a subsector in the covariant $\bar{\mu}$-scheme effective dynamics. Section \ref{Application II} discusses the non-singular black hole solution from the covariant $\bar{\mu}$-scheme effective dynamics. 
Section \ref{More on the relation with mubar scheme} discusses the consistency between the effective dynamics in cosmology and black hole. The consistency picks up a unique choice of free parameters in the covariant $\bar{\mu}$-scheme. Section \ref{Mimetic-CGHS model and light-cone effective dynamics} formulates the covariant $\bar{\mu}$-scheme effective dynamics in the light-cone gauge and applies it to the mimetic-CGHS model to study the 2d black hole in presence of the null shell.

\section{Covariant $\bar{\mu}$-scheme effective dynamics of spherical symmetric quantum gravity}\label{effective dynamics of spherical symmetric quantum gravity}

In this section, we focus on the sector of spherical symmetrical degrees of freedom in LQG in the canonical formulation. The spacetime manifold is assumed to admit a 3+1 decomposition $\sm_4\simeq \R\times \cs$, where $\cs\simeq \R\times S^2$. We define the spherical coordinate ${\sig}=(x,\theta,\phi)$ on the spatial slice $\cs$. The global time coordinate is denoted by $t$. The classical phase space for LQG has the canonical variables $A_a^i,E^a_i$ ($i=1,2,3$, $a=x,\theta,\varphi$), where $A_a^i$ is the Ashtekar-Barbero connection and $E^a_i$ is the densitized triad. In spherically symmetric spacetimes, we only consider $(A_a^i,E^a_i)$ that are invariant under rotations up to gauge transformations \cite{Ashtekar:2005qt,Bojowald:2005cb,Chiou:2012pg,Gambini:2013hna,Zhang:2021xoa,Han:2020uhb} 
\be
A^j\tau_j&=&A_1(x)\tau_1\dd x+( A_2(x)\tau_2+ A_3(x)\tau_3)\dd\theta+( A_2(x)\tau_3-A_3(x)\tau_2)\sin(\theta)\dd\varphi\nonumber\\
&&+\cos(\theta)\tau_1\dd\varphi,\label{eq:AEexpress}\\
E_j\tau^j&=&E^1(x)\sin(\theta)\tau_1\partial_x+( E^2(x)\tau_2+ E^3(x)\tau_3)\sin(\theta)\partial_\theta+( E^2\tau_3-E^3\tau_2)\partial_\varphi,\nonumber
\ee
where $\tau_j=-\frac{i}{2}\sigma_j$ with $\sigma_j$ denoting Pauli matrices. The symplectic form $\Omega$ on the phase space reduces to
\be
\Omega(\delta_1,\delta_2)&=&-\frac{1}{8\pi G\beta}\int\dd^3x\, \delta_1 A_a^j(x)\wedge\delta_2 E^a_j(x)\nonumber\\
&=&-\frac{1}{2G\beta}\int\rmd x\left[\delta_1A_1(x)\wedge \delta_2 E^1(x)+2\delta_1 A_2(x)\wedge\delta_2 E^2(x)+2\delta_1A_3(x)\wedge\delta_2 E^3(x) \right],\label{eq:symplecticform}
\ee
where $\delta_1$ and $\delta_2$ are differentials on the phase space. The symmetry-reduced theory is an (1+1)-dimensional field theory with the infinite-dimensional phase space. 

The SU(2) Gauss constraint is reduced to only one constraint:
\begin{equation}\label{eq:gaussian}
G[\lambda]=4\pi\int \rmd x\,\lambda(x)\left[2A_2(x)E^3(x)-2A_3(x)E^2(x)+\partial_xE^1(x)\right].
\end{equation}
while other two components become trivial. $G[\l]$ can generate gauge transformation to make $E^3$ vanish, and thus we gauge fix 
\begin{equation}
E^3(x)=0.\label{gaugefix}
\end{equation}
Correspondingly, the Gauss constraint \eqref{eq:gaussian} is solved for $A_3(x)$
\begin{equation}
A_3(x)=\frac{\partial_xE^1(x)}{2E^2(x)}.\label{A3}
\end{equation}
Therefore $(A_3,E^3)$ is removed from the canonical pairs. Following \cite{Han:2020uhb,Zhang:2021xoa,Gambini:2013hna}, we introduce
\begin{equation}
K_x(x):=\frac{1}{2\beta}A_1(x),\quad K_\varphi(x):=\frac{1}{\beta}A_2(x),\quad  E^x(x)=E^1(x),\quad E^\varphi(x)=E^2(x).
\end{equation}
Recall that the Ashtekar-Barbero connection $A=\Gamma+\b K$, the above relation between $K$ and $A$ are due to the vanishing Levi-Civita connection $\G$ for these components. $K_x$ has been rescaled by a factor of 2 in order to make the Poisson brackets uniform. 
\begin{equation}
\label{Poisson1}
\{K_j(x),E^k(x')\}=G\delta^k_j\delta(x,x'),\quad j,k=x,\varphi.
\end{equation}
In terms of $E^x$ and $E^\varphi$, the spherical symmetric metric is given by
\be
\label{metric1}
\rmd s^2=-N(t,x)^2\rmd t^2+\frac{E^\varphi(t,x)^2}{|E^x(t,x)|}\lt[\rmd x+N^x(t,x)\rmd t\rt]^2+|E^x(t,x)|\rmd\Omega^2,
\ee
where the angular part is given by $\rmd\Omega^2=\rmd\theta^2+\sin^2\theta \rmd\varphi^2$.

The Hamiltonian $H$ of classical gravity reduced to the spherical symmetrical sector reads 
\be
H_0 &=&\int \rmd x \lt[N(t,x)\mathcal{C}(t,x) + N^x(t,x)\cc_x(t,x)\rt],\label{H0CBDY}\\
\mathcal{C} &=& \frac{1}{4G\sqrt{E^{x}}}\left(-\frac{2 E^{x} E^{x \prime} E^{\varphi \prime}}{E^{\varphi 2}}+\frac{4 E^{x} E^{x \prime \prime}+E^{x \prime 2}}{2 E^{\varphi}}-8 E^{x} K_{x} K_{\varphi}-2 E^{\varphi}\left[K_{\varphi}^{2}+1\right]\right).\nonumber\\
\cc_x&=&E^{\varphi} K_{\varphi}^{\prime}-K_{x} E^{x \prime}\nonumber
\ee
where $E^x{}'=\partial_x E^x$. Both $\cc$ and $\cc_x$ are 1st-class constraints for pure gravity. However, when we couple gravity to Gaussian dust fields and formulate the theory in the reduced phase space \cite{Han:2020uhb,Giesel:2012rb,Kuchar:1990vy}, the dust fields defines the material reference frame, and $H_0$ with $N=1,N^x=0$ is the physical Hamiltonian for the dust-time. In this case, neither $\cc$ nor $\cc_x$ is a constraint.

In LQG, the $\bar{\mu}$-scheme effective dynamics is generated by the modification of $H_0$ in terms of the $\bar{\mu}$-scheme holonomies. In the case of the spherical symmetric quantum gravity, the $\bar{\mu}$-scheme holonomies are two types of U(1) holonomies \cite{Chiou:2012pg,Gambini:2013hna,Zhang:2021xoa,Han:2020uhb}
\be
h_x&=& e^{2i\bar{\mu}_xK_x}\simeq e^{\int_{e_1}\rmd x\,A_1},\quad\bar{\mu}_x=\frac{\b\sqrt{\Delta} \sqrt{E^x}}{E^\varphi},\label{mubarh1}\\
h_\theta&=&e^{i\bar{\mu}_\theta K_\varphi}=e^{\int_{e_2}\rmd\theta\, A_2},\qquad\bar{\mu}_\theta=\frac{\b\sqrt{\Delta} }{\sqrt{E^x}}\\
h_\varphi&=&e^{\int_{e_3}\rmd\varphi\, A_2\sin(\theta)}=h_\theta.\label{mubarh3}
\ee
The $\bar{\mu}$-scheme holonomies $h_{x},h_\theta,h_\varphi$ know both $A$ and $E$. These holonomies are along the edges $e_1,e_2,e_3$ of the fixed geometrical length $\sqrt{\Delta}$ in the $x,\theta,\varphi$ directions. Indeed, assuming $A_1$ and $\bar{\mu}_x$ to be approximately constant along $e_1$, $e^{2i\bar{\mu}_xK_x}\simeq e^{\int_0^1\rmd u\bar{\mu}_xA_1}= e^{\int_{e_1}\rmd x\,A_1}$ holds in \eqref{mubarh1}, if $\int_{e_1}\rmd x\cdots =\int_0^1\rmd u\, \bar{\mu}_x \cdots$, and thus the length of $e_1$ is fixed by $\sqrt{\Delta}$
\be  
\int_{e_1}\rmd x\sqrt{g_{xx}}=\int_0^1\rmd u\, \bar{\mu}_x\sqrt{g_{xx}}=\sqrt{\Delta},
\ee
where the metric $g_{\mu\nu}$ is given by \eqref{metric1}. Similarly the length of $e_1$ and $e_3$ are also fixed by $\sqrt{\Delta}$
\be
&&\int_{e_2}\rmd\theta \sqrt{g_{\theta\theta}}=\int_0^1\rmd u\,\bar{\mu}_\theta \sqrt{g_{\theta\theta}}=\sqrt{\Delta}\\
&&\int_{e_3}\rmd\varphi \sqrt{g_{\varphi\varphi}}=\int_0^1\rmd u\,\bar{\mu}_\theta \sin(\theta) \sqrt{g_{\varphi\varphi}}=\sqrt{\Delta}
\ee
In LQG, $\Delta$ is identify to the minimal nonzero eigenvalue of the area operator. $A_1$ and $\bar{\mu}_x$ has been assumed to be approximately constant along $e_1$. It means that the effective theory neglects the fluctuation of $A_1$ in any $x$-interval of Planck length. The modification of $H_0$ in terms of the $\bar{\mu}$-scheme holonomies is often called the $\bar{\mu}$-scheme polymerization. The modified Hamiltonian is called $\bar{\mu}$-scheme effective Hamiltonian.

The simplest $\bar{\mu}$-scheme effective Hamiltonian, denoted by $H_{\rm simple}$, is obtained by applying the following simple replacement rule to $\cc$ \cite{Chiou:2012pg,Han:2020uhb}
\be
K_\varphi\to \frac{\sqrt{E^x}}{\b\sqrt{\Delta}}\sin\lt[\frac{\b\sqrt{\Delta}}{\sqrt{E^x}} K_\varphi\rt], \quad
K_x\to \frac{E^\varphi}{2\b\sqrt{\Delta}\sqrt{E^x}}\sin\lt[\frac{\b\sqrt{\Delta}\sqrt{E^x}}{{E^\varphi}}2K_x\rt].
\ee
The resulting $H_{\rm simple}$ (with $N=1,N^x=0$) as the physical Hamiltonian on the reduced phase space has been studied extensively in \cite{Han:2020uhb}. $H_{\rm simple}$ relates to the full SU(2) theory by
\be
H_{\rm simple}&=&\frac{2}{\b^2\kappa\Delta}\int\rmd^3x\sum_{j,k}e(\Box_{jk})\mathrm{Tr}\lt(h_{\Delta}(\Box_{jk})\frac{[E^j,E^k]}{\sqrt{\det(q)}}\rt)+\text{terms independent of}\ K.
\ee
where $\Box_{jk}$ denotes the plaquette of the fixed geometrical area $\Delta$ in the $(j,k)$-plane. $e(\Box_{jk})$ denotes the area element on $\Box_{jk}$. $h_{\Delta}(\Box_{jk})= h_{\Delta}^{(j)}h_{\Delta}^{(k)}(h_{\Delta}^j)^{-1}(h_{\Delta}^{(k)})^{-1}$, $j,k=x,\theta,\varphi$, is the SU(2) loop holonomy around $\Box_{jk}$. The loop holonomy regularizes the curvature of the Ashtekar-Barbero connection by $ F_{jk}\simeq\frac{1}{\Delta}[h_{\Delta}(\Box_{jk})-1]$. $h_{\Delta}^{(j)} \in \Su$ are the representation of $h_j\in \mathrm{U}(1)$ acting on the fundamental representation of SU(2):
\be
&&h_{\Delta}^{(x)}= e^{2\bar{\mu}_x K_x\t_1},\quad h_{\Delta}^{(\theta)}=e^{\bar{\mu}_\theta K_\varphi\t_2},\quad h_{\Delta}^{(\varphi)} =e^{\bar{\mu}_\theta K_\varphi \t_3}.
\ee
It is manifest that $\Delta\to 0$ reduces $H_{\rm simple}$ to $H_0$ with $N=1,N^x=0$. The correction in $H_{\rm simple}$ to $H_0$ is called the holonomy correction.

This paper mainly focuses on the new $\bar{\mu}$-scheme polymerization called the covariant $\bar{\mu}$-scheme polymerization. This polymerization gives the effective Hamiltonian $H$,
\be
H=\int\rmd x\lt[ N(t)\cc_{\Delta}(t,x)+N^x(t,x) \cc_x(t,x)\rt]
\ee
where the lapse function $N$ only depends on $t$ and $\cc_\Delta$ reads
\be
\cc_\Delta&=&\frac{\sqrt{E^x} E^\varphi}{2G \b^2\Delta}\lt[{ \sin ^2\left(\frac{2\b  \sqrt{\Delta } \sqrt{E^x} K_x }{   E^\varphi }\right)}-4\,{ \sin ^2\left(\frac{\b\sqrt{\Delta }\sqrt{E^x}}{ E^\varphi }K_x+\frac{ \b \sqrt{\Delta } }{2\sqrt{E^x} }K_\varphi\right)}\rt]\nonumber\\
&&+\, \frac{1}{4G\sqrt{E^x}}\Bigg(-\frac{2 E^{x} E^{x \prime} E^{\varphi \prime}}{E^{\varphi 2}}+\frac{4 E^{x} E^{x \prime \prime}+E^{x \prime 2}}{2 E^{\varphi}}-2 E^{\varphi}\Bigg).\label{ccDelta00}
\ee
There exists a covariant Lagrangian behind the Hamiltonian $H$, so the effective dynamics generated by $H$ is covariant. This is the reason why it is called the covariant $\bar{\mu}$-scheme. The covariant Lagrangian is the mimetic gravity Lagrangian with the prescribed higher derivative interactions. The field content in the Lagrangian includes a scalar field $\phi$ in addition to the gravitational field. The scalar field $\phi$ serves as the physical time, and correspondingly $\cc_\Delta$ is not a constraint and $H$ is the physical Hamiltonian. The discussion of the mimetic gravity and the derivation of $H$ from the Lagrangian are given in Sections \ref{Mimetic gravity in four dimensions}, \ref{Spherical symmetry reduction and 2d mimetic-dilaton-gravity models}, and \ref{Hamiltonian dynamics}. The covariant $\bar{\mu}$-scheme Hamiltonian $H$ gives further correction in terms of the holonomies $h_x,h_\theta,h_\varphi$ in addition to the holonomy correction in $H_{\rm simple}$. This correction is necessary to make the effective dynamics covariant.

Note that the simple $\bar{\mu}$-scheme effective dynamics with $H_{\rm simple}$ studied in \cite{Han:2020uhb} is also manifestly covariant, since it is formulated in the reduced phase space and in terms of the Dirac observables. But a covariant Lagrangian is missing for $H_{\rm simple}$. In contrast,  $H$ with \eqref{ccDelta00} has the advantage of having a covariant Lagrangian, which turns out to be useful for going beyond the canonical formulation of the effective dynamics (see Section \ref{Mimetic-CGHS model and light-cone effective dynamics}).

The spacetime manifold in this paper has the boundary at infinity, so the boundary conditions and boundary terms in $H$ need to be discussed. The boundary term in terms of Ashtekar variables in the case of asymptotically flat spacetimes has been discussed in the literature e.g. \cite{thiemann1995generalized,Corichi:2013zza,Campiglia:2014yja}. In the following we briefly discuss how the boundary term for $H$ can be obtained. We set $N^x\to 0$ at the boundary. The procedure and result are similar to the discussion in \cite{Han:2020uhb} for $H_{\rm simple}$. Indeed, when deriving EOMs from $H$, the variation $\delta H$ and the integration by part result in the following boundary terms
\begin{eqnarray}
N(t)\lt[\frac{8 \pi  E^x{} \delta E^x{}'{}}{\kappa  \sqrt{\left| E^x{}\right| } \left| E^\varphi{}\right| }-\frac{8 \pi  E^x{} \delta E^\varphi {} \left| E^\varphi{}\right|  E^x{}'{}}{\kappa  E^\varphi{}^3 \sqrt{\left| E^x{}\right| }}\rt].\label{bdyterm}
\end{eqnarray}
The following boundary conditions will play the roles in our analysis.

\begin{itemize}

\item When we study the dynamics of spherical symmetric black hole in Section \ref{Application II}, we consider $E^x,E^\varphi$ to behave asymptotically as the Schwarzschild geometry in the Lema\^{\i}tre coordinates as $x\to\infty$ \footnote{The Schwarzschild spacetime in the Lema\^{\i}tre coordinates $(t,x,\theta,\varphi)$ is given by \eqref{metric1} with $E^x= \left(\frac{3}{2} \sqrt{R_s}\, (x-t)\right)^{4 / 3},\  E^\varphi=\sqrt{{R_s}} \lt(\frac{3}{2}{\sqrt{{R_s}}\, (x-t)}\rt)^{1/3}$. }:
\begin{eqnarray}
E^x|_{bdy}\sim \left(\frac{3}{2} \sqrt{R_s}\, x\right)^{4 / 3},\quad E^\varphi|_{bdy}\sim  \sqrt{{R_s}} \lt(\frac{3}{2}{\sqrt{{R_s}}\, x}\rt)^{1/3}, \label{bdyschw1}
\end{eqnarray}
where $R_s$ is the Schwarzschild radius. The boundary condition satisfies $E^x{}'=2E^\varphi$ and thus $\delta E^x{}'=2\delta E^\varphi$ asymptotically. The boundary term \eqref{bdyterm} vanishes at $x\to\infty$.

\item The Neumann boundary condition $E^x{}'|_{bdy}= 0,\ \delta E^x{}' |_{bdy}=0 $ appears in Section \ref{Application II} as $x\to-\infty $. Both terms in \eqref{bdyterm} vanish by this boundary condition.

\end{itemize}

\section{Mimetic gravity in four dimensions}\label{Mimetic gravity in four dimensions}

The mimetic gravity provides the manifestly covariant Lagrangian for the covariant $\bar{\mu}$-scheme effective Hamiltonian. The field content of the mimetic gravity has the gravity $g_{\mu\nu}$ and a scalar field $\phi$, as well as a lagrangian multiplier $\l$. The extended mimetic gravity action on a 4-manifold $\sm_4$ reads
\begin{eqnarray}
\label{GA}
S[g_{\mu\nu},\phi,\lambda] =\frac{1}{8\pi G} \int_{\sm_4} \rmd^4x \, \sqrt{-g} \, \left[ \frac{f(\phi)}{2} \, {\cal R}^{(4)} + 
\, L_\phi(\phi,\chi_1,\cdots,\chi_p) \, + \, \lambda(\mathscr{X} + 1)\right]\, ,
\end{eqnarray}
where
\bea
\label{chin}
\mathscr{X}=\phi_\mu\phi^\mu,\quad\chi_n \equiv \sum_{\mu_1,\cdots,\mu_n} \phi_{\mu_1}^{\mu_2} \, \phi_{\mu_2}^{\mu_3} \cdots \phi_{\mu_{n-1}}^{\mu_n} \, \phi_{\mu_n}^{\mu_1},\quad\phi_\mu=\nabla_\mu\phi,\quad  \phi_{\mu\nu}=\nabla_\mu\nabla_\nu\phi.
\eea

The variation with respect to $\l$ gives the mimetic constraint
\be
\delta_\l S=0\quad \Longleftrightarrow \quad \mathscr{X}+1=0.\label{eomX}
\ee
For all $\phi$ satisfying the mimetic constraint, the constant $\phi$ surfaces are all spacelike. If the manifold $\sm_4$ admits a global foliation such that $\phi$ is constant on every slice, $\phi$ is a global time function on $\sm_4$. Then $\phi$ can serve as a clock field defining the internal time of the system, similar to the situation of deparametrizing gravity by coupling to dust or scalar fields \cite{Giesel:2007wi,Domagala:2010bm,Giesel:2012rb,Brown:1994py}. Indeed if $f(\phi)=1$ and $L_\phi(\phi,\chi_1,\cdots,\chi_p)=0$, $S$ reduces to the case of the Einstein gravity coupled to a single component dust field (comparing to e.g. \cite{Giesel:2007wi}).

The mimetic potential $L_\phi(\phi,\chi_1,\cdots,\chi_p)$ gives the higher-derivative coupling between $g_{\mu\nu}$ and $\phi$. Here $\phi$ plays the dual role of (1) being the clock field and (2) modifying the Einstein gravity by adding higher-derivative interactions, which turns out to result in the covariant $\bar{\mu}$-scheme polymerization at the Hamiltonian level.

Here we make the following choice for simplification: 
\be
f(\phi)=1, \quad L_{\phi}=L_{\phi}\lt(\chi_1,\chi_2\rt), \quad \chi_1=\Box\phi,\quad \chi_2=\phi_{\mu\nu}\phi^{\mu\nu}. 
\ee
That $L_\phi$ only depends on $\chi_1$ and $\chi_2$ turns out to be a convenient choice for the spherical symmetric dynamics. The higher-derivation coupling in $L_\phi$ turns out to be responsible for the covariant $\bar{\mu}$ polymerization. As is shown below, $\chi_1,\chi_2$ relates to two independent components of extrinsic curvatures of the constant $\phi$ slice in the spherical symmetric spacetime. We leave $L_\phi$ as an arbitrary function at this moment, and its explicit expression will be determined later.

Given the above simplification, the variational principle $\delta S=0$ gives the following equations, in addition to the mimetic constraint
\be
\delta_\phi S=0,\quad  &\Longleftrightarrow&\quad-2\nabla_{\mu}\lambda\phi^{\mu}+\nabla^{\mu}\nabla_{\mu}\frac{\partial L_{\phi}}{\partial\chi_{1}}+2\nabla_{\mu}\nabla_{\nu}\left(\frac{\partial L_{\phi}}{\partial\chi_{2}}\phi^{\mu\nu}\right)=0,\label{eomphi}\\
\delta_g S=0\quad  &\Longleftrightarrow&\quad G_{\mu\nu}+2\lambda\phi_{\mu}\phi_{\nu}- T^{\phi}_{\mu\nu}=0,\label{eomgrav}
\ee
where 
\be
T^{\phi}_{\mu\nu}&=&g_{\mu\nu}L_{\phi}+\left(-2\frac{\partial L_{\phi}}{\partial\chi_{1}}\nabla_{(\mu}\phi_{\nu)}+\nabla^{\alpha}\left[\frac{\partial L_{\phi}}{\partial\chi_{1}}\left(g_{\alpha\mu}\phi_{\nu}+g_{\alpha\nu}\phi_{\mu}-g_{\mu\nu}\phi_{\alpha}\right)\right]\right)\nonumber\\
&&+2\left(-2\frac{\partial L_{\phi}}{\partial\chi_{2}}\phi_{\mu}^{\alpha}\phi_{\alpha\nu}+\nabla^{\alpha}\left[\frac{\partial L_{\phi}}{\partial\chi_{2}}\left(\phi_{\alpha\mu}\phi_{\nu}+\phi_{\alpha\nu}\phi_{\mu}-\phi_{\mu\nu}\phi_{\alpha}\right)\right]\right).
\ee
The trace of Eq.\eqref{eomgrav} can be used for solving $\l$
\be
\l=-\frac{1}{2}\lt(R+T^\phi\rt),\label{lambdaRT}
\ee
where $T^\phi$ is the trace of $T^\phi_{\mu\nu}$. The equation of motion for $\phi$ in \eqref{eomphi} is not independent, but is implied by the Einstein equation \eqref{eomgrav}, $\nabla^\mu G_{\mu\nu}=0$, and the mimetic constraint. The independent equations from $\delta S=0$ are the mimetic constraint \eqref{eomX} and the Einstein equation \eqref{eomgrav} with \eqref{lambdaRT} inserted.

\section{Spherical symmetry reduction and 2d mimetic-dilaton-gravity models}\label{Spherical symmetry reduction and 2d mimetic-dilaton-gravity models}

\subsection{Symmetry reduction}

In this paper, we mainly focus on gravity with spherical symmetry. We assume $\sm_4=\sm_2\times S^2$ and the general spherical symmetric metric reads
\be
\rmd s^{2}=h_{ij}(t,x)\rmd x^i\rmd x^j+E^{x}(t,x) \lt( \rmd \theta^{2}+\sin ^{2} \theta \rmd \varphi^{2}\rt)
\ee
We denote by $h_{ij}$ the 2d metric  
\be
h_{ij}\rmd x^i\rmd x^j=-N(t,x) ^{2} \rmd t^{2}+\frac{E^{\varphi}(t,x)^{2}}{E^{x}(t,x) }\left[\rmd x+N^{x}(t,x)  \rmd t\right]^{2}
\ee
The fields $E^x,E^\varphi,N,N^x$, as well as the $\phi,\l$ in the mimetic action, are assumed independent of $\theta,\varphi$. 

We introduce the dilaton field $\psi=\frac{1}{2}\log\lt(E^x\rt)$ The symmetry reduction of $S$ gives the following 2d action
\be
S_2&=&\frac{1}{4G}\int_{\sm_2} \rmd^2 x\,\sqrt{-h}\lt\{ e^{2\psi}\left(R_{h}+2h^{ij}\partial_{i}\psi\partial_{j}\psi\right)+2+\, e^{2\psi}\lt[ L_{\phi}\left(\chi_1,\chi_2\right)+\lambda\left(\mathscr{X}+1\right)\rt]\rt\},
\ee
where $R_h$ is the 2d scalar curvature, and 
\be
\mathscr{X}=\phi_{j}\phi^{j},\quad
\chi_{1}=\Box_{h}\phi+2h^{ij}\partial_{i}\psi\partial_{j}\phi,\quad \chi_{2}=\phi_{ij}\phi^{ij}+2\left[h^{ij}\partial_{i}\psi\partial_{j}\phi\right]^{2},\label{chi2d}
\ee

Eqs.\eqref{chi2d} relates $\chi_1,\chi_2$ to three 2d quantities $\Box_{h}\phi$, $h^{ij}\partial_{i}\psi\partial_{j}\phi$, and $\phi_{ij}\phi^{ij}$. However, we show that $\phi_{ij}\phi^{ij}=(\Box_{h}\phi)^2$ on the constraint surface $\mathscr{X}+1=0$, so $\chi_1,\chi_2$ are functions of only $\Box_{h}\phi$ and $h^{ij}\partial_{i}\psi\partial_{j}\phi$. Indeed, we check the relation $\phi_{ij}\phi^{ij}=(\Box_{h}\phi)^2$ explicitly in the light-cone coordinate $(u,v)$, where the 2d metric is written as $h_{ij}\rmd x^i\rmd x^j= -e^{2 \omega (u,v)}\text{d}u \text{d}v$. In this coordinate, $\mathscr{X}+1=0$ is solved by $e^{2\omega (u,v)}=4 \partial_v\phi\, \partial_u\phi$. Apply this relation to compute $\Box_{h}\phi$ and $\phi_{ij}\phi^{ij}$, we obtain
\be
\phi_{ij}\phi^{ij}=\lt(\Box_{h}\phi\rt)^2=\lt(\frac{\partial_u\partial_v\phi }{\partial_v\phi \,\partial_u\phi }\rt)^2.
\ee
Since both $\phi_{ij}\phi^{ij}$ and $\Box_{h}\phi$ are scalars, whose values are coordinate independent, the validity of the relation $\phi_{ij}\phi^{ij}=\lt(\Box_{h}\phi\rt)^2$ is coordinate-independent.

We have $\chi_1,\chi_2$ as functions of two 2d quantities $\Box_{h}\phi$ and $h^{ij}\partial_{i}\psi\partial_{j}\phi$, in particular
\be
\chi_{2}=\lt(\Box_{h}\phi\rt)^2+2\left(h^{ij}\partial_{i}\psi\partial_{j}\phi\right)^{2}.\label{chi2in2d}
\ee
By this relation and \eqref{chi2d}, $L_\phi$ in the 2d action can be understood as a function of $\Box_{h}\phi$ and $h^{ij}\partial_{i}\psi\partial_{j}\phi$:
\be
L_\phi(\chi_1,\chi_2)=L'_\phi(\Box_{h}\phi,\, h^{ij}\partial_{i}\psi\partial_{j}\phi).
\ee
Although any function $L_\phi(\chi_1,\chi_2)$ can be understood as a function of $\Box_{h}\phi$ and $h^{ij}\partial_{i}\psi\partial_{j}\phi$, the inverse is nontrivial, because the squares in \eqref{chi2in2d} result in that solving $\Box_{h}\phi$ and $h^{ij}\partial_{i}\psi\partial_{j}\phi$ as functions of $\chi_1,\chi_2$ involves in square-roots and non-unique solutions. 
\be
\Box_{h}\phi=\frac{1}{3} \left(\chi_1-\sqrt{2} \sqrt{3 \chi_2-\chi_1^2}\right),\quad h^{ij}\partial_{i}\psi\partial_{j}\phi=\frac{1}{6} \left(2 \chi_1+\sqrt{2} \sqrt{3 \chi_2-\chi_1^2}\right),\\
\text{or}\quad \Box_{h}\phi=\frac{1}{3} \left(\chi_1+\sqrt{2} \sqrt{3 \chi_2-\chi_1^2}\right),\quad h^{ij}\partial_{i}\psi\partial_{j}\phi=\frac{1}{6} \left(2 \chi_1-\sqrt{2} \sqrt{3 \chi_2-\chi_1^2}\right).
\ee
The space of $\Box_{h}\phi$ and $h^{ij}\partial_{i}\psi\partial_{j}\phi$ is the double-cover of the space of $\chi_1,\chi_2$. So the space of functions $L_\phi(\chi_1,\chi_2)$ is not equivalent to the space of $L'_\phi(\Box_{h}\phi,\, h^{ij}\partial_{i}\psi\partial_{j}\phi)$, which is defined on the double-cover.

In either 4d or 2d, we can lift the mimetic potential to the double-cover of $\chi_1,\chi_2$ and consider $L'_\phi$ in the action instead of $L_\phi$. In 2d, we have the explicit parametrization of the double-cover by $\Box_{h}\phi$ and $h^{ij}\partial_{i}\psi\partial_{j}\phi$, so $L'_\phi=L'_\phi(\Box_{h}\phi,\, h^{ij}\partial_{i}\psi\partial_{j}\phi)$. By this setup, the 2d action of the spherical symmetric mimetic gravity is given by 
\be
S_2=\frac{1}{4G}\int_{\sm_2} \rmd^2 x &\sqrt{-h}&\Big\{ e^{2\psi}\left(R_{h}+2h^{ij}\partial_{i}\psi\partial_{j}\psi\right)+2\nonumber\\
&&\quad +\ e^{2\psi}\lt[ L'_\phi(\Box_{h}\phi,\, h^{ij}\partial_{i}\psi\partial_{j}\phi)+\lambda\left(\mathscr{X}+1\right)\rt]\Big\}.\label{S2000}
\ee

We introduce the variables $X,Y$ which relate $\Box_{h}\phi,\, h^{ij}\partial_{i}\psi\partial_{j}\phi$ by
\be
X=-\Box_{h}\phi-h^{ij}\partial_{i}\psi\partial_{j}\phi, \qquad Y=-h^{ij}\partial_{i}\psi\partial_{j}\phi.\label{XYab}
\ee
The space of functions of $\Box_{h}\phi,\, h^{ij}\partial_{i}\psi\partial_{j}\phi$ are equivalent to the space of functions of $X,Y$. We set
\be
L'_\phi(\Box_{h}\phi,\, h^{ij}\partial_{i}\psi\partial_{j}\phi)=\tilde{L}(X,Y).
\ee

\subsection{2d mimetic-dilaton-gravity models}

$S_2$ in \eqref{S2000} is a 2d dilaton-gravity model with the mimetic scalar field $\phi$ and the higher-derivative coupling, although it is derived from the 4d mimetic gravity. Here we modify \eqref{S2000} by including a 1-parameter deformation labelled by $\eta$
\be
S^{(\eta)}_2=\frac{1}{4G}\int_{\sm_2} \rmd^2 x &\sqrt{-h}&\Big\{ e^{2\psi}\left[R_{h}+2(1+\eta)h^{ij}\partial_{i}\psi\partial_{j}\psi\right]+2(1+\eta)e^{2\eta\psi}\nonumber\\
&&\quad +\ e^{2\psi}\lt[\tilde{ L}(X,Y)+\lambda\left(\mathscr{X}+1\right)\rt]\Big\}.\label{dilatongravity}
\ee
When $\tilde{L}=\lambda=0$, the parameter $\eta$ labels the continuous deformation from the spherical symmetric 4d gravity at $\eta=0$ to the 2d CGHS dilaton-gravity model at $\eta=1$. $S^{(\eta)}_2$ at $\eta=1$ couples the mimetic scalar field $\phi$ to the CGHS model in 2d, with the mimetic constraint $\mathscr{X}+1=0$ and the higher-derivative coupling $\tilde{L}(X,Y)$. $S^{(\eta)}_2$ defines a continuous family of 2d mimetic-dilation-gravity models deforming from the symmetry reduction of 4d mimetic gravity to the mimetic-CGHS model. Our following discussion mostly keeps $\eta$ arbitrary, and we fix $\eta=0$ or $\eta=1$ only when extracting results specifically for spherical symmetric 4d gravity or the mimetic-CGHS model.

The lagrangian analysis of $S_2^{(\eta)}$ closely resembles the mimetic gravity in 4 dimensions. The follows are equations of motion from the variational principle
\be
\delta_\l S_2^{(\eta)}=0\quad &\Longleftrightarrow&\quad\nabla_j\phi\nabla^j\phi+1=0,\label{mimeticin2d}\\
\delta_\phi S_2^{(\eta)}=0\quad &\Longleftrightarrow&\quad \Box\xi_{1}-\nabla_{j}\left(\xi_{2}\partial^{j}\psi\right)-2\nabla_{j}\left(\lambda e^{2\psi}\phi^{j}\right)=0,\label{variphi2d}\\
\delta_\psi S_2^{(\eta)}=0\quad &\Longleftrightarrow&\quad 2e^{2\psi}\left[R_{h}-2\left(1+\eta\right)\partial^{j}\psi\partial_{j}\psi-2\left(1+\eta\right)\Box\psi+2\eta\left(1+\eta\right)e^{2\left(\eta-1\right)\psi}+L'_\phi \right]\nonumber\\
&&\qquad-\,\nabla_{j}\left(\xi_{2}\phi^{j}\right)=0,\label{varipsi2d}\\
\delta_{h^{ij}} S_2^{(\eta)}=0\quad &\Longleftrightarrow&\quad e^{2\psi}\left[-2\nabla_{i}\nabla_{j}\psi+2h_{ij}\Box\psi+\left(3-\eta\right)h_{ij}\partial^{k}\psi\partial_{k}\psi-\left(1+\eta\right)e^{2\left(\eta-1\right)\psi}h_{ij}\right]\nonumber\\
&&\qquad +\,2 e^{2\psi}\left(\eta-1\right)\partial_{i}\psi\partial_{j}\psi+e^{2\psi}\lambda\phi_{i}\phi_{j}-\frac{1}{2}h_{ij}e^{2\psi}L'_\phi+\xi_{2}\partial_{(i}\psi\partial_{j)}\phi\nonumber\\
&&\qquad\,-\frac{1}{2}\left\{-2\xi_{1}\nabla_{(i}\phi_{j)}+\nabla^{k}\left[\xi_{1}\left(h_{ki}\phi_{j}+h_{kj}\phi_{i}-h_{ij}\phi_{k}\right)\right]\right\}=0,\label{varihij}
\ee
where the covariant derivatives are in 2d, and we have defined $\xi_1=e^{2\psi}L'_{\phi}{}^{(1,0)}$, $\xi_2=e^{2\psi} L'_{\phi}{}^{(0,1)}$. Eqs.\eqref{varipsi2d} and \eqref{varihij} reduces the Einstein equation to 2d by spherical symmetry when $\eta=0$. Eq.\eqref{variphi2d} from the variation of $\phi$ is again redundant, because it is implied by $\nabla^i$ acting on \eqref{varihij} (contracting $i$ index) and \eqref{varipsi2d}, as well as the mimetic constraint.

\subsection{Gauge fixing and foliation}\label{Gauge fixing and foliation}

Recall that the mimetic constraint implies the the constant-$\phi$ slice is spacelike, and thus the mimetic scalar field $\phi$ can serve as the clock field defining the internal time. Reducing to $2d$, we assume there exists a foliation $\sm_2\simeq \Sig\times \R$, such that $\phi$ is constant on every 1d curve $\Sig$. Then generally $\phi=\phi(t)$ where $t$ is any global time function associated to the foliation. In this foliation, $\mathscr{X}+1=0$ implies $\phi_j=(N,0)$, where the lapse function $N=N(t)=\dot{\phi}(t)$ is a function of $t$ only. We are allowed to set the time function $t=\phi$, then the lapse function $N=1$.

The condition $\phi=\phi(t)$, as a gauge fixing for the diffeomorphism invariant in either 2d or 4d, does not restrict any physical degrees of freedom. Indeed, given any globally smooth field $\phi$ (in particular $\nabla_\mu\phi$ is defined globally), the foliation can always be obtained by defining the $\Sigma$ to have constant $\phi$. Since the equation of motion for $\phi$, \eqref{eomphi} or \eqref{variphi2d}, is redundant, $\phi$ is only involved in the mimetic constraint and the Einstein equation. The restriction of $\phi$ is mild. Indeed, we can insert any $\phi=\phi(t)$ into the Einstein equation to solve for $g_{\mu\nu}$. This is also equivalent to inserting $\phi=\phi(t)$ in the action $S_2^{(\eta)}$ to reduce $S_2^{(\eta)}$ to the gauge-fixed action $\tilde{S}_2^{(\eta)}$, then performing the variation of $\tilde{S}_2^{(\eta)}$ and solving $\delta\tilde{S}_2^{(\eta)}=0$.

Let us derive the gauge-fixed action $\tilde{S}_2^{(\eta)}$. Firstly, The gauge-fixing condition reduces Eqs.\eqref{XYab} to the following simple relations
\be
X=\frac{\dot{E}^{\varphi}-(N^x {{E}^{\varphi}})'}{NE^{\varphi}}, \qquad Y=\frac{\dot{E}^{x}-N^x {{E}^{x}}'}{2 N E^x},\label{XYEN1}
\ee
The right-hand sides relates to the extrinsic curvatures of the constant-$\phi$ slice. Eqs.\eqref{XYEN1} shows that $X,Y$ are the same as the ones in \cite{BenAchour:2017ivq} (see Eqs.(4.29) there). It is useful to solve for $\dot{E}^x,\dot{E}^\varphi$
\be
\dot{E}^x= E^x{}' N^x+{2}N E^x  Y,\qquad \dot{E}^\varphi=(E^\varphi{} N^x)'+N E^\varphi  X.\label{Edot}
\ee
We insert the gauge-fixing condition $\phi=\phi(t)$ in ${S}_2^{(\eta)}$. The relations $\phi_j=(N,0)$, \eqref{XYEN1} and \eqref{Edot} reduce $S_2$ to the following expression
\be
\tilde{S}^{(\eta)}_2&=&\frac{1}{2G}\int \rmd t\rmd x \, N \, E^{\varphi}\sqrt{E^x} \left\{ -\lt[2 XY-(1-\eta)Y^2\rt] + \tilde{L}(X,Y)+\frac{1}{2} R_{\eta}^{(3)} \right\},\label{GAA}
\ee
Here $N=\dot{\phi}(t)$ must be understood as the external field in $\tilde{S}_2$, since it is determined by the gauge-fixing condition. $X,Y$ are understood as \eqref{XYEN1} in $\tilde{S}_2$. The dynamical fields in $\tilde{S}_2$ are $E^x,E^\varphi,N^x$. $R^{(3)}_{\eta}$ depends only on $E^x,E^\varphi$ and their spatial derivatives
\be
R^{(3)}_{\eta}=\frac{2 E^x{}' E^\varphi{}'}{E^\varphi{}^3}-\frac{(1-\eta ) E^x{}^\prime{}^2}{2 E^x E^\varphi{}^2}-\frac{2 E^x{}^{\prime\prime}}{E^\varphi{}^2}+\frac{2 (\eta +1) }{E^x{}^{1-\eta}}
\ee
and $R^{(3)}_{\eta=0}$ is the scalar curvature of the 3d spatial metric
\be
\rmd s_{(3)}^{2}=\frac{\left(E^{\varphi}\right)^{2}}{E^{x}}\rmd x^{2}+E^{x} \lt( \rmd \theta^{2}+\sin ^{2} \theta \rmd \varphi^{2}\rt).
\ee

One can check explicitly that the variations of $\tilde{S}_2$ with respect to the dynamical variables $E^x,E^\varphi,N^x$ reproduce the same equations of motion as from variating $S_2$ followed by the gauge-fixing,
\be
\delta_{E^x,E^\varphi,N^x} \tilde{S}^{(\eta)}_2=\delta_{E^x,E^\varphi,N^x} S^{(\eta)}_2\Big|_{\phi=\phi(t),\phi_j=(N,0)}\ .\label{eomS2p}
\ee
Namely the gauge-fixing $\phi=\phi(t)$ commutes with the variation of the action with respect to $E^x,E^\varphi,N^x$. This is a consequence from the redundancy of $\delta_\phi S_2^{(\eta)}$.

$N$ is not dynamical in $\tilde{S}^{(\eta)}_2$, so $\delta_N S^{(\eta)}_2=0$ cannot be reproduced from $\tilde{S}^{(\eta)}_2$, but before the gauge-fixing, $\delta_N S^{(\eta)}_2=0$ is only used to solve the lagrangian multiplier $\l$, while $\tilde{S}^{(\eta)}_2$ is independent of $\l$. It is closely related to the fact that the trace of the Einstein equation is used to solve for $\l$ (see \eqref{lambdaRT}), and there is no Hamiltonian constraint after the gauge-fixing, as to be seen in a moment.

$\tilde{S}^{(\eta)}_2$ is not manifestly covariant, simply because it is based on the gauge-fixing $\phi=\phi(t)$. But the equations of motion from $\tilde{S}^{(\eta)}_2$ are identical to the ones from $S^{(\eta)}_2$, which is manifestly generally covariant in 2d. The equations based on the foliation with $\phi=\phi(t)$ does not contradict with the fact that the theory is generally covariant.

\section{Hamiltonian formulation of $\bar{\mu}$-scheme effective dynamics}\label{Hamiltonian dynamics}

\subsection{Legendre transformation and the construction of mimetic potential}

We apply the Hamiltonian analysis to $\tilde{S}_2^{(\eta)}$. The Hamiltonian equations reduce the 2nd order equations of motion from the Lagrangian theory to a set of 1st order differential equations, which are suitable for the initial value problem. 

In order to perform the Legendre transformation, we obtain the momenta conjugated to $E^{x}$, $E^{\varphi}$, and $N^x$ by
\begin{eqnarray}
  \pi_{\varphi} & = & 2G \frac{\delta S_2'}{\delta \dot{E}^{\varphi}} =  \sqrt{E^x} \lt( \partial_X\tilde{L}-2 Y\rt) \, ,\label{mom1}\\
  \pi_x & = &2G \frac{\delta S_2'}{\delta \dot{E}^{x}} =\frac{E^\varphi }{2 \sqrt{E^x}}\lt( \partial_{Y}\tilde{L}-2X+2(1-\eta)Y\rt) \, ,\label{mom2}\\ 
  \pi_{N^x}&=&  \frac{\delta S_2'}{\delta \dot{N}^{x}} =0\, .\label{mom3}
  \end{eqnarray}
The non-vanishing Poisson brakets are
\begin{eqnarray}
\{ E^x(x), \pi_x(x') \} & = & \{ E^{\varphi}(x), \pi_{\varphi}(x') \} =2G\delta(x,x')
\end{eqnarray}
The vanishing $\pi_{N^x}$ in \eqref{mom3} gives the primary constraint.

The Legrandre transformation is the inverse of \eqref{mom1} and \eqref{mom2} and expresses $X,Y$ in terms of $\pi_\varphi,\pi_x$, and it needs the explicit expression of $\tilde{L}$. In the following, we construct $\tilde{L}$ that corresponds to the covariant $\bar{\mu}$-scheme: Firstly, we introduce the following matrix notations 
\be
\mathbf{p}=\left(\begin{array}{c}
  \frac{1}{\sqrt{E^{x}}}\pi_{\varphi}\\
  \frac{2\sqrt{E^{x}}}{E^{\varphi}}\pi_{x}
  \end{array}\right),\quad \mathbf{q}=\left(\begin{array}{c}
    X\\
    Y
    \end{array}\right),\quad \mathbf{A}=\left(\begin{array}{cc}
      0 & 2\\
      2 & 2(\eta-1)
      \end{array}\right)
\ee
Eqs.\eqref{mom1} and \eqref{mom2} can be written as 
\be
\mathbf{p}=\nabla_{\bf q}\tilde{L}-\mathbf{A}\cdot\mathbf{q}
\ee
We consider the linear transformation $\mathbf{B}\in\mathrm{GL}(2,\R)$ acting on $q^a$
\be
\mathbf{q}\mapsto \mathbf{u}=\mathbf{B}\cdot\mathbf{q}\ ,\qquad \mathbf{B}=\left(\begin{array}{cc}
  a & b\\
  c & h
  \end{array}\right),
\ee
where $a,b,c,h$ are parameters that are constant on the spacetime. Our aim is to find $\mathbf{B}$ making \eqref{mom1} and \eqref{mom2} decouple. Indeed, the transformation leads to
\be
(\mathbf{B}^{-1})^{T}{\bf p}=\nabla_{\bf u}\tilde{L}-(\mathbf{B}^{-1})^{T}\cdot \mathbf{A}\cdot \mathbf{B}^{-1}\cdot {\bf u}.\label{BpABu}
\ee
Two equations in \eqref{BpABu} decouple when $(\mathbf{B}^{-1})^{T}\cdot \mathbf{A}\cdot \mathbf{B}^{-1}$ is a diagonal matrix, which occurs when
\be
b=a\left(-\frac{h}{c}+\eta-1\right).
\ee
In this case, we denote the diagonals by $\gamma_1$ and $\g_2$
\be
(\mathbf{B}^{-1})^{T}\cdot \mathbf{A}\cdot \mathbf{B}^{-1}=\left(\begin{array}{cc}
  \gamma_{1} & 0\\
  0 & \gamma_{2}
  \end{array}\right),\quad \gamma_{1}=\frac{2c}{a^{2}(c(\eta-1)-2h)},\quad\gamma_{2}=\frac{-2}{c(c(\eta-1)-2h)}
\ee
and we denote by
\be
&&(\mathbf{B}^{-1})^{T}{\bf p}=\left(\begin{array}{c}
  P_U\\
  P_V
  \end{array}\right),\qquad \mathbf{u}=\left(\begin{array}{c}
    U\\
    V
    \end{array}\right),\\
&& U=a X+a\left(-\frac{h}{c}+\eta-1\right)Y,\qquad V=c X+hY,\label{UVXY}\\
&& P_U=-\frac{h \pi_\varphi }{\sqrt{E^x} a (c(\eta-1)-2h)} +\frac{2 c \sqrt{E^x} \pi_x}{E^\varphi a (c(\eta-1)-2h)},\label{PU0}\\
&& P_V=-\frac{2  \sqrt{E^x} \pi_x}{E^\varphi (c(\eta-1)-2h)}+\frac{ \pi_\varphi  \left(-\frac{h}{c}+\eta -1\right)}{\sqrt{E^x} (c(\eta-1)-2h)}.\label{PV0}
\ee
The transformation from $(\mathbf{q},\mathbf{p})$ to $(U,V,P_U,P_V)$ is a $4\times 4$ symplectic matrix. The transformation results in that Eq.\eqref{BpABu} becomes decoupled
\be
P_U=\partial_U\tilde{L}-\gamma_1 U,\qquad P_V=\partial_V\tilde{L}-\gamma_2 V\label{PUPV}
\ee
In the limit that the higher-derivative coupling in the mimetic action is turned off: $\tilde{L}\to 0$, we have
\be
P_U\to-\gamma_1 U,\qquad P_V\to-\gamma_2 V.\label{classicallimit}
\ee
These relates are deformed when turning on nontrivial $\tilde{L}$. Given any expressions of $P_U,P_V$ as functions of $U,V$, $\tilde{L}$ can be constructed (up to integration constants) by solving \eqref{PUPV}. The covariant $\bar{\mu}$-scheme effective dynamics corresponds to 
\be
P_U=\frac{\sin^{-1}\left(-2\gamma_{1}\alpha_{1}\sqrt{\Delta}\,U\right)}{2\alpha_{1}\sqrt{\Delta}},\qquad P_V=\frac{\sin^{-1}\left(-2\gamma_{2}\alpha_{2}\sqrt{\Delta}\,V\right)}{2\alpha_{2}\sqrt{\Delta}},\label{PUPV1}
\ee
where $\a_1,\a_2$ are free parameters that are constant on the spacetime, and the factor of $2$ is conventional. When we relate the construction to LQG, $\Delta$ should relate to the minimal nonzero eigenvalue in the LQG area spectrum. From the perspective of mimetic gravity, $\Delta$ is the coupling constant for the higher-derivative couplings in $\tilde{L}$. The limit \eqref{classicallimit} is recovered by $\Delta\to0$. The expression of $\tilde{L}$ is obtained by solving \eqref{PUPV} and requiring $\lim_{\Delta\to0}\tilde{L}=0$:
\be
\tilde{L}(U,V)&=&L_1(U)+L_2(V),\label{tildeLUV}\\
L_1(U)&=&\frac{1}{4 \alpha _1^2 \gamma _1 \Delta }-\frac{\sqrt{1-4 \alpha _1^2 \gamma _1^2 \Delta  U^2}}{4 \alpha _1^2 \gamma _1 \Delta }+\frac{\gamma _1 U^2}{2}-\frac{U \sin ^{-1}\left(2 \alpha _1 \gamma _1 \sqrt{\Delta } U\right)}{2 \alpha _1 \sqrt{\Delta }},\label{tildeLUV1}\\
L_2(V)&=&\frac{1}{4 \alpha _2^2 \gamma _2 \Delta }-\frac{\sqrt{1-4 \alpha _2^2 \gamma _2^2 \Delta  V^2}}{4 \alpha _2^2 \gamma _2 \Delta }+\frac{\gamma _2 V^2}{2}-\frac{V \sin ^{-1}\left(2 \alpha _2 \gamma _2 \sqrt{\Delta } V\right)}{2 \alpha _2 \sqrt{\Delta }}.\label{tildeLUV2}
\ee
$\tilde{L}(X,Y)$ is obtained by applying the relation \eqref{UVXY}.

The inverse of \eqref{PUPV1} gives
\be
\g_1 U=-\frac{\sin \left(2 \alpha _1 \sqrt{\Delta }\, P_U\right)}{2 \alpha _1 \sqrt{\Delta }},\quad \g_2 V=-\frac{\sin \left(2 \alpha _2 \sqrt{\Delta }\, P_V\right)}{2 \alpha _2 \sqrt{\Delta }}
\ee
The Legandre transformation as the inverse of \eqref{mom1} and \eqref{mom2} is obtained by applying the relations \eqref{UVXY} - \eqref{PV0}.

As a remark, $\tilde{L}$ may be defined as a multi-valued function by replacing $\sin^{-1}$ in $L_1(U)$ and $L_2(V)$ by $\sin_{(k)}^{-1}$ and $\sin_{(m)}^{-1}$ ($k,m\in\mathbb{Z}$) respectively. $\sin^{-1}_{(k)}$ is defined by 
\be
\sin^{-1}_{(k)}(\psi)&=&(-1)^k\arcsin(\psi)+k\pi\in \lt[-\frac{\pi}{2}+k\pi,\frac{\pi}{2}+k\pi\rt],\quad k\in\mathbb{Z},\ \psi\in[-1,1].
\ee
and $\sin^{-1}_{(m)}$ is similar. The space of $(\pi_x,\pi_\varphi)$ (or equivalently ($P_U,P_V$)) is the cover space of the space of $(X,Y)$ (or equivalently ($U,V$)). The quotient from the space of $(\pi_x,\pi_\varphi)$ to the space of $(X,Y)$ is given by the ``gauge invariance''
\be
P_U
\sim (-1)^k P_U+\frac{k\pi}{2\alpha_1\sqrt{\Delta}}, \quad
P_V
\sim(-1)^m P_V+\frac{m\pi}{2\alpha_2\sqrt{\Delta}},
\ee
with $k,m\in\mathbb{Z}$.  $\tilde{L}$ is single-valued on the phase space although it is multi-valued in $X,Y$:
\be
\tilde{L}(U,V,P_U,P_V)&=&L_1(U,P_U)+L_2(V,P_V),\\
L_1(U,P_U)&=&\frac{1}{4 \alpha _1^2 \gamma _1 \Delta }-\frac{\sqrt{1-4 \alpha _1^2 \gamma _1^2 \Delta  U^2}}{4 \alpha _1^2 \gamma _1 \Delta }+\frac{\gamma _1 U^2}{2}+P_U U,\\
L_2(V,P_V)&=&\frac{1}{4 \alpha _2^2 \gamma _2 \Delta }-\frac{\sqrt{1-4 \alpha _2^2 \gamma _2^2 \Delta  V^2}}{4 \alpha _2^2 \gamma _2 \Delta }+\frac{\gamma _2 V^2}{2}+P_V V.
\ee

\subsection{The Hamiltonian}

The primary Hamiltonian from $\tilde{S}_2^{(\eta)}$ is given by
\be
H&=&\frac{1}{2G}\int \rmd x\,\lt(-N E^{\varphi} \sqrt{E^{x}}\left\{\tilde{L}-\left[2 X Y-(1-\eta)Y^{2}\right]+\frac{1}{2} R_\eta^{(3)}\right\}+\pi_x\dot{E}^x+\pi_\varphi\dot{E}^\varphi+\L \pi_{N^x}\rt)\nonumber\\
&=&\int\rmd x \lt(N\cc_\Delta+N^x\cc_x+\L \pi_{N^x}\rt)
\ee
where
\be
\cc_\Delta&=&\frac{1}{2 G}\lt(- \sqrt{E^x} E^\varphi \left\{\tilde{L}-\left[2 X Y-(1-\eta)Y^{2}\right]+\frac{1}{2} R_\eta^{(3)}\right\}+ {2}Y E^x \pi_x + X E^\varphi \pi_\varphi\rt)\nonumber\\
\cc_x&=&\frac{1}{2 G}\lt(E^x{}' \pi_x-E^\varphi  \pi_\phi'\rt).
\ee
It is important that here $\cc_\Delta$ is not a constraint since $N=\dot{\phi}(t)$ is regarded as an external field in $\tilde{S}_2^{(\eta)}$. 

Expressing $\cc_\Delta$ on the phase space gives
\be
\cc_\Delta&=&\frac{E^\varphi \sqrt{E^x}  [c (1-\eta )+2 h] }{8 c \Delta {G}}\left[{\frac{a^2}{\a_1^2}\sin ^2\left(\a_1\sqrt{\Delta}\,P_U\right)}-{\frac{c^2}{\a_2^2} \sin ^2\left(\a_2\sqrt{\Delta}\,P_V\right)}\right]\nonumber\\
&&-\frac{E^\varphi \sqrt{E^x} }{4 G} \left(-\frac{2 E^x{}''}{E^\varphi{}^2}+\frac{2 E^x{}' E^\varphi{}'}{E^\varphi{}^3}-\frac{(1-\eta ) E^x{}'{}^2}{2 E^x E^\varphi{}^2}+2 (\eta +1)(E^x)^{\eta -1}\right)\label{ccDelta1}
\ee
where $P_U,P_V$ are given by \eqref{PU0} and \eqref{PV0}. 
To relate to conventions and notations in some early literatures e.g. \cite{Han:2020uhb,Chiou:2012pg,Gambini:2013hna}, we introduce $K_x,K_\varphi$ and change variables  
\be
\pi_x=-2K_x \qquad \pi_\varphi=-2K_\varphi.
\ee
The Poisson brackets between $K_x,K_\varphi$ and $E^x,E^\varphi$ are the same as \eqref{Poisson1}. $P_U$ and $P_V$ are given by
\be
&& P_U=\frac{-2h K_\varphi }{\sqrt{E^x} a (c(1-\eta)+2h)}+\frac{4 c \sqrt{E^x} K_x}{E^\varphi a (c(1-\eta)+2h)},\label{PU01}\\
&& P_V=\frac{-4  \sqrt{E^x} K_x}{E^\varphi  (c(1-\eta)+2h)}+\frac{ 2K_\varphi  \left(-\frac{h}{c}+\eta -1\right)}{\sqrt{E^x}  (c(1-\eta)+2h)}.\label{PV01}
\ee
$H$ is the covariant $\bar{\mu}$-scheme effective Hamiltonian of the spherical symmetric LQG. The $\bar{\mu}$-scheme holonomies can be extracted from $\sin \left(\a_1\sqrt{\Delta}\,P_U\right)$ and $\sin \left(\a_2\sqrt{\Delta}\,P_V\right)$ in $\cc_\Delta$:
\be
h_x=e^{i\frac{\sqrt{\Delta}\sqrt{E^x}}{E^\varphi}A_1},\qquad h_{\theta}=h_\varphi= e^{i \frac{\sqrt{\Delta}}{\sqrt{E^x}}A_2}.\label{mubarholonomy}
\ee
with certain choice of the parameters $a,c,h,\a_1,\a_2$. For example, a convenient choice is $a=c= 1,\ h= -1,\ \alpha_2 = 2 \alpha_1=\b$, which leads to \eqref{ccDelta00} mentioned in Section \ref{effective dynamics of spherical symmetric quantum gravity}. As we see in Section \ref{Application I}, the cosmological effective dynamics gives the restriction to the parameters. We are going to discuss in Section \ref{More on the relation with mubar scheme} about further restricting the parameters by other considerations.

In the limit $\Delta\to0$ that removes higher derivative couplings, $K_x,K_\varphi$ relate to the components of the extrinsic curvature of the constant-$\phi$ slice, and $H$ recovers the classical Hamiltonian of the dilaton-gravity models by
\be
\cc_\Delta&\to& 
\frac{1}{4G\sqrt{E^x}} \bigg\{-\frac{2 E^{x} E^{x \prime} E^{\varphi \prime}}{E^{\varphi 2}}+\frac{4 E^{x} E^{x \prime \prime}+(1-\eta)E^{x \prime 2}}{2 E^{\varphi}}-8 E^x K_x K_{\varphi} \nonumber\\
&&\quad  -2 E^{\varphi}\lt[(1-\eta)K_\varphi^2+(1+\eta)(E^x)^\eta\rt]\bigg\}.
\ee
All the parameters $a,c,h,\a_1,\a_2$ disappear in the limit. $\eta=0$ recovers the Hamiltonian of the spherical symmetry reduction of 4d gravity.

Continuing of the Hamiltonian analysis, the dynamical stability of the primary constraint $\pi_{N^x}\approx 0$ gives the diffeomorphism constraint as the secondary constraint
\be
\cc_x=-\frac{1}{ G}\lt(E^x{}' K_x-E^\varphi  K_\varphi'\rt)\approx 0
\ee
Furthermore we have the conservation law 
\be
\{\cc_x(x),H\}\approx 0
\ee
Thus the dynamical stability of $\cc_x$ does not give any further constraint.

The equations of motion of the mimetic-dilation-gravity models $S_2^{(\eta)}$ becomes 4 Hamiltonian equations 
\be
\dot{f}=\{f,H\},\qquad f=E^x,E^\varphi,K_x,K_\varphi
\ee 
subject to the constraint $\cc_x=0$. The Hamiltonian equations are partial differential equations, which are first order in $t$ and second order in $x$.

\section{Homogeneous and isotropic bouncing cosmology}\label{Application I}

As the first application of the equations of motion, we assume the spatial homogeneity in addition to the spherical symmetry on the spatial slices in $\sm_4$. The assumption applies to the homogeneous-isotropic cosmology. In this cases, the metric ansatz in 4d is 
\begin{eqnarray}
    \label{os metric}
    \rmd s^{2}=-\rmd  t^{2}+\frac{\fa(t)^{2}}{1-kx^2} \rmd  x^{2}+x^{2} \fa(t)^{2} \lt( \rmd \theta^{2}+\sin ^{2} \theta \rmd \varphi^{2}\rt).
\end{eqnarray}  
where $\fa(t)$ denotes the scale factor, and the spatial geometry is flat, spherical, hyperbolic for $k=0,1,-1$. We have set $\phi=t$ and the lapse function $N=\dot{\phi}=1$, i.e. $\phi_\mu=(1,0,0,0)$, as well as the shift vector $N^x=0$. The same metric also applies to the Oppenheimer-Snyder model homogeneous gravitational collapse inside the black hole. We set $\eta=0$ in this section, since we focus on the symmetry reduction of 4d gravity.

We include a massless scalar field for the discussion of cosmology. The scalar field modify $\cc_\Delta$ and $\cc_x$ in $H$ by
\be
\cc_\Delta &\to&\cc_\Delta+\frac{\Pi ^2}{8 \pi  \sqrt{E^x} E^\varphi}+\frac{2 \pi  E^x{}^{3/2} \Phi '{}^2}{E^\varphi}\\
\cc^x &\to& \cc^x+\Pi\Phi'
\ee
We look for the solution satisfying the symmetry to the Hamiltonian equations $\dot{f}=\{f,H\}$, where $f=E^x,E^\varphi,K_x,K_\varphi,\Phi,\Pi$ and $H=\int\rmd x\, \cc_\Delta$. We insert the following ansatz in the Hamiltonian equations,
\be
&& E^x(t,x)= x^2 \fa(t)^2,\quad E^\varphi (t,x)= \frac{x \fa(t)^2}{\sqrt{1-kx^2}},\\
&&K_\varphi (t,x)= 2 x K(t),\quad K_x(t,x)= \frac{K(t)}{\sqrt{1-kx^2}},\\
&&\Pi (t,x)= \frac{4 \pi  x^2 \Pi (t)}{\sqrt{1-kx^2}},\quad \Phi (t,x)= \Phi (t).
\ee
The ansatz respects the symmetry and the metric \eqref{os metric}. Inserting the ansatz reduces the Hamiltonian equations from partial differential equations to ordinary differential equations. Moreover the diffeomorphism constraint $\cc_x=0$ is satisfied by the ansatz.

Since the ansatz relates both $E^x,E^\varphi$ to single $\fa(t)$ and relation both $K_x,K_\varphi$ to single $K(t)$, the Hamiltonian equations give a consistency condition
\be
\frac{a (2 c+h) \sin \left(\frac{8 \alpha _1 \sqrt{\Delta } (c-h) K(t)}{a \fa(t) (c+2 h)}\right)}{\alpha _1 c}=\frac{(c-h) \sin \left(\frac{8 \alpha _2 \sqrt{\Delta } (2 c+h) K(t)}{c \fa(t) (c+2 h)}\right)}{\alpha _2}.\label{consistencycondi}
\ee
The homogeneous and isotropic symmetries suppose to reduce the Hamiltonian equations to evolution equations of $\fa(t)$ and $K(t)$. This consistency condition must be satisfied identically without imposing any restriction to $\fa(t)$ and $K(t)$. Then it gives the restriction to the parameter $a,c,h,\a_1,\a_2$. Here we choose
\be
a=c= 1,\quad h= -1,\quad \alpha_2 = 2 \alpha_1.\label{choiceparameter}
\ee
In order to compare the equations to LQC, We consider the following change of variables from $(\fa,K)$ to $(b,V)$:
\be
\fa(t)= {V(t)}^{1/3},\quad K(t)\to \frac{b(t) V(t)^{1/3}}{2},
\ee
where $V(t)$ is the spatial volume element. We also define
\be
\beta=2\a_2=4\a_1.\label{definebeta}
\ee
By the change of variables, the equations of motion reduces to the $\bar{\mu}$-scheme effective equations of LQC, 
\be
\dot{V}(t)&=& \frac{3 V(t) \sin (2\b \sqrt{\Delta}\, b(t))}{2 \b  \sqrt{\Delta }},\label{cosm1}\\
\dot{b}(t)&=& -\frac{3 \sin ^2(\b\sqrt{\Delta}\, b(t))}{2 \b^2 \Delta }-\frac{kV(t)^{4/3}+4 \pi G\, \Pi (t)^2}{2 V(t)^2},\\
\dot{\Phi} (t)&=& \frac{\Pi (t)}{V(t)},\qquad \dot{\Pi}(t)= 0.\label{cosm2}
\ee
The above equations coincide with the effective equations of LQC with the $K$-quantization (see e.g. \cite{Singh:2013ava}) by identifying $\b$ to be the Barbero-Immirzi parameter. In particular, the LQC holonomy corrections given by the sine functions are reproduced by the mimetic gravity with our proposed $L_\phi$.

Coming back to the choice of parameters, the condition \eqref{consistencycondi} can be solved either by demanding both sides of \eqref{consistencycondi} to vanish, or by equating up to sign both quantities inside and outside the sine functions. We consider the following solutions\footnote{When $2c+h\neq 0$ and $c\neq h$, Eq.\eqref{consistencycondi} can be written as $\frac{\sin\left(\mathcal{X}\mathcal{F}(t)\right)}{\mathcal{X}}=\frac{\sin\left(\mathcal{Y}\mathcal{F}(t)\right)}{\mathcal{Y}}$ where $\frac{\alpha_{1}(c-h)}{a}=\mathcal{X},\ \frac{\alpha_{2}(2c+h)}{c}=\mathcal{Y},\ \frac{8\sqrt{\Delta}K(t)}{\mathfrak{a}(t)(c+2h)}=\mathcal{F}(t)$. It implies $\mathcal{X}=\pm\mathcal{Y}$ or $\mathcal{X}=\pm\mathcal{Y}+\mathcal{P}(t)$. In the second possibility, $\mathcal{P}(t)$ is due to the oscillation of $\frac{\sin\left(\mathcal{X}\mathcal{F}(t)\right)}{\mathcal{X}}$ in $\cx$. $\mathcal{P}(t)$ depends on $t$ because the oscillation depends on $\cf(t)$. Here we only consider the first possibility, since $\cx,\cy$ are $t$-independent.}
\begin{enumerate}

  \item $c=h$: In this case, the definition of $\b$ in \eqref{definebeta} is replaced by $\b =2 \a_2/h$.
  
  \item $c=-h/2$: $\b$ is defined by $\b=2\a_1/a$.
  
  \item $h=\frac{c (\a_1 c-2 \alpha_2  a)}{\alpha_2  a+\a_1 c}$: The above choice of parameters \eqref{choiceparameter} is a special case of this solution. In this case, we define $\b=\frac{2 \alpha_2  \a_1}{\alpha_2 a-\a_1 c}$ correspondingly.
  
  \item $h=\frac{c (\a_1 c+2 \alpha_2  a)}{-\alpha_2  a+\a_1 c}$: This flips $a\to -a$ of the above case. We have $\b=\frac{2 \alpha_2  \a_1}{\alpha_2 a+\a_1 c}$ correspondingly.
  
\end{enumerate}
These solutions with the corresponding definition of $\b$ lead to the same effective equations as \eqref{cosm1} - \eqref{cosm2}. So all these choices are allowed for the cosmological effective dynamics. We will come back to these solutions in Section \ref{More on the relation with mubar scheme} and consider the restriction of parameter beyond the cosmological effective dynamics.

We may introduce an effective Hamiltonian $H_{\rm eff}$ and an effective Poisson bracket $\{\ ,\ \}_{\rm eff}$ of the homogeneous-isotropic cosmology
\be
H_{\rm eff}=-\frac{3 V \sin ^2(\b \sqrt{\Delta}\, b)}{8 \pi  \b ^2 \Delta  G}-\frac{3kV^{1/3}}{8\pi G }+\frac{\Pi ^2}{2 V},\qquad \{b,V\}_{\rm eff}=4\pi G,\quad \{\Phi,\Pi\}_{\rm eff}=1
\ee
$H_{\rm eff}$ and $\{\ ,\ \}_{\rm eff}$ coincide to the Hamiltonian and Poisson bracket in LQC. The equations \eqref{cosm1} - \eqref{cosm2} are equivalent to 
\be
\dot{f}=\{f,H_{\rm eff}\}_{\rm eff},\quad f=V,b,\Phi,\Pi.
\ee

\begin{figure}[h]
  \begin{subfigure}{0.5\textwidth}
  \includegraphics[width = 1\textwidth]{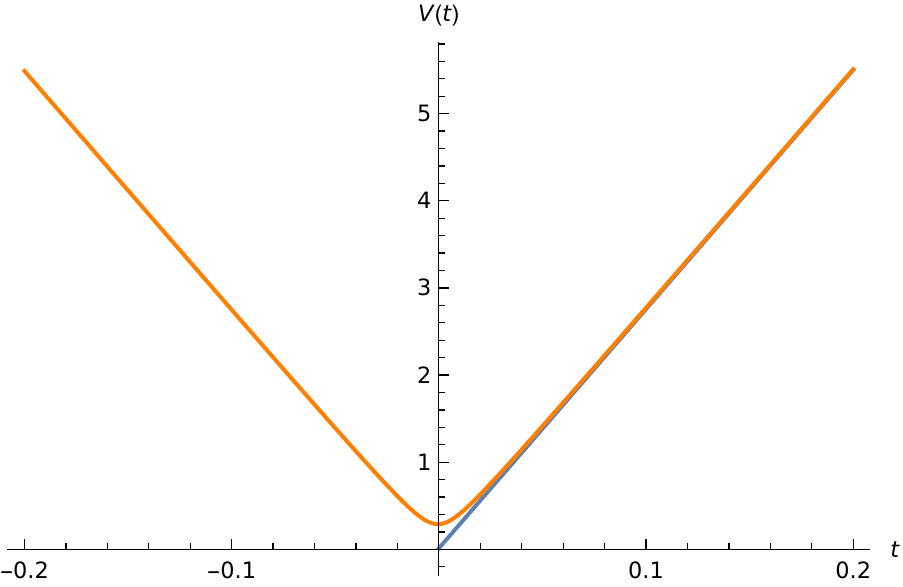} 
  \caption{}
  \label{cosmV}
   \end{subfigure}
   \begin{subfigure}{0.5\textwidth}
    \includegraphics[width = 1\textwidth]{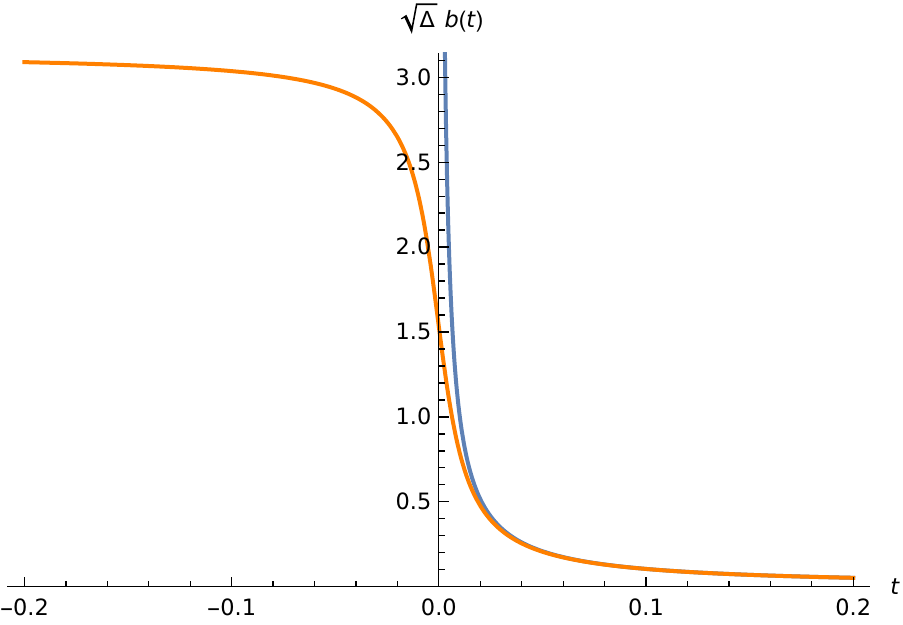} 
    \caption{}
    \label{cosmb}
     \end{subfigure}
   \caption{The figures plot the solution (orange curves) of \eqref{cosm1} - \eqref{cosm2} with $k=0$ (and $\b=1$), and compare to the classical FRW cosmology (blue curves). The bouncing time is at $t_c\simeq 0 $.}
  \end{figure}

We consider the cosmological evolution and set the initial time $t_0$ to be nowadays. For the initial condition, it is reasonable to assume that the mimetic scalar and higher-derivative coupling should have negligible contribution nowadays, so that the initial data $V(t_0),b(t_0),\Phi(t_0),\Pi(t_0)$ gives $H_{\rm eff}=0$ same as the Hamiltonian constraint. $H_{\rm eff}$ is conserved in the time evolution. The time evolution from the initial data gives the solution illustrated in FIGs.\ref{cosmV} and \ref{cosmb}. The dynamics resolves the big-bang singularity with a non-singular bounce. The bounce is symmetric in time-reversal. 

Recall the Einstein equation of mimetic gravity \eqref{eomgrav}. We extract the stress-energy tensor $T'_{\mu\nu}=T^\phi_{\mu\nu}+(R+T^\phi)\phi_\mu\phi_\nu$ of the mimetic field by compute the 4d Einstein tensor and applying the equations of motion:
\be
G_{\mu\nu}=8\pi G T^{\rm scalar}_{\mu\nu}+T'_{\mu\nu}
\ee
We still assume $H_{\rm eff}=0$, and we obtain
\be
\!\!\! && T^{\rm scalar}_{\mu\nu}=(\rho_{s}+p_s) \phi_\mu\phi_\nu+p_s g_{\mu\nu}\ ,\qquad \rho_s=p_s=\frac{\Pi^2}{2 V^2}=\frac{3 \sin ^2(\beta  \sqrt{\Delta } b)}{8 \pi  \beta ^2 \Delta  G}+\frac{3 k}{8 \pi  G V^{2/3}},\nonumber\\
\!\!\! &&  T'_{\mu\nu}=(\rho'+p') \phi_\mu\phi_\nu+p' g_{\mu\nu}\ ,\qquad \rho'=-\frac{3 \sin ^4(\beta  \sqrt{\Delta }b)}{ \beta ^2 \Delta },\quad p'=-\frac{9 \sin ^4(\beta  \sqrt{\Delta }b)}{\beta ^2 \Delta }-\frac{8k \sin ^2(\beta  \sqrt{\Delta } b)}{ V^{2/3}}.\nonumber
\ee
$T'_{\mu\nu}$ behaves effectively as a perfect fluid with the density $\rho'$ and pressure $p'$. Both $\rho'$ and $p'$ are of $O(\Delta)$. From the viewpoint of the effective dynamics of LQG, $T'_{\mu\nu}$ is the effective stress-energy tensor counting the quantum correction to the Einstein equation, while it is also the stress-energy tensor of the mimetic field from the mimetic-gravity point of view.

The bounce is at the time $t_c$ where
\be
\dot{V}(t_c)=0,\qquad b(t_c)=\frac{\pi}{2\b\sqrt{\Delta}}
\ee
we obtain the critical densities and pressures
\be
&&\rho_s(t_c)=p_s(t_c)=\frac{3}{8 \pi  \beta ^2 \Delta  G}+\frac{3 k}{8 \pi  G V(t_c)^{2/3}},\\
&&\rho'(t_c)=-\frac{3}{ \beta ^2 \Delta },\qquad p'(t_c)=-\frac{9}{ \beta ^2 \Delta  }-\frac{8k}{ V(t_c)^{2/3}}.
\ee
The Kretschmann invariant at the bounce is given by
\be
\ck(t_c)=R_{\mu\nu\rho\sig}(t_c)R^{\mu\nu\rho\sig}(t_c)=\frac{108}{ \beta ^4 \Delta ^2}+\frac{60 k^2}{V(t_c)^{4/3}}+\frac{144 k}{\beta ^2 \Delta  V(t_c)^{2/3}}.
\ee
At the bounce, both the critical densities and the Kretschmann invariant are Planckian when $\Delta\sim\ell_P^2$ relates to the minimal nonzero eigenvalue of the LQG area operator. 

Based on $\Delta\sim\ell_P^2$, the cosmic bounce is a result of the quantum effect from the LQG viewpoint. From the mimetic gravity viewpoint, the same effect is formulated as resulting from the higher-derivative coupling with the mimetic scalar $\phi$. In our opinion, this two viewpoints are not contradicting but closely related. The key point is that the LQC holonomy corrections, which is responsible for the bounce, can be reproduced at the Hamiltonian level by the mimetic gravity with our proposed $L_\phi$. It provides an evidence supporting our proposal that the mimetic gravity should be a candidate of the quantum effective theory for LQG, and the equations of motion of the mimetic gravity lagrangian should capture quantum effects in LQG.

\section{Non-singular spherical symmetric black hole}\label{Application II}

\subsection{Non-singular black hole solution and asymptotic $\mathrm{dS_2}\times S^2$}\label{Non-singular black hole and asymptotic Nariai geometry}

We remove the assumption of the spatial homogeneity but still assume the spherical symmetry. To be consistent with the discussion of cosmology, we still use the choice of parameters $\eta=0$ and \eqref{choiceparameter}. We still define the Barbero-Immirzi parameter by $\b=4\a_1=2\a_2$ as in cosmology, 

We again choose $\phi(t)=t$ so that $N=\dot{\phi}=1$. The Hamiltonian is given by $H=\int \rmd x\, \cc_\Delta$, where
\be
\cc_\Delta&=&\frac{\sqrt{E^x} E^\varphi}{2G \Delta}\lt[{ \sin ^2\left(\frac{2 \b \sqrt{\Delta } \sqrt{E^x} K_x }{   E^\varphi }\right)}-4\,{ \sin ^2\left(\frac{\b\sqrt{\Delta }\sqrt{E^x}}{ E^\varphi }K_x+\frac{ \b\sqrt{\Delta } }{2 \sqrt{E^x} }K_\varphi\right)}\rt]\nonumber\\
&&+\, \frac{1}{4G\sqrt{E^x}}\Bigg(-\frac{2 E^{x} E^{x \prime} E^{\varphi \prime}}{E^{\varphi 2}}+\frac{4 E^{x} E^{x \prime \prime}+E^{x \prime 2}}{2 E^{\varphi}}-2 E^{\varphi}\Bigg).\label{ccDelta111}
\ee
and we further fix $\b=1$ for the following numerical study of the equations of motion.

The Hamiltonian $H$ generates the dynamics of the (1+1)d canonical fields $E^x,E^\varphi,K_x,K_\varphi$, subject to the constraint $\cc_x=0$. The spacetime metric is given by
\be
\rmd s^{2}=-\rmd t^{2}+\frac{E^{\varphi}(t,x)^{2}}{E^{x}(t,x)}\rmd x^{2}+E^{x}(t,x) \lt( \rmd \theta^{2}+\sin ^{2} \theta \rmd \varphi^{2}\rt),\label{N=1metric}
\ee
which provides the geometrical interpretation to the solution.

We would like to study the spherical symmetric black hole solution and compare to the Schwarzschild black hole. The Schwarzschild spacetime in the Lema\^{\i}tre coordinates $(t,x,\theta,\varphi)$ is given by \eqref{N=1metric} with 
\be
E^x= \left(\frac{3}{2} \sqrt{R_s}\, (x-t)\right)^{4 / 3},\quad  E^\varphi=  \sqrt{{R_s}} \lt(\frac{3}{2}{\sqrt{{R_s}}\, (x-t)}\rt)^{1/3}
\ee 
where $R_s$ is the Schwarzschild radius.

As the boundary condition for $H$, we consider $E^x,E^\varphi$ to behave asymptotically as the Schwarzschild geometry in the Lema\^{\i}tre coordinates as $x\to\infty$:
\begin{eqnarray}
E^x\sim \left(\frac{3}{2} \sqrt{R_s}\, x\right)^{4 / 3},\quad E^\varphi\sim  \sqrt{{R_s}} \lt(\frac{3}{2}{\sqrt{{R_s}}\, x}\rt)^{1/3}, \label{bdyschw1}
\end{eqnarray}
Under this boundary condition, $H$ does not need a boundary term to make $\delta H$ well-defined \cite{Han:2020uhb,Giesel:2022rxi}.

The Hamiltonian equations $\partial_t{f}=\{f,H\}$ give a set of 4 partial differential equations (PDEs). The set of PDEs are 1st-order in $t$ and 2nd-order in $x$. We introduce the following change of variables in order to make the formulae compact
\be
K_x= \frac{1}{8} \zeta _2 e^{\xi },\qquad K_\varphi= -\frac{1}{4} \left(\zeta _1+\zeta _2\right) e^{\psi },\qquad
E^\varphi= e^{\xi +\psi},\qquad E^x= e^{2 \psi }
\ee
One can check that $\zeta_1=2P_U,\ \zeta_2=2P_V$. The Hamiltonian equations in terms of $\zeta_1,\zeta_2,\xi,\psi$ are given by
\be
\dot{\zeta} _1&=& \frac{1}{2  \Delta }\Bigg[8  \Delta  e^{-2 \xi } \xi ^{\prime} \psi ^{\prime}-8  \Delta  e^{-2 \xi } \psi ^{\prime}{}^2-8 \Delta  e^{-2 \xi } \psi ^{\prime\prime}+ \sqrt{\Delta } \zeta _1 \sin \left(\frac{\sqrt{\Delta } \zeta _2}{2} \right)\nonumber\\
&&+2 \sqrt{\Delta } \zeta _2 \sin \left(\frac{\sqrt{\Delta } \zeta _2}{2} \right)-16 \cos \left(\frac{\sqrt{\Delta } \zeta _1}{4}\right)+4 \cos \left(\frac{\sqrt{\Delta } \zeta _2}{2} \right)+12\Bigg],\label{eomzeta1}\\
\dot{\zeta} _2 &=& \frac{1}{\Delta }\Bigg[-4  \Delta  e^{-2 \xi } \xi ^{\prime} \psi ^{\prime}+2 \Delta  e^{-2 \xi } \psi ^{\prime}{}^2+4 \Delta  e^{-2 \xi } \psi ^{\prime\prime}+2  \Delta  e^{-2 \psi }+  \sqrt{\Delta } \zeta _1 \sin \left(\frac{\sqrt{\Delta } \zeta _1}{4}\right)\nonumber\\
&&+2  \sqrt{\Delta } \zeta _2 \sin \left(\frac{\sqrt{\Delta } \zeta _1}{4} \right)+4 \cos \left(\frac{\sqrt{\Delta } \zeta _1}{4}\right)-\cos \left(\frac{\sqrt{\Delta } \zeta _2}{2} \right)-3\Bigg],\label{eomzeta2}\\
\dot{\xi}&=& \frac{1}{2 \sqrt{\Delta }}\sin \left(\frac{\sqrt{\Delta } \zeta _2}{2}\right),\label{eomxi}\\
\dot{\psi} &=& -\frac{1}{2 \sqrt{\Delta }}\Bigg[\sin \left(\frac{\sqrt{\Delta } \zeta _2}{2}\right)+2 \sin \left(\frac{\sqrt{\Delta } \zeta _1}{4}\right)\Bigg].\label{eompsi}
\ee
where $\dot{f}=\partial_t f$ and $f'=\partial_x f$ for $f=\zeta_1,\zeta_2,\xi,\psi$.

To simplify the equations, we apply the following ansatz as in \cite{Han:2020uhb}
\be
f(t,x)=f(z)\ ,\quad z=x-t\ ,\quad f=\zeta_1,\zeta_2,\xi,\psi.\label{ansatzfz} 
\ee
The ansatz is inspired by the Schwarzschild geometry in the Lema\^{\i}tre coordinates, and it assumes the killing symmetry generated by $\xi=\partial_t+\partial_x$. The ansatz reduces the PDEs to four 1st-order ordinary differential equations (ODEs) 
\be
\frac{\rmd f(z)}{\rmd z}=\cf_f(z),\qquad f=\zeta_1,\zeta_2,\xi,\psi.\label{ODEs0}
\ee
The explicit expressions of the ODEs are given below
\be
\frac{\rmd \zeta _1}{\rmd z}&=& \frac{e^{-2 (\xi +\psi )} }{4 \Delta  \left(\cos \left(\frac{\sqrt{\Delta } \zeta _2}{2}\right)-\cos \left(\frac{\sqrt{\Delta } \zeta _1}{4}\right)+e^{2 \xi }\right)}\Bigg[8 e^{2 (\xi +\psi )} \sin ^2\left(\frac{\sqrt{\Delta } \zeta _2}{2}\right)+16 e^{2 (\xi +\psi )} \sin ^2\left(\frac{\sqrt{\Delta } \zeta _1}{4}\right)\nonumber\\
&&+24 e^{2 (\xi +\psi )} \sin \left(\frac{\sqrt{\Delta } \zeta _2}{2}\right) \sin \left(\frac{\sqrt{\Delta } \zeta _1}{4}\right)-4 e^{2 (\xi +\psi )} \cos ^2\left(\frac{\sqrt{\Delta } \zeta _2}{2}\right)-12 e^{2 (\xi +\psi )} \cos \left(\frac{\sqrt{\Delta } \zeta _2}{2}\right)\nonumber\\
&&-8 e^{4 \xi +2 \psi } \cos \left(\frac{\sqrt{\Delta } \zeta _2}{2}\right)+16 e^{2 (\xi +\psi )} \cos \left(\frac{\sqrt{\Delta } \zeta _1}{4}\right) \cos \left(\frac{\sqrt{\Delta } \zeta _2}{2}\right)+32 e^{4 \xi +2 \psi } \cos \left(\frac{\sqrt{\Delta } \zeta _1}{4}\right)\nonumber\\
&&-\sqrt{\Delta } \zeta _1 e^{2 (\xi +\psi )} \left(2 e^{2 \xi } \sin \left(\frac{\sqrt{\Delta } \zeta _2}{2}\right)+\sin \left(\sqrt{\Delta } \zeta _2\right)+4 \sin \left(\frac{\sqrt{\Delta } \zeta _1}{4}\right) \cos \left(\frac{\sqrt{\Delta } \zeta _2}{2}\right)\right)\nonumber\\
&&-2 \sqrt{\Delta } \zeta _2 e^{2 (\xi +\psi )} \left(2 e^{2 \xi } \sin \left(\frac{\sqrt{\Delta } \zeta _2}{2}\right)+\sin \left(\sqrt{\Delta } \zeta _2\right)+4 \sin \left(\frac{\sqrt{\Delta } \zeta _1}{4}\right) \cos \left(\frac{\sqrt{\Delta } \zeta _2}{2}\right)\right)\nonumber\\
&&-8 \Delta  e^{2 \xi } \cos \left(\frac{\sqrt{\Delta } \zeta _2}{2}\right)+e^{2 \psi } \sin \left(\frac{\sqrt{\Delta } \zeta _2}{2}\right) \sin \left(\sqrt{\Delta } \zeta _2\right)+4 e^{2 \psi } \sin \left(\frac{\sqrt{\Delta } \zeta _1}{4}\right) \sin \left(\sqrt{\Delta } \zeta _2\right)\nonumber\\
&&+8 e^{2 \psi } \sin ^2\left(\frac{\sqrt{\Delta } \zeta _1}{4}\right) \cos \left(\frac{\sqrt{\Delta } \zeta _2}{2}\right)-24 e^{4 \xi +2 \psi }\Bigg],\label{ODEz1}\\
\frac{\rmd \zeta _2}{\rmd z}&=& \frac{e^{-2 (\xi +\psi )}}{4 \Delta  \left(\cos \left(\frac{\sqrt{\Delta } \zeta _2}{2}\right)-\cos \left(\frac{\sqrt{\Delta } \zeta _1}{4}\right)+e^{2 \xi }\right)} \Bigg[-16 e^{2 (\xi +\psi )} \sin \left(\frac{\sqrt{\Delta } \zeta _1}{4}\right) \sin \left(\frac{\sqrt{\Delta } \zeta _2}{2}\right)\nonumber\\
&&+12 e^{2 (\xi +\psi )} \cos \left(\frac{\sqrt{\Delta } \zeta _1}{4}\right)-16 e^{4 \xi +2 \psi } \cos \left(\frac{\sqrt{\Delta } \zeta _1}{4}\right)+4 e^{2 (\xi +\psi )} \cos \left(\frac{\sqrt{\Delta } \zeta _2}{2}\right) \cos \left(\frac{\sqrt{\Delta } \zeta _1}{4}\right)\nonumber\\
&&-4 e^{2 (\xi +\psi )} \cos \left(\frac{\sqrt{\Delta } \zeta _1}{2}\right)+4 e^{4 \xi +2 \psi } \cos \left(\frac{\sqrt{\Delta } \zeta _2}{2}\right)+3 e^{2 (\xi +\psi )} \cos \left(\sqrt{\Delta } \zeta _2\right)\nonumber\\
&&-2 \sqrt{\Delta } \zeta _1 e^{2 (\xi +\psi )} \left(2 e^{2 \xi } \sin \left(\frac{\sqrt{\Delta } \zeta _1}{4}\right)-\left(\sin \left(\frac{\sqrt{\Delta } \zeta _2}{2}\right)+2 \sin \left(\frac{\sqrt{\Delta } \zeta _1}{4}\right)\right) \cos \left(\frac{\sqrt{\Delta } \zeta _1}{4}\right)\right)\nonumber\\
&&-4 \sqrt{\Delta } \zeta _2 e^{2 (\xi +\psi )} \left(2 e^{2 \xi } \sin \left(\frac{\sqrt{\Delta } \zeta _1}{4}\right)-\sin \left(\frac{\sqrt{\Delta } \zeta _1}{2}\right)-\sin \left(\frac{\sqrt{\Delta } \zeta _2}{2}\right) \cos \left(\frac{\sqrt{\Delta } \zeta _1}{4}\right)\right)\nonumber\\
&&+8 \Delta  e^{2 \xi } \cos \left(\frac{\sqrt{\Delta } \zeta _1}{4}\right)-2 e^{2 \psi } \sin \left(\frac{\sqrt{\Delta } \zeta _1}{4}\right) \sin \left(\frac{\sqrt{\Delta } \zeta _1}{2}\right)-4 e^{2 \psi } \sin \left(\frac{\sqrt{\Delta } \zeta _1}{2}\right) \sin \left(\frac{\sqrt{\Delta } \zeta _2}{2}\right)\nonumber\\
&&+4 e^{2 \psi } \cos ^3\left(\frac{\sqrt{\Delta } \zeta _1}{4}\right)-5 e^{2 \psi } \cos \left(\frac{\sqrt{\Delta } \zeta _1}{4}\right)+e^{2 \psi } \cos \left(\sqrt{\Delta } \zeta _2\right) \cos \left(\frac{\sqrt{\Delta } \zeta _1}{4}\right)\nonumber\\
&&-8 \Delta  e^{4 \xi }-15 e^{2 (\xi +\psi )}+12 e^{4 \xi +2 \psi }\Bigg],\label{ODEz2}\\
\frac{\rmd \xi}{\rmd z}&=& -\frac{1}{2 \sqrt{\Delta }}\sin \left(\frac{\sqrt{\Delta } \zeta _2}{2}\right),\label{ODEz3}\\
\frac{\rmd \psi}{\rmd z}&=& \frac{1}{2 \sqrt{\Delta }}\Bigg[\sin \left(\frac{\sqrt{\Delta } \zeta _2}{2}\right)+2 \sin \left(\frac{\sqrt{\Delta } \zeta _1}{4}\right)\Bigg].\label{ODEz4}
\ee

As the initial condition of the ODEs, we require the $E^x,E^\varphi,K_x,K_\varphi$ reduces asymptotically to the Schwarzschild as $z\to\infty$. By the Schwarzschild metric in the Lema\^{\i}tre coordinates, we have for $z=z_0\gg 1$
\be
&&E^x(z_0)=  \left(\frac{3}{2}\sqrt{R_s}
   z_0\right)^{4/3},\quad E^\varphi(z_0)=  \sqrt{R_s}
   \lt(\frac{3}{2}\sqrt{R_s} z_0\rt)^{1/3},\label{bc0}\\
 &&  K_x(z_0)= \frac{R_s}{3\times 2^{2/3} {3}^{1/3}
   \left(\sqrt{R_s} z_0\right)^{4/3}}\ ,\quad K_\varphi(z_0)=-\frac{\lt(\frac{2}{3}\rt)^{1/3}
   \sqrt{R_s}}{\lt({\sqrt{R_s} z_0}\rt)^{1/3}}\ .\label{bc1}
\ee  
Translating the initial condition to $\zeta_1,\zeta_2,\xi,\psi$ gives
\be
\zeta_1(z_0)=\zeta_2(z_0)=\frac{4}{3z_0},\qquad \psi ( z_0)=\frac{1}{6} \log \left(\frac{81 R_s^2  z_0^4}{16}\right),\qquad \xi ( z_0)= \frac{1}{3} \log \left(\frac{2 R_s}{3  z_0}\right).\label{bc2}
\ee
Then the set of ODEs can be solved numerically by assigning numerical values to parameters $R_s,\Delta,z_0$ and imposing the above initial condition at $z_0$.

\begin{figure}[h]
  \begin{subfigure}{0.5\textwidth}
    \includegraphics[width = 1\textwidth]{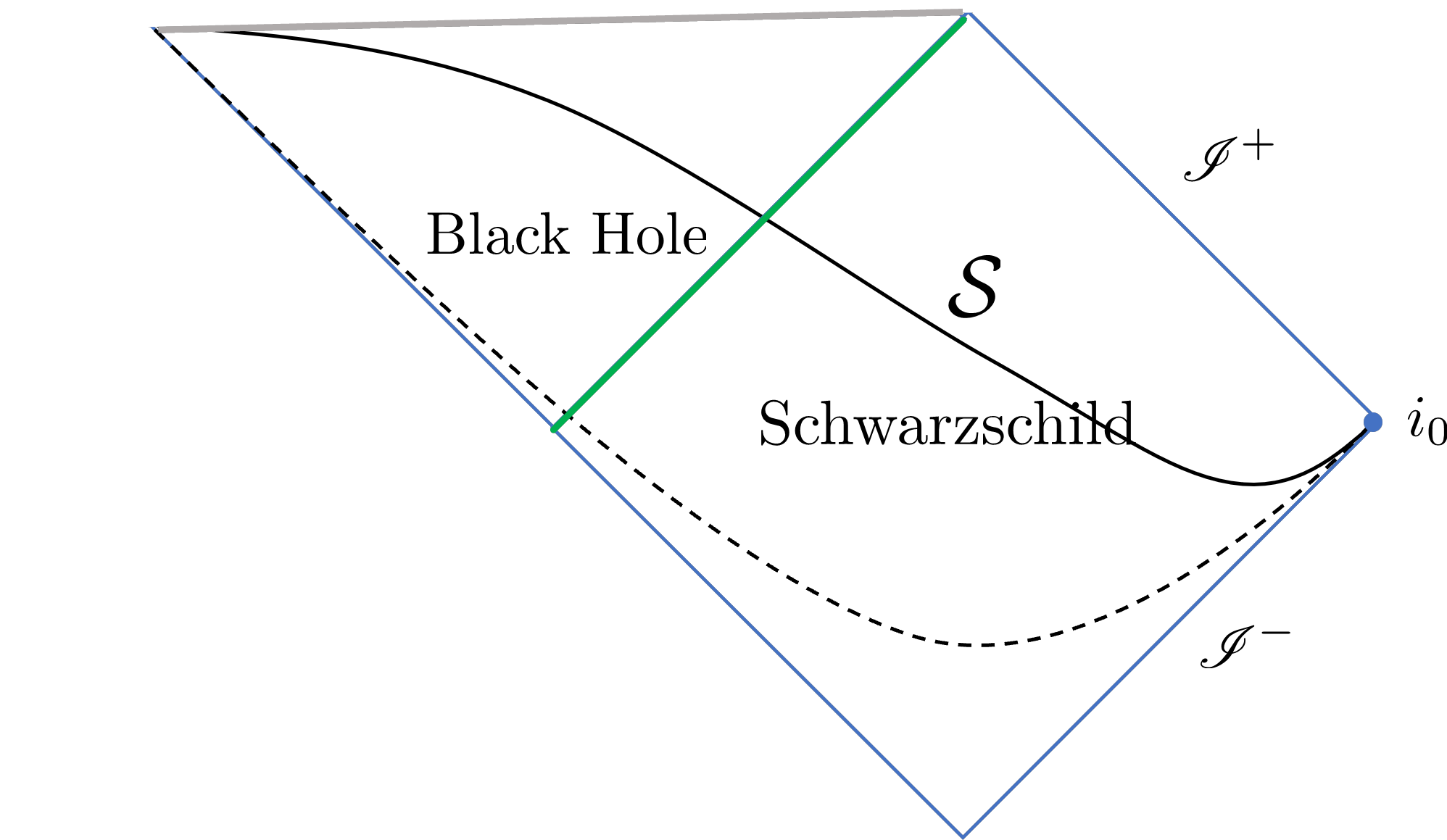} 
    \caption{Constant $t$ slices}
  \end{subfigure}
  \begin{subfigure}{0.5\textwidth}
    \includegraphics[width =1.1\textwidth]{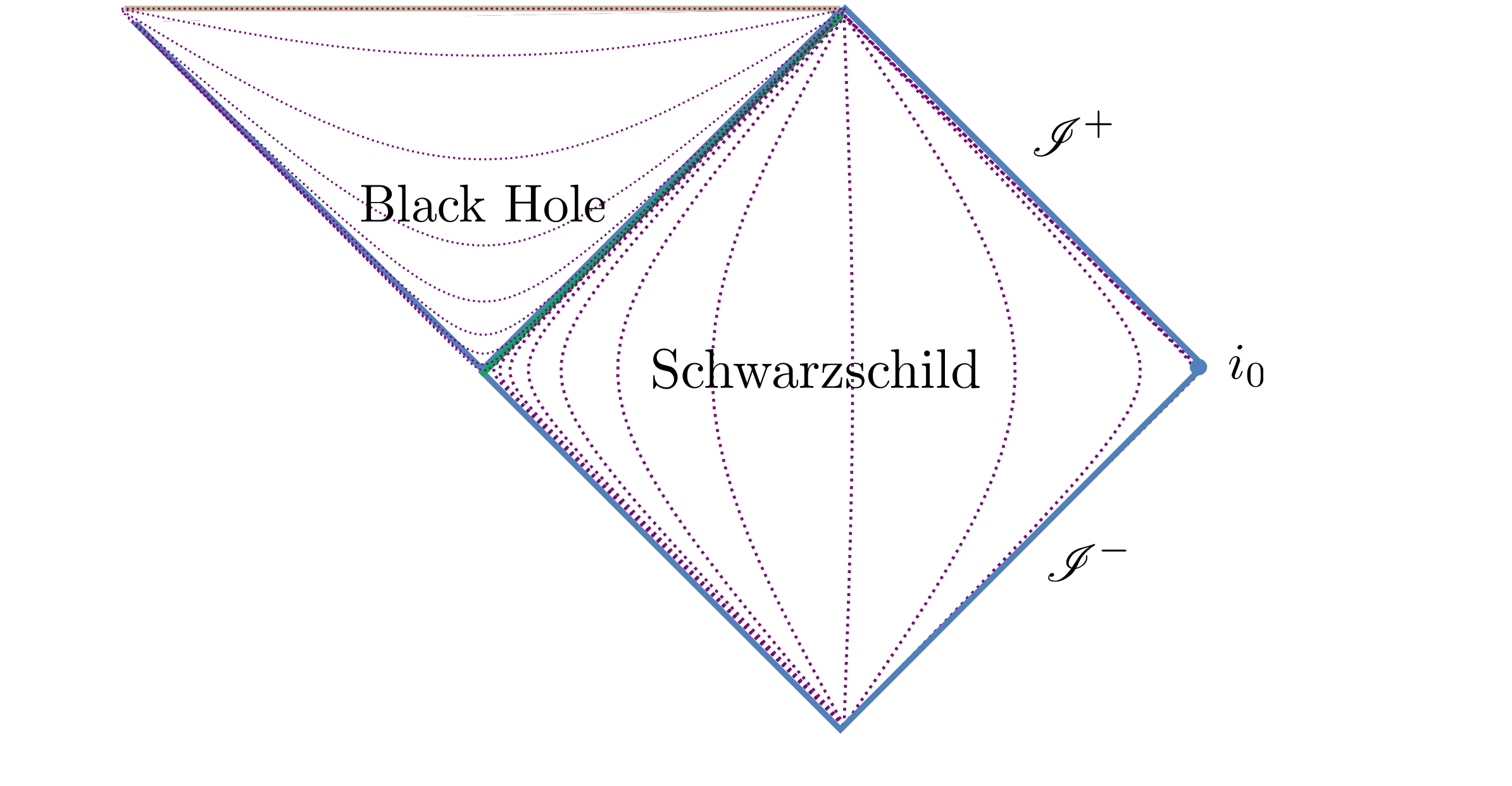} 
    \caption{Constant $z$ slices}
  \end{subfigure}
\caption{}\label{tandzslices}
\end{figure}

Eqs.\eqref{ODEs0} describe an evolution of fields in $z$. Although introducing $z$ and the ansatz \eqref{ansatzfz} are understood as the trick to simplify the PDEs, it is still interesting to understand the constant $z$ slices in the spacetime and obtain a picture of the evolution. Indeed by $\xi^\mu\nabla_\mu z=0$, the killing vector $\xi$ is tangent to the constant $z$ slice. The constant $z$ slices are precisely the Kantowski-Sachs foliation commonly used in earlier studies of LQG black holes e.g. \cite{Ashtekar:2018lag,Ashtekar:2020ckv,Assanioussi:2019twp11,Bohmer:2007wi}. The constant $z$ slices is timelike in the exterior and spacelike in the interior of the black hole. See FIG.\ref{tandzslices} for the illustration in the Schwarzschild spacetime and comparing to the constant $t$ slices in the Lemaitre coordinates. The initial condition of the ODEs \eqref{bc0} and \eqref{bc1} are imposed on the timelike constant $z$ slice far away from the black hole. 

In the earlier work of LQG black holes with the Kantowski-Sachs foliation, one has to treat the black hole interior and exterior separately, with 2 different set of effective equations. In our formulation, the evolutions in the exterior and interior are unified in one set of equations \eqref{ODEs0}, since it is derived from the PDEs in the $(x,t)$ coordinate, which covers both the exterior and interior. Moreover, the dynamics is covariant and independent of the choice of foliations, since it is derived from the manifestly covariant lagrangian.

The set of ODEs \eqref{ODEs0} can be solved numerically with the initial condition \eqref{bc2}. The numerical solution is computed  with high numerical precision using the computing software \texttt{Julia}. The resulting numerical solution satisfies the ODEs up to the maximal error bounded by $\sim 10^{-35}$. 

The geometrical interpretation of the solution is obtained by inserting the solution in the metric \eqref{N=1metric}. The spacetime geometry given by the solution has the following key features: 

\begin{itemize}
  
\item The spacetime geometry is well-approximated by the Schwarzschild geometry for large $z>0$, and it gives corrections to the Schwarzschild spacetime as $z$ becomes small. 

\item The black hole singularity, which happens at $z=0$ in the Schwarzschild spacetime, is resolved. The curvature is always finite.  

\item The spacetime extends smoothly to $z<0$ due to the singularity resolution. For negative $z$ and large $|z|$, the spacetime approach asymptotically to the Nariai geometry ${\rm dS_2}\times S^2$. 

\end{itemize}
The same features also appear in the black hole effective model in \cite{Han:2020uhb,Zhang:2021xoa}, where the model is constructed by applying the simplest $\bar{\mu}$-scheme polymerization to the reduced phase space physical Hamiltonian from gravity coupled to dust.

\begin{figure}[h]
  \begin{center}
  \includegraphics[width = 0.6\textwidth]{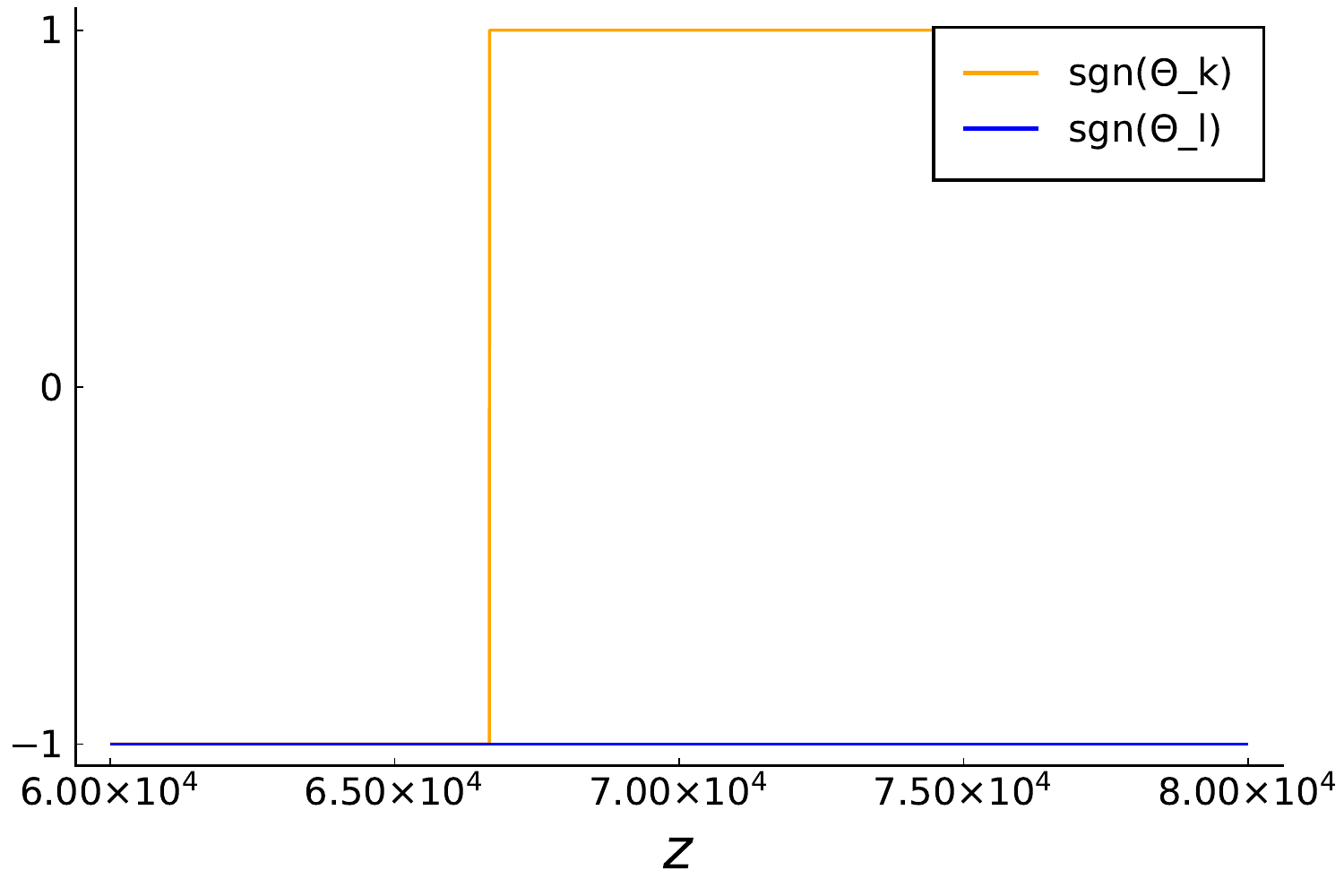} 
   \caption{Plots of $\sgn(\Theta_k)$ (orange) and $\sgn(\Theta_l)$ (blue) of the numerical solution with $z_0=3\times10^8$, $\Delta=0.01$, $\b=1$, $R_s=10^5$. $\Theta_k=0$ is at $z_H\simeq 6.67\times 10^4\simeq \frac{2}{3}R_s$. The correction ${|z_H-2R_s/3|}\sim 10^{-9}$.}
  \label{horizon}
   \end{center}
  \end{figure}

Let us discuss in more detail about these features. Firstly, in the regime of large $z>0$, the correction to the standard Einstein gravity is negligible, and $E^x(z),E^\varphi(z),K_x(z),K_\varphi(z)$ of the solution are well-approximated by \eqref{bc0} and \eqref{bc1} with $z_0$ replaced by $z$. In particular, the quantum correction at the event horizon $z_H = 2R_s/3$ is negligible. The solution gives that the marginal trapped surface locals at $z\simeq z_H$ with negligible correction. The location of the marginal trapped surface is given by $\Theta_k = 0$ and $\Theta_l < 0$ where $\Theta_k $ and $\Theta_l$ are outward and inward null expansions (see FIG.\ref{horizon}). The marginal trapped surface corresponds to the killing horizon in the spacetime, although it is not the event horizon since the singularity is resolved.

\begin{figure}[h]
  \begin{center}
  \includegraphics[width = 0.6\textwidth]{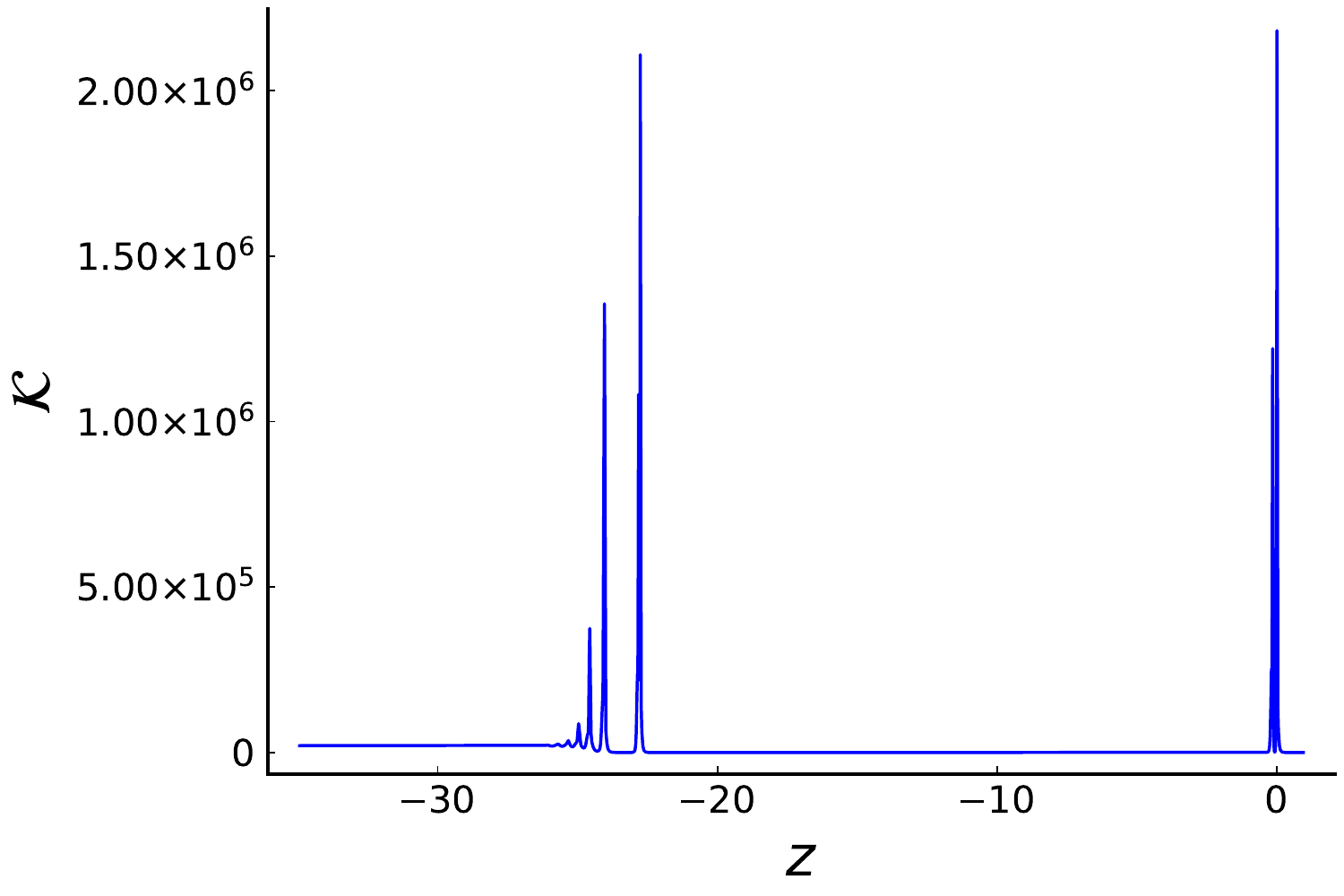} 
   \caption{Plot of Kretschmann invariant $\ck$ of the numerical solutions with $z_0=3\times10^8$, $\Delta=0.01$, $R_s=10^5$.}
  \label{krets}
   \end{center}
  \end{figure}

\begin{figure}[h]
    \begin{subfigure}{0.5\textwidth}
      \includegraphics[width = 1\textwidth]{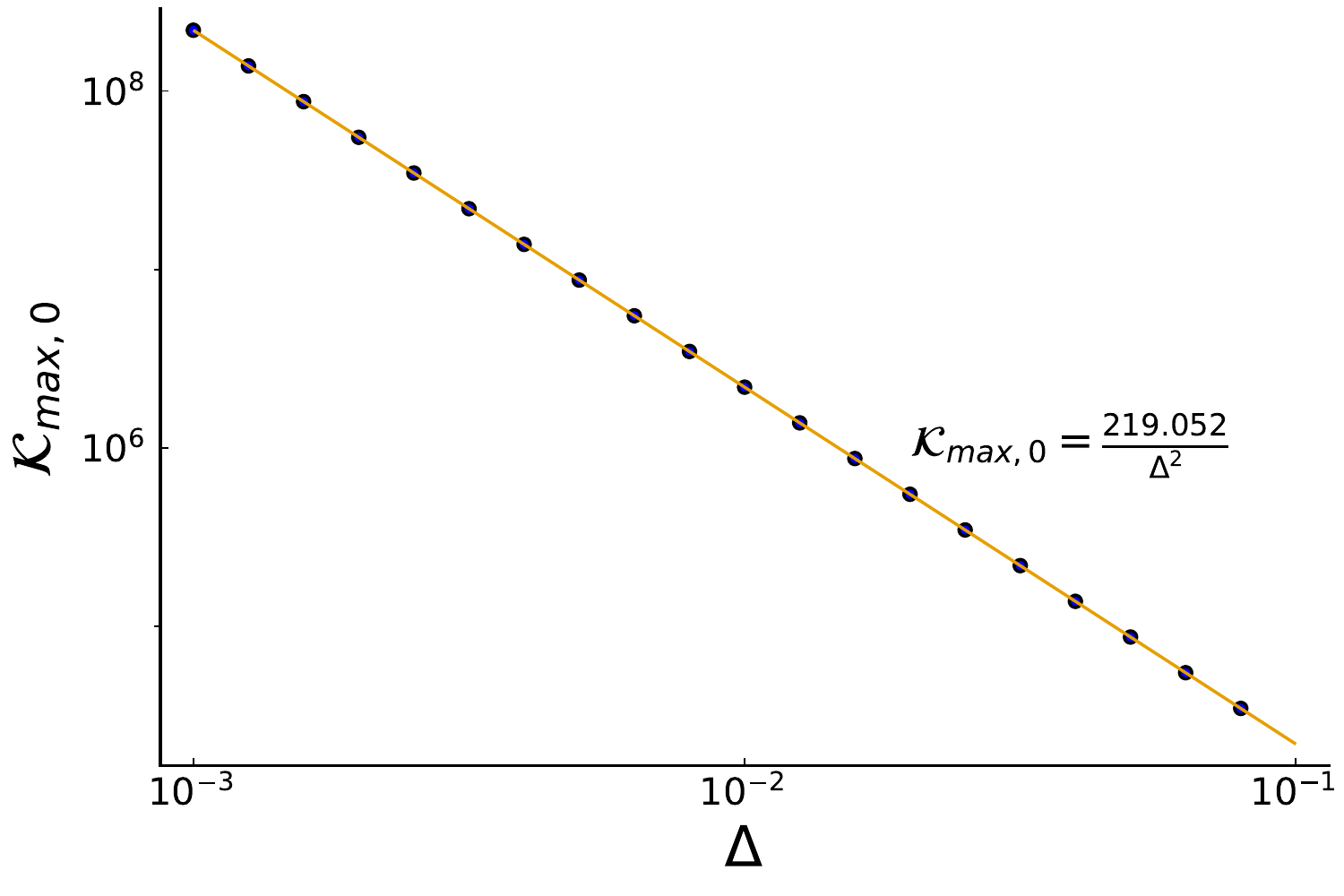} 
      \caption{}
    \end{subfigure}
    \begin{subfigure}{0.5\textwidth}
      \includegraphics[width = 1\textwidth]{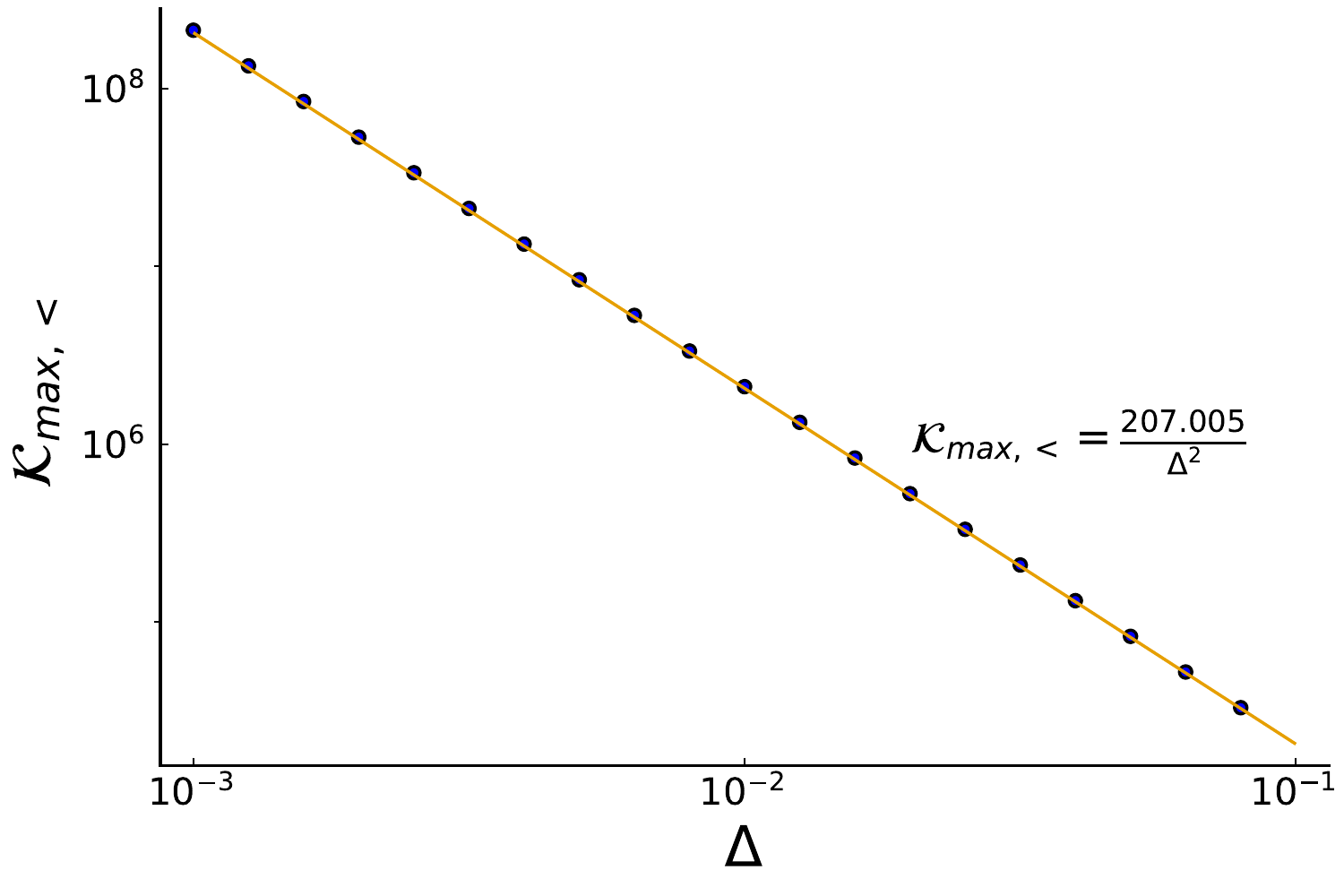} 
      \caption{}
    \end{subfigure}
  \caption{(a) $\Delta$ versus the maximum $\ck_{max,0}$, for $z$ inside the neighborhood $N_0$. (b) $\Delta$ versus the maximum of $\ck_{max,<}$, for $z$ inside the neighborhood $N_<$.}\label{Kfit}
  \end{figure}

\begin{figure}[h]
    \begin{subfigure}{0.5\textwidth}
    \includegraphics[width = 1\textwidth]{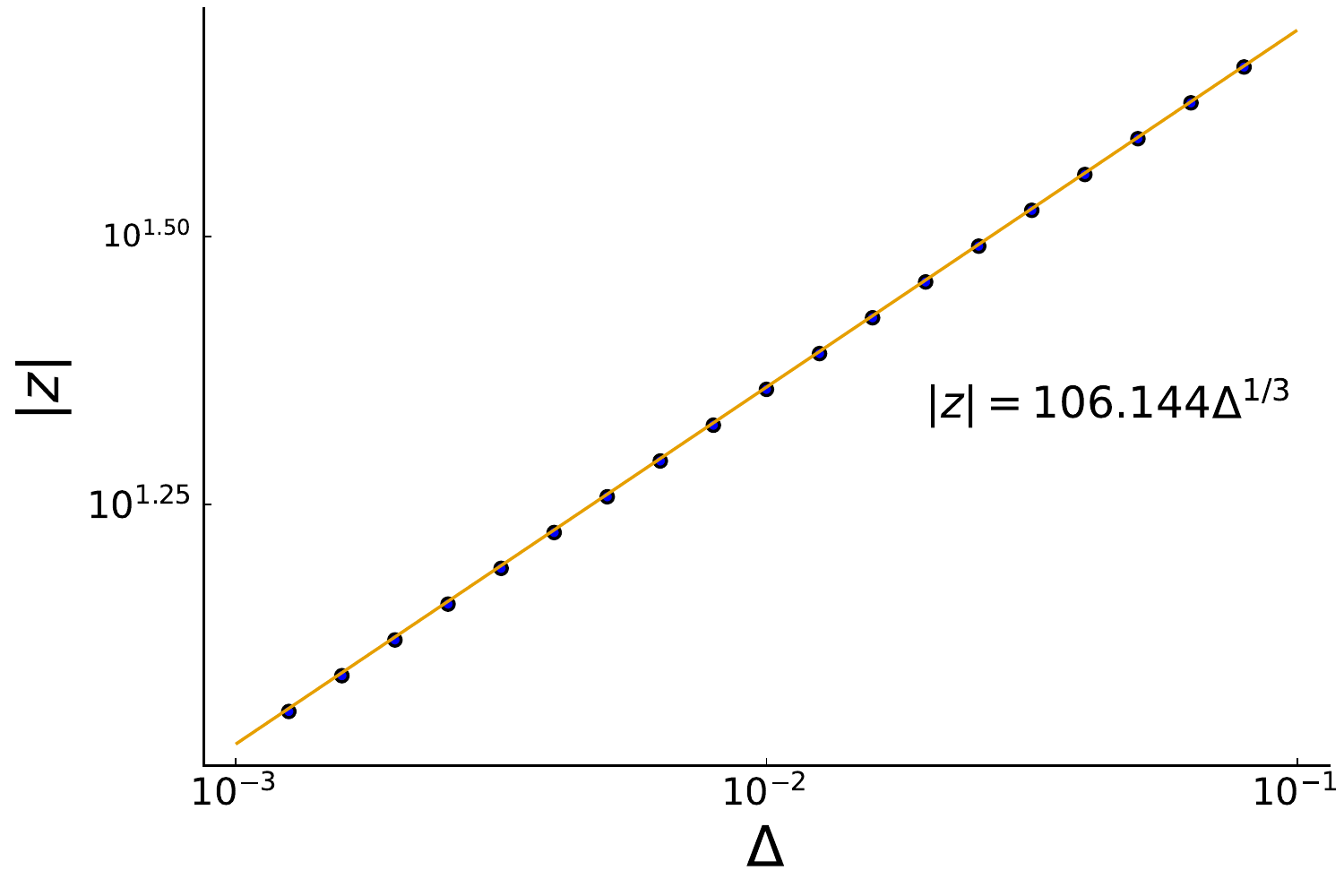} 
    \caption{}
  \end{subfigure}
  \begin{subfigure}{0.5\textwidth}
    \includegraphics[width = 1\textwidth]{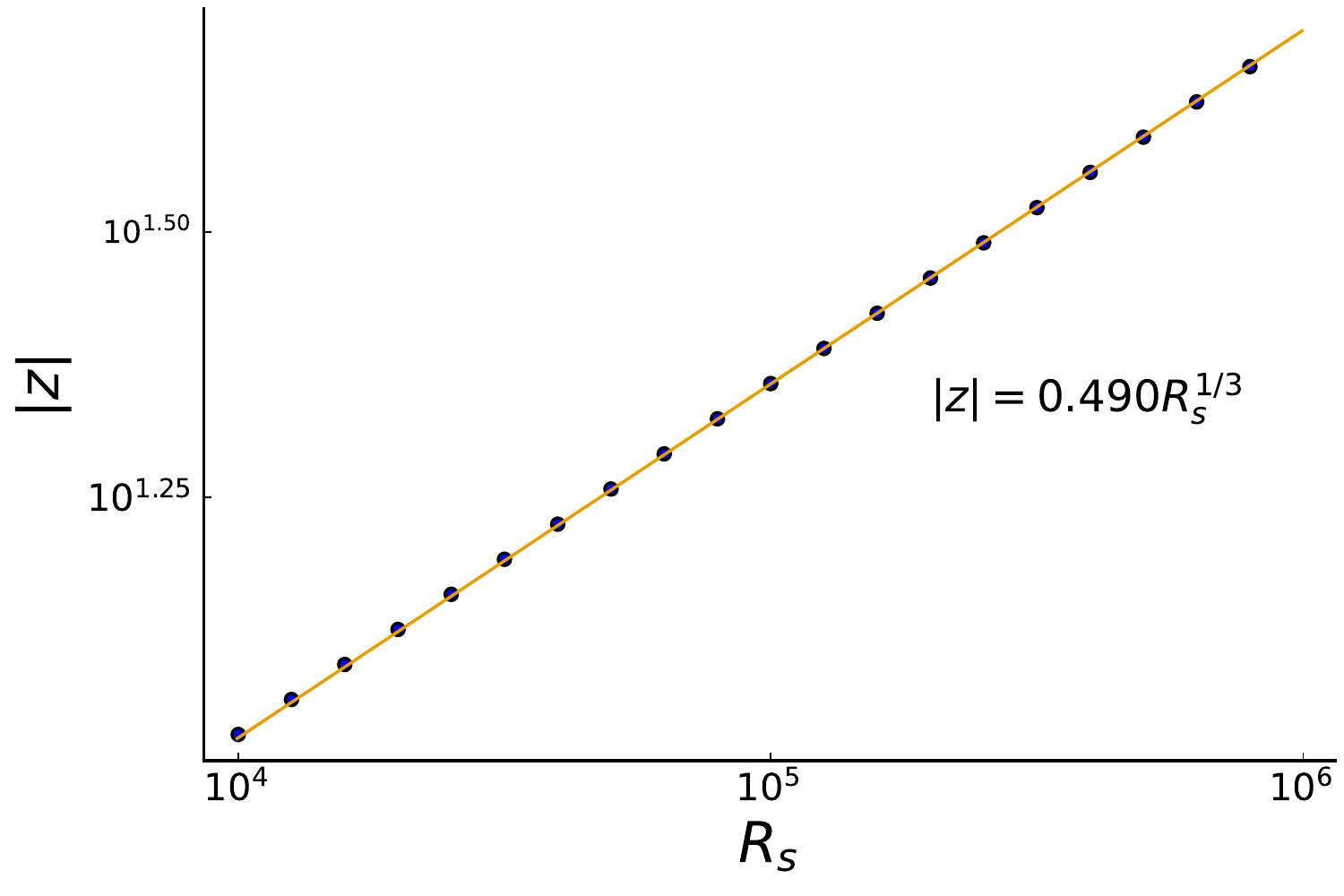} 
    \caption{}
  \end{subfigure}
    \caption{The distance $|z|$ between the locations of two $\ck$ maxima in $N_0$ and $N_<$ and the relations with $\Delta$ and $R_s$.  }
    \label{z-delta}
  \end{figure}

The Kretschmann invariant $\ck=R^{\mu\nu\rho\sigma}R_{\mu\nu\rho\sigma}$ is bounded, as shown in FIG.\ref{krets}. The black hole singularity at $z=0$ is resolved. The spacetime geometries extend smoothly to $z<0$. It demonstrates 2 groups of local maxima of $\ck$ located respectively in the neighborhood $N_0$ of $z=0$ and in a neighborhood $N_<$ of $z<0$. We compute the maximal value of $\ck$ in $N_0$ and $N_<$ respectively, denote them by $\ck_{max,0}$ and $\ck_{max,<}$, and test their dependence on $\Delta$. The numerics demonstrate that both $\ck_{max,0}$ and $\ck_{max,<}$ are proportional to $\Delta^{-2}$  (see FIG \ref{Kfit}). If $\Delta\sim\ell_P^2$ relates to the minimal area gap in LQG, The behavior of Kretschmann scalar $ \ck_{max,0},\ck_{max,<}\sim \Delta^{-2}$ indicates that the singularity resolution happens at the Planckian curvature. The distance $|z|$ between the locations of two maxima $\ck_{max,0}$ and $\ck_{max,<}$ relates to both $\Delta$ and $R_s$ and behaves as $|z|\sim R_s^{1/3}\Delta^{1/3}$, see FIG.\ref{z-delta}. Asymptotically for large negative $z$, $\ck$ approaches to be $z$-independent constant, whose dependence on $\Delta$ is still $\sim \Delta^{-2}$. We come back to this asymptotic behavior shortly. 

\begin{figure}[h]
  \begin{subfigure}{0.5\textwidth}
    \includegraphics[width = 1\textwidth]{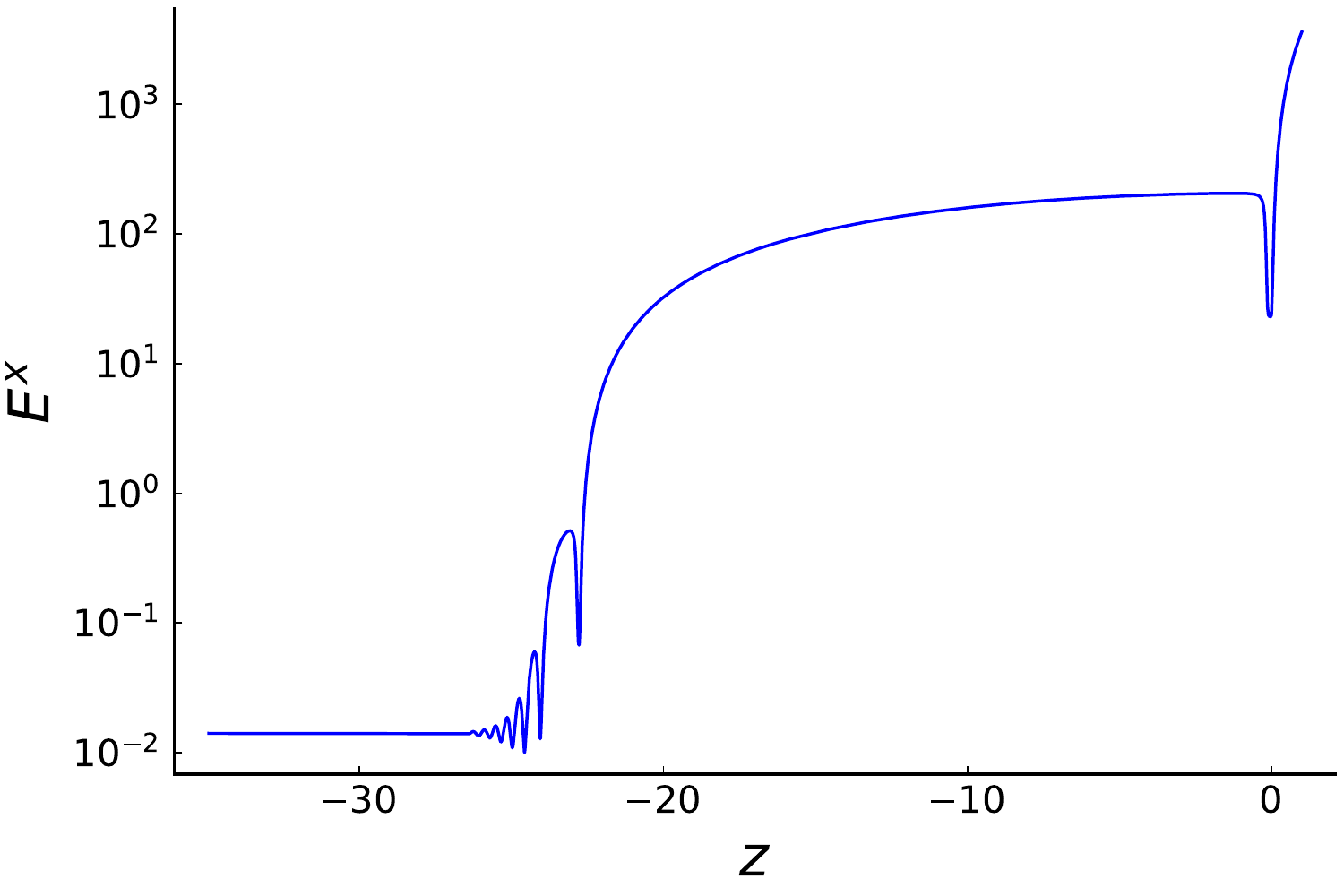} 
    \caption{}\label{plotEx}
  \end{subfigure}
  \begin{subfigure}{0.5\textwidth}
    \includegraphics[width = 1\textwidth]{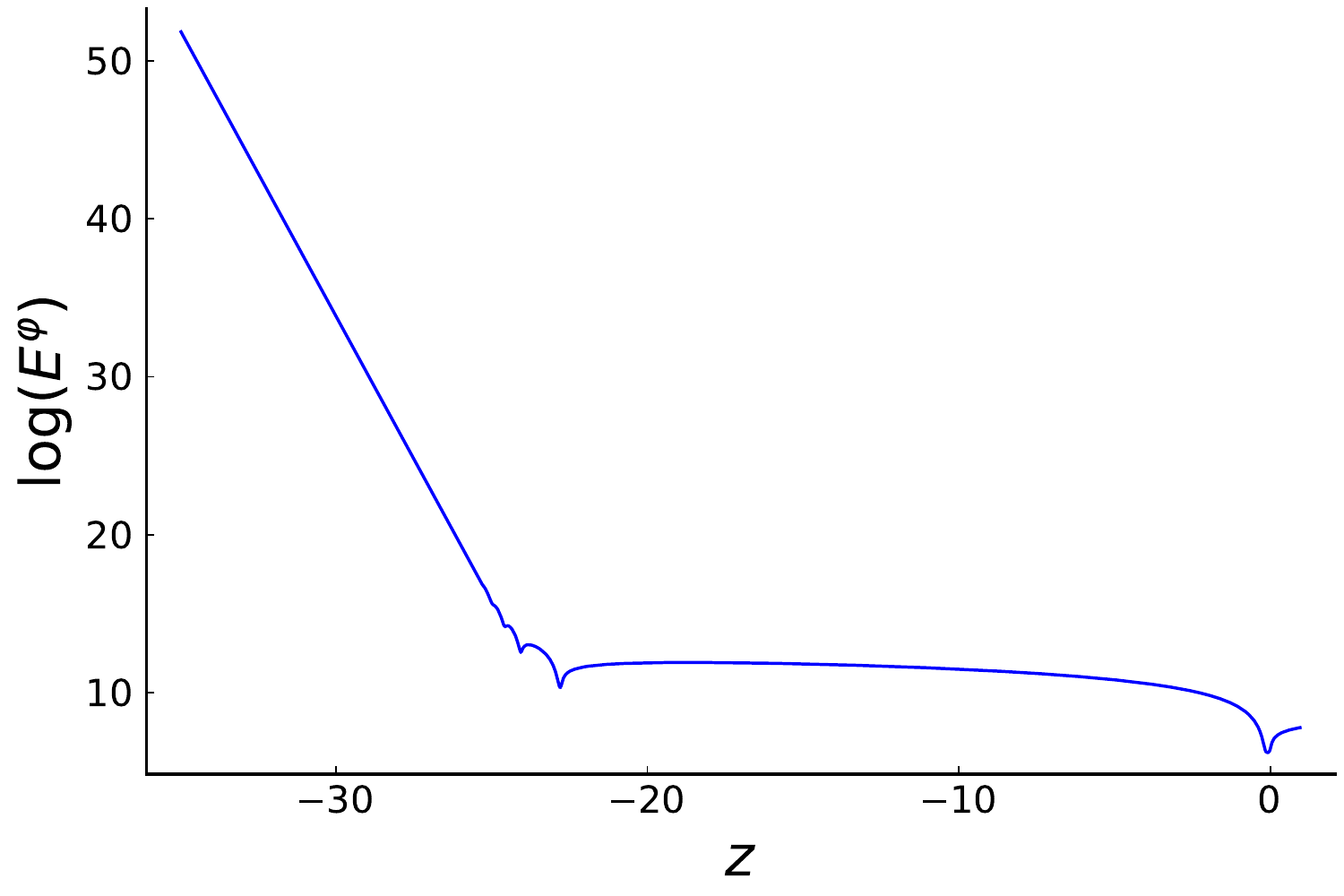} 
    \caption{}\label{plotLogEphi}
  \end{subfigure}\\
  \begin{subfigure}{0.5\textwidth}
    \includegraphics[width = 1\textwidth]{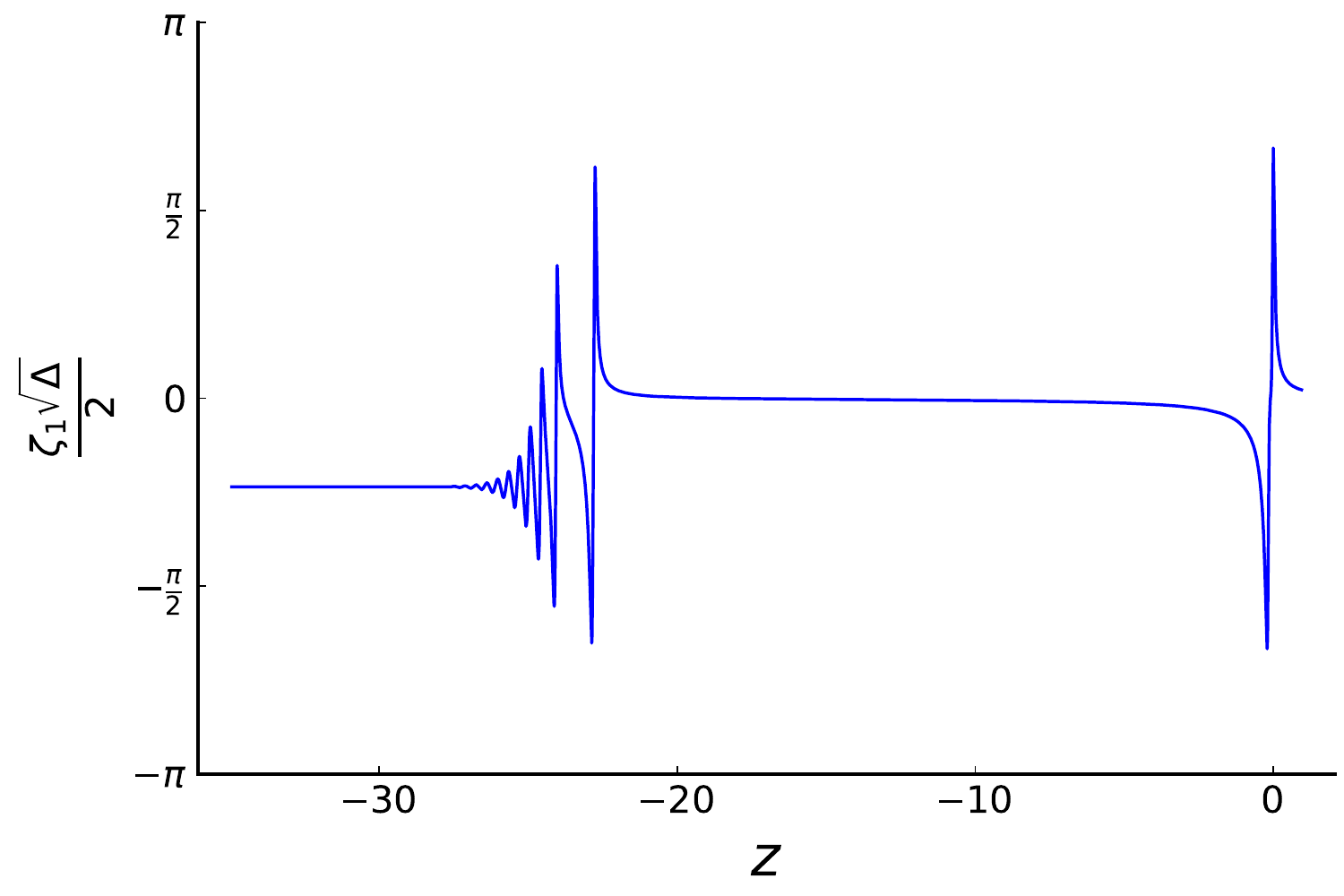} 
    \caption{}\label{plotK1}
  \end{subfigure}
  \begin{subfigure}{0.5\textwidth}
    \includegraphics[width =1\textwidth]{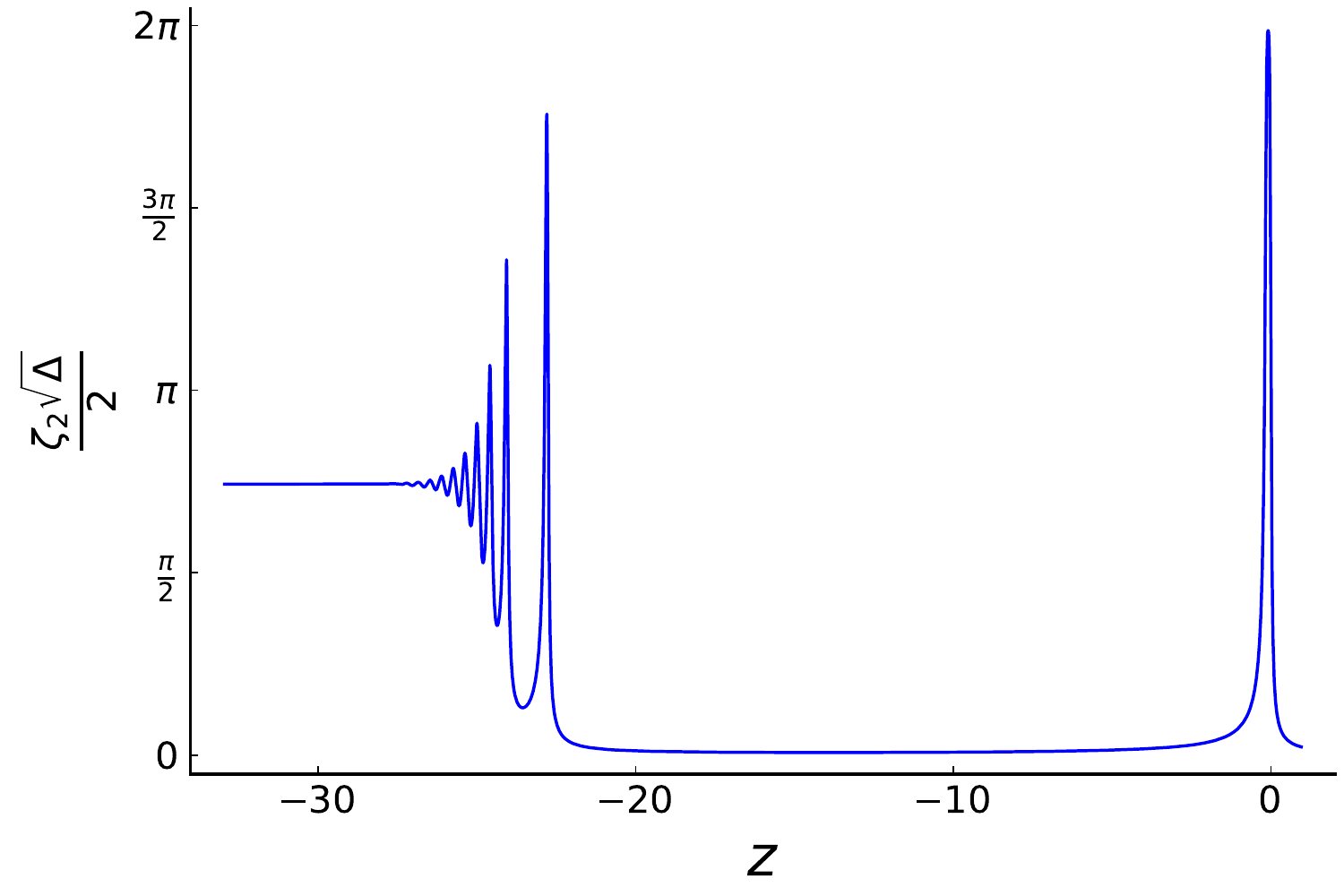} 
    \caption{}\label{plotK2}
  \end{subfigure}
\caption{Plots of $E^x=e^{2\psi},\ \log(E^\varphi)=\psi+\xi,\ \zeta_1\sqrt{\Delta},\ \zeta_2\sqrt{\Delta}$ of the numerical solutions in the regime $z<1$ with $z_0=3\times10^8$, $\Delta=0.01$, $R_s=10^5$.}\label{plotsol}
\end{figure}

\begin{figure}[h]
  \begin{subfigure}{0.5\textwidth}
    \includegraphics[width = 1\textwidth]{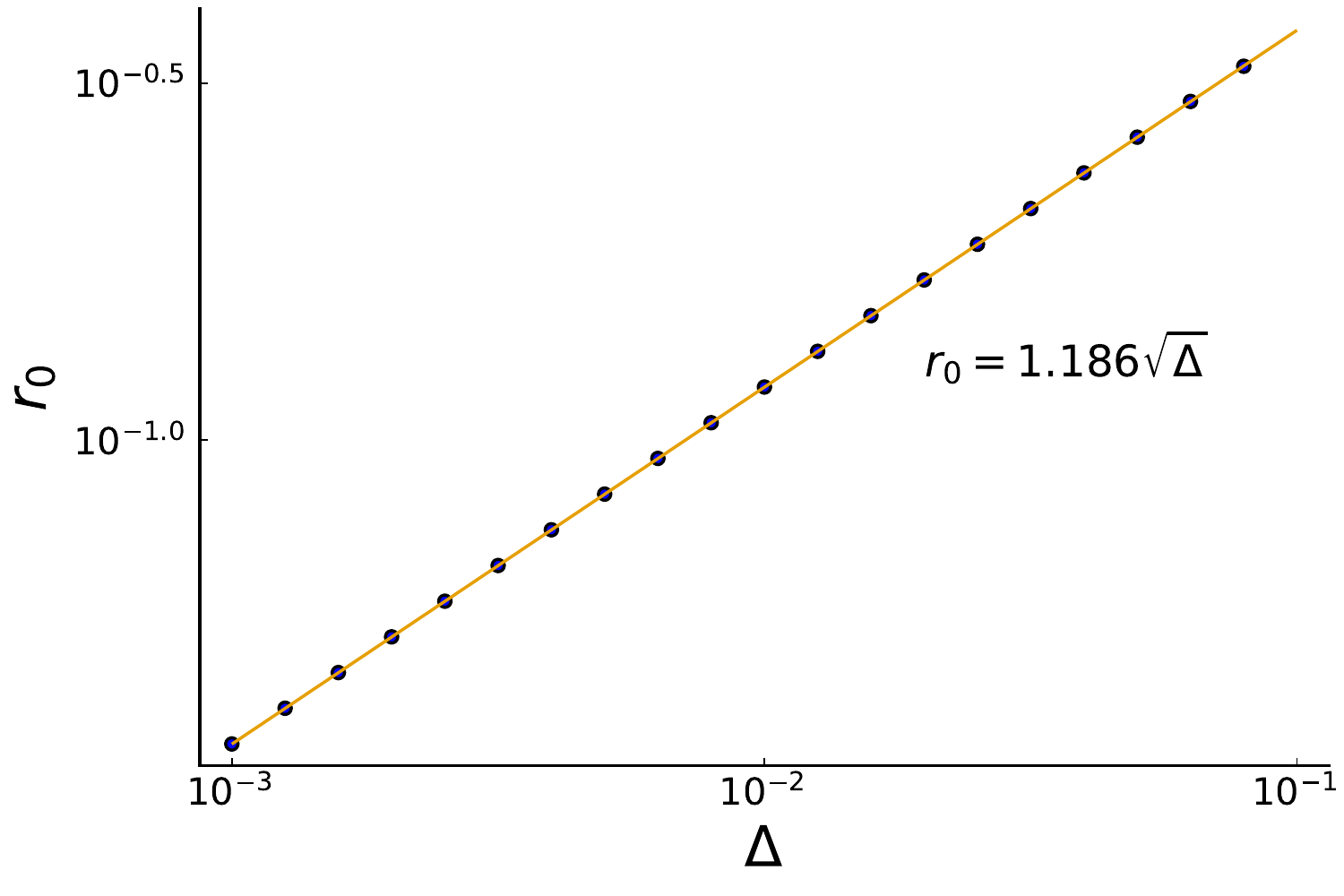} 
    \caption{}\label{r0fit}
  \end{subfigure}
  \begin{subfigure}{0.5\textwidth}
    \includegraphics[width = 1\textwidth]{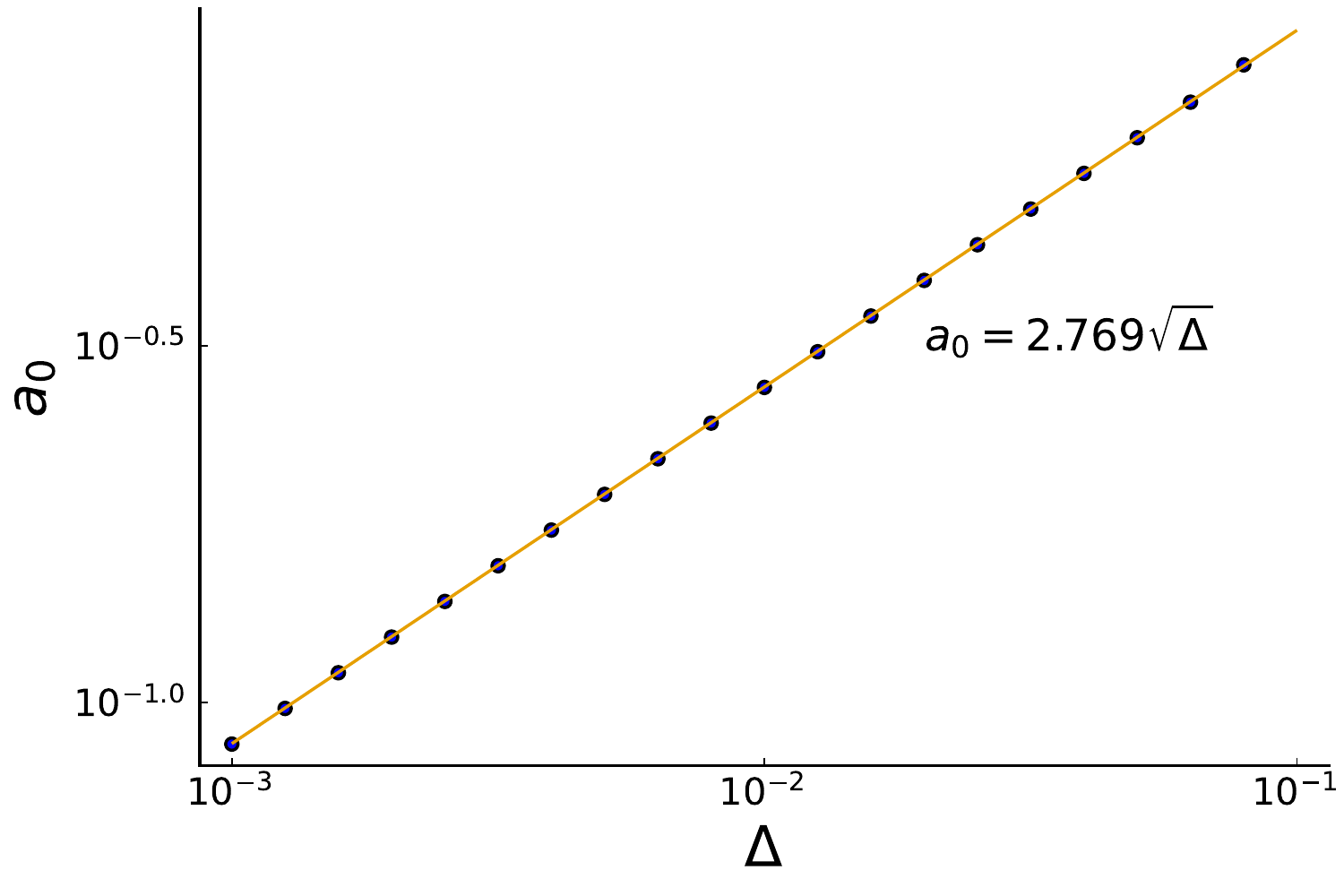} 
    \caption{}\label{a0fit}
  \end{subfigure}
\caption{The blue dots are the numerical values of $r_0,a_0$ from the solutions at different values of $\Delta\in[10^{-3},10^{-1}]$. The orange lines plot the best fit functions. Other parameters used in the numerics are $z_0=3\times 10^8$, $R_s=10^5$.}\label{r0a0}
\end{figure}

FIG.\ref{plotsol} demonstrates $E^x,\ \log(E^\varphi),\ \zeta_1\sqrt{\Delta},\ \zeta_2\sqrt{\Delta}$ of the numerical solution, when evolving smoothly across the Schwarzschild singularity at $z=0$ and extending to $z<0$. From FIG.\ref{plotEx}, the evolution of $E^x$ shows that the radial coordinate $r=\sqrt{E^x}$ is not a good coordinate anymore when extending the spacetime to $z<0$, since $r=\sqrt{E^x}$ is not monotonic in the evolution. $(t,x)$ are good coordinates for the extended spacetime due to the regularity of the metric components. As $z\to -\infty$, $E^x, \zeta_1, \zeta_2$ approach to constants while $E^\varphi \sim e^{-z/a_0-a_1}$ with $a_0,a_1>0$. We denote by $E^x\sim r_0^2$ the constant as $z\to -\infty$. The solution indicates that the spacetime geometry approaches asymptotically the following metric as $z\to -\infty$
\be
\rmd s^2\sim -\rmd t^2+r_0^{-2}e^{2(t-x)/a_0-2a_1}\rmd x^2+r_0^2(\rmd\theta^2+\sin ^2\theta\rmd\varphi^2).\label{ds2s2}
\ee
This metric is the $\rmd S_2\times S^2$ geometry with $\rmd S$-radius $a_0$ and $S^2$-radius $r_0$. The geometry is also known as the Nariai geometry \cite{Bousso:1996pn,Hawking:1995ap}. Here the existence of $\rmd S_2\times S^2$ is a consequence of the covariant $\bar{\mu}$-scheme Hamiltonian, which comes from the higher derivative coupling in the mimetic gravity. Indeed both $r_0$ and $a_0$ depend on $\Delta$. The dependence of $r_0$ and $a_0$ on $\Delta$ is analyzed numerically, and the results are shown in FIG.\ref{r0a0}. The results indicate the following scaling properties of $r_0$ and $a_0$:
\be
r_0,a_0\propto\sqrt{\Delta}.
\ee
Both the sphere radius $r_0$ and the effective cosmological constant $1/a_0^2\sim \Delta^{-1}$ relate to the quantum effect. In $\rmd S_2\times S^2$, the Kretschmann invariant does not depend on $z$: $\ck_{\rmd S_2\times S^2}=4(a_0^{-4}+r_0^{-4})\sim \Delta^{-2}$. This explains the asymptotically constant behavior of $\ck$ as $z\to-\infty$ in FIG.\ref{krets}. The ratio between $\ck_{max,0}$ and $\ck_{\rmd S_2\times S^2}$ is approximately $104.556$.

Recall that the 4d spacetime manifold is a product $\sm_4\simeq \sm_2\times S^2$, when we study the spherical symmetric gravity. If we suppress the $S^2$ factor and focus on the geometry on the 2d manifold $\sm_2$, the 2d spacetime $h_{ij}\rmd x^i\rmd x^j=-\rmd t^2+\frac{(E^\varphi)^2}{E^x}\rmd x^2$ given by the solution leads to the conformal diagram as in FIG.\ref{spacetime}. The maximal extension is given in FIG.\ref{maxextension} (see Appendix \ref{conformal diagram} for the conformal factor used for the diagram). The entire 2d spacetime is non-singular, and has the complete null infinity $\mathscr{I}^+$. A part of $\mathscr{I}^+$ is spacelike as the null infinity of the asymptotic $\mathrm{dS}_2$, while the other part of $\mathscr{I}^+$ is null as the null infinity of the asymptotic Schwarzschild geometry. The point in the conformal diagram where the spacelike and null parts of $\mathscr{I}^+$ meet is the timelike infinity $i_+$ for spacetime region outside the black hole, and it corresponds the spatial infinity $i_0$ for the asymptotic $\mathrm{dS}_2$, although it is not the spatial infinity of the entire spacetime.

There is no event horizon due to the singularity resolution. $z\simeq z_H$ foliated by the marginal trapped surfaces is a killing horizon\footnote{The killing vector has the norm $g_{\mu\nu}\xi^\mu \xi^\nu=-1+e^{2\xi(z)}$. Comparing to the expansions $\Theta_k=(e^{-\xi (z)} -1)\frac{\psi '(z)}{\sqrt{2}}$ and $\Theta_l=-(e^{-\xi (z)} +1)\frac{\psi '(z)}{\sqrt{2}}$ shows that the killing vector is null when $\Theta_k=0$ and $\Theta_l< 0$. }. We have called the region inside the killing horizon the black hole interior.  

FIG.\ref{spacetime} is the conformal diagram for the 2d spacetime rather than the full 4d spacetime, because the 4-dimensional $\mathrm{dS}_2\times S^2$ metric dividing the conformal factor gives vanishing $S^2$ radius at $\mathscr{I}^+$ \footnote{We thank Abhay Ashtekar for pointing this out.}. Note that this conformal diagram is the same as the one obtained in \cite{Han:2020uhb}.

\begin{figure}[h]
  \begin{center}
  \includegraphics[width = 0.6\textwidth]{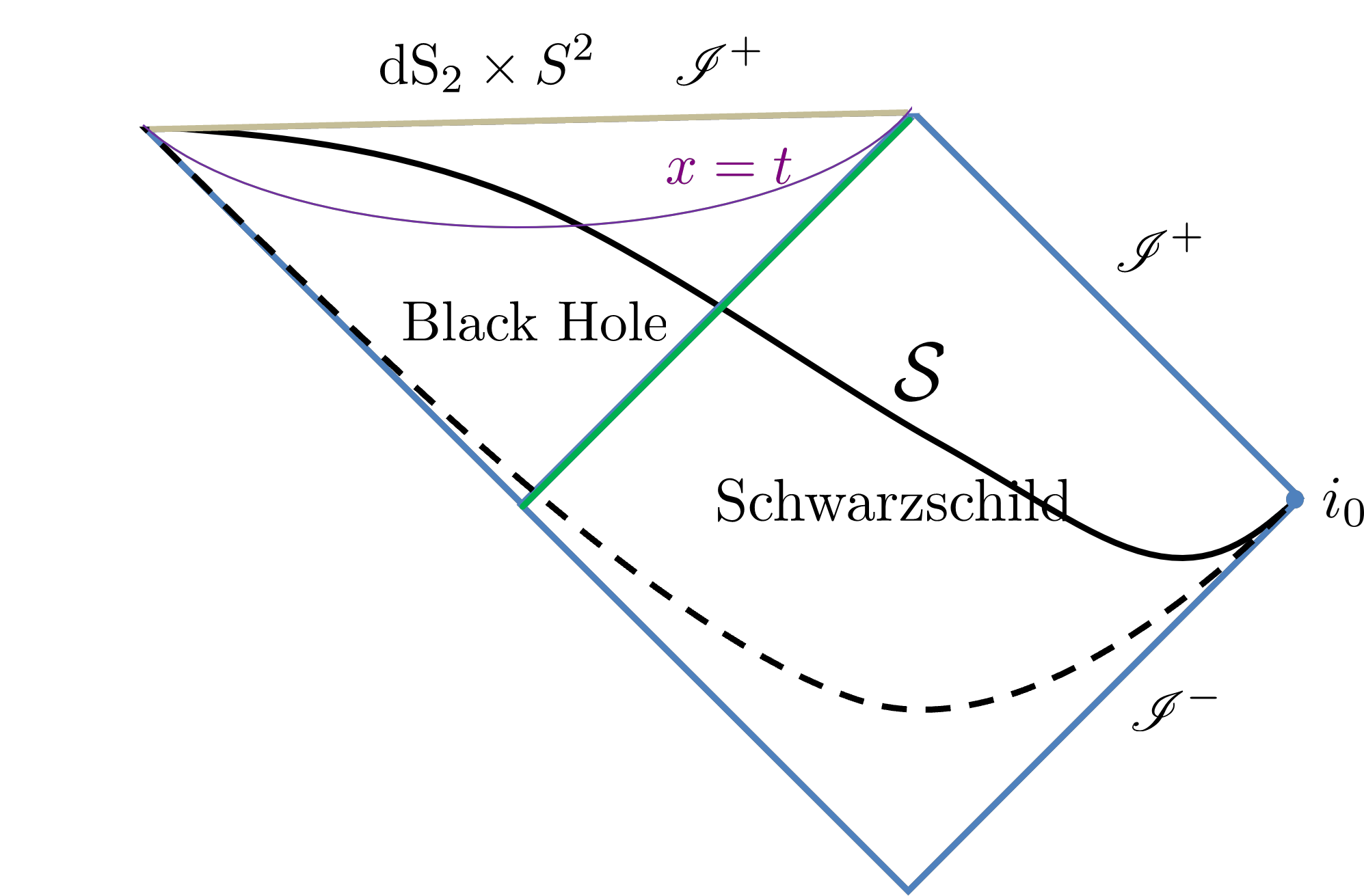} 
   \caption{The conformal diagram of the non-singular black hole spacetime reduced to 2d covered by $(t,x)$ coordinate. $\cs$ (black curve) is a typical spatial slice with constant $t$. Dashed curves are another spatial slice in the far past. The green line illustrate the killing horizon. Near the future infinity, the 4d asymptotic geometry is ${\rm dS}_2\times S^2$ with Planckian radii.}
  \label{spacetime}
   \end{center}
\end{figure}

\begin{figure}[h]
  \begin{center}
  \includegraphics[width = 0.6\textwidth]{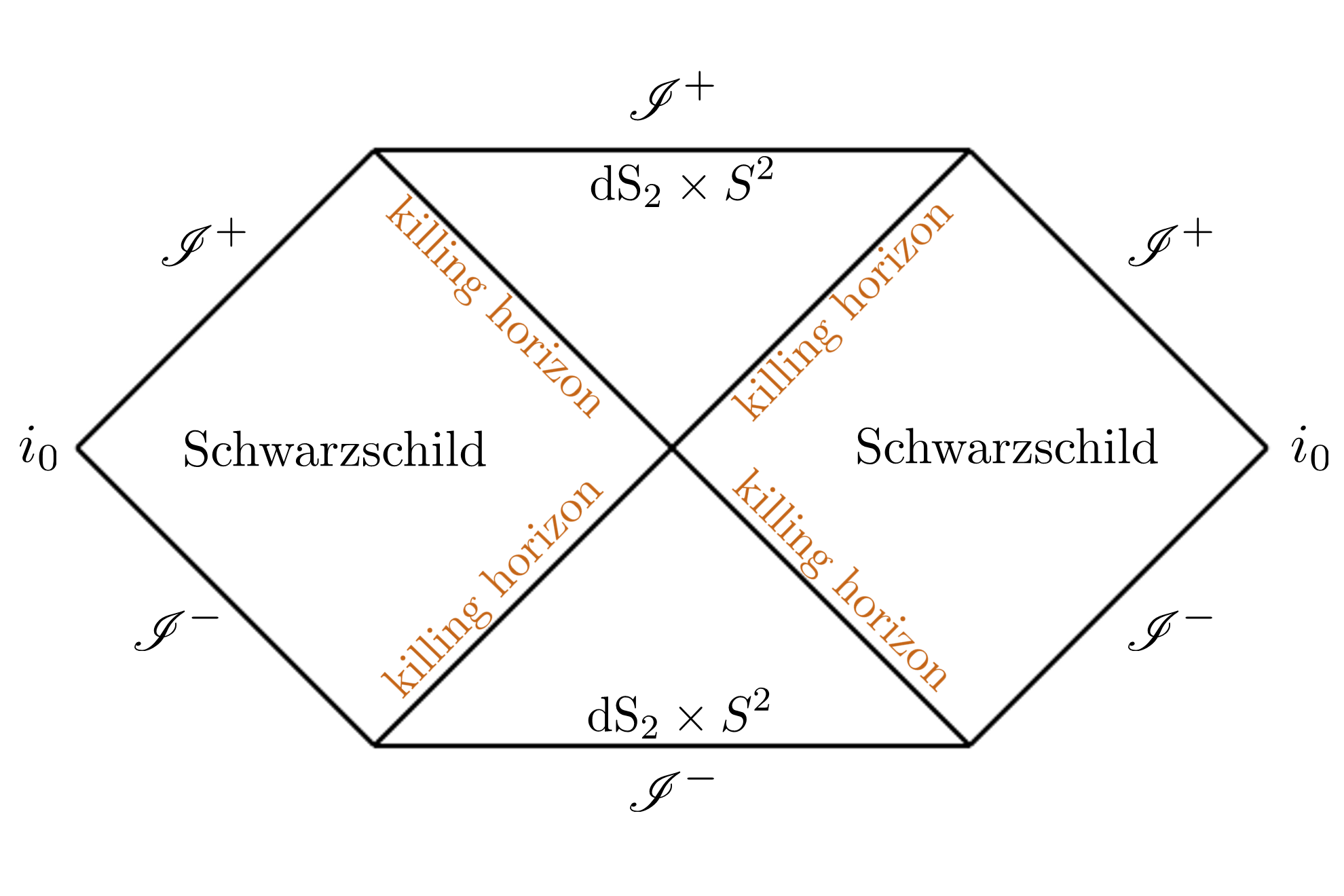} 
   \caption{The conformal diagram of the 2d maximal extension.}
  \label{maxextension}
   \end{center}
\end{figure}

\begin{figure}[h]
  \begin{center}
  \includegraphics[width = 0.7\textwidth]{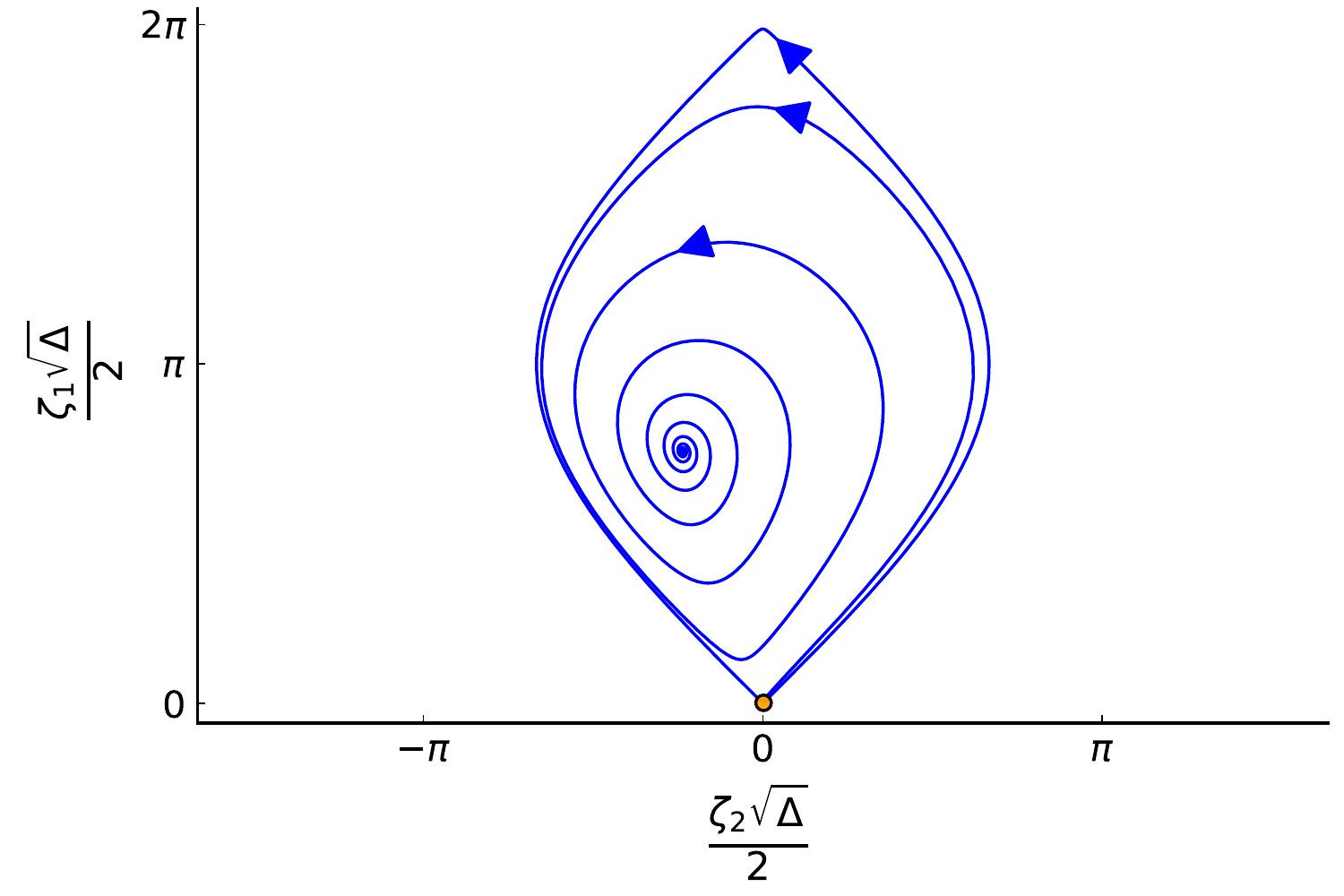} 
  \end{center}
  \caption{This figure plots the trajectory of the $z$-evolution in the $(\zeta_1,\zeta_2)$-space. The arrow indicates the direction of the evolution from $z>0$ to $z<0$. The orange dot indicates the initial value of $\zeta_1,\zeta_2$ at $z_0$. The attractor is the $\mathrm{dS}_2\times S^2$ geometry.}
  \label{attractor}
\end{figure}

An interesting feature of the black hole solution is demonstrated by plotting the trajectory of the $z$-evolution in the $(\zeta_1,\zeta_2)$-space, shown in FIG.\ref{attractor}. The evolutions of $\zeta_1,\zeta_2$ are the keys of the $\bar{\mu}$-scheme dynamics, because they determines $\psi,\xi$ thus the metric by \eqref{eomxi} and \eqref{eompsi}. The $z$-evolution begins with $(0,0)$ in the $(\zeta_1,\zeta_2)$-space and gives a spiral curve falling into the attractor. The covariant $\bar{\mu}$-scheme effective equations \eqref{ODEz1} - \eqref{ODEz4} have 2 types of sine/cosine functions of $\zeta_1$ and $\zeta_2$ respectively, so the trajectory in the $(\zeta_1,\zeta_2)$-space is bounded. Moreover, viewing $\zeta_1,\zeta_2$ as the subsystem and $\xi,\psi$ as the ``environment'', the coupling between $\zeta_1,\zeta_2$ and $\xi,\psi$ leads to the ``dissipation'' causing the radius of the circular trajectory to shrink during the evolution thus resulting in the spiral curve. The trajectory converges to the attractor that corresponds to the $\mathrm{dS}_2\times S^2$ geometry. This phenomena is common in the dissipative dynamical systems \cite{blanchard2012differential}. FIG.\ref{attractor} suggests that there should be an ``basin of attraction'', inside which any initial value of $(\zeta_1,\zeta_2)$ evolves convergently to the $\mathrm{dS}_2\times S^2$ attractor.

\subsection{Null expansion, stress-energy tensor, and quasi-normal oscillation near $\mathrm{dS_2}\times S^2$}

\begin{figure}[h]
  \begin{subfigure}{0.5\textwidth}
    \includegraphics[width = 1\textwidth]{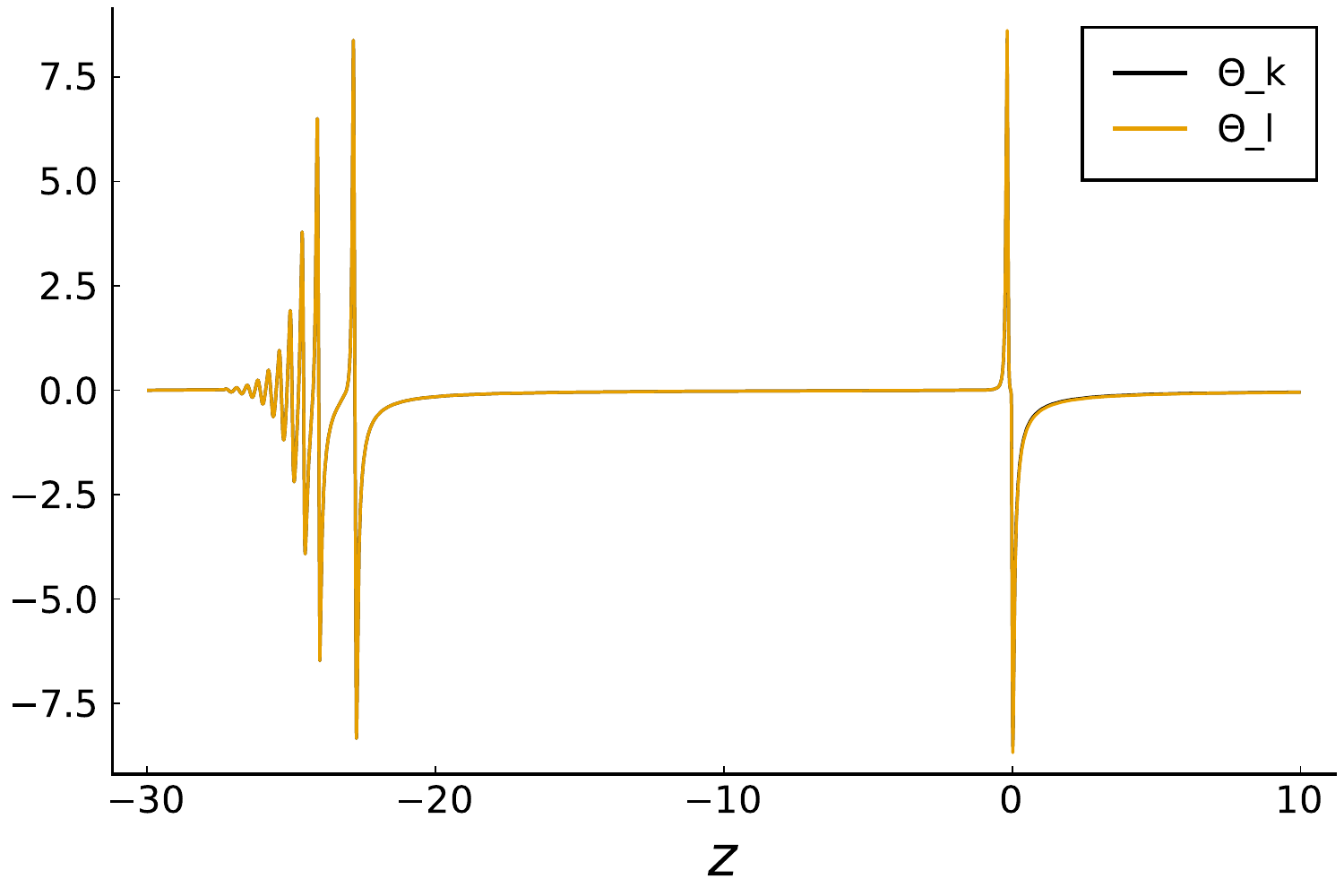} 
    \caption{}\label{expansionkl0}
  \end{subfigure}
  \begin{subfigure}{0.5\textwidth}
    \includegraphics[width = 1\textwidth]{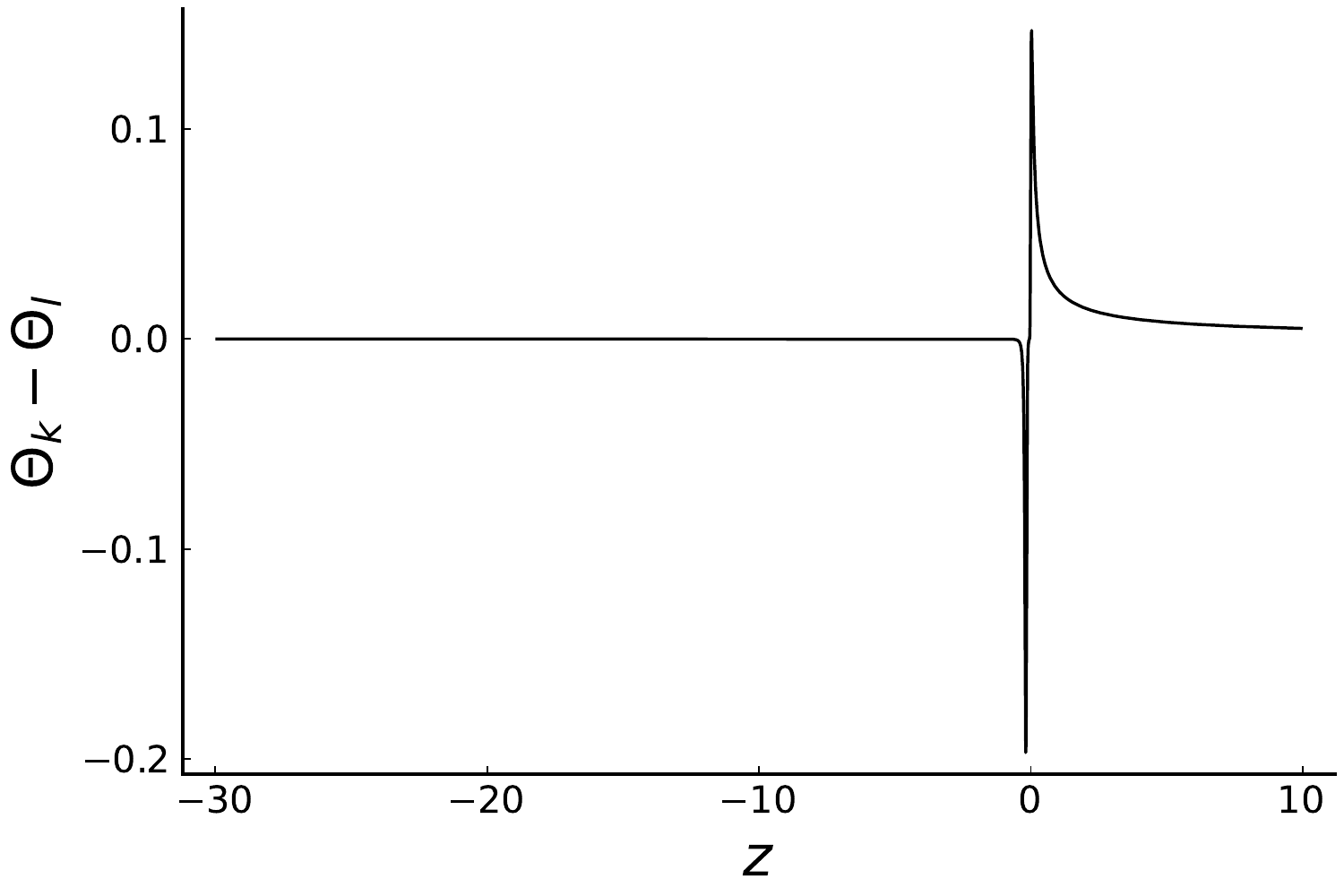} 
    \caption{}\label{expansion_diff}
  \end{subfigure}
\caption{(a) The plots of expansions $\Theta_k$ and $\Theta_l$ for $z\leq 10$. (b) The plots of the difference $\Theta_k-\Theta_l$.}
\label{expansionkl}
\end{figure}

In the sperical symmetric spacetime \eqref{N=1metric}, the outward and inward null geodesic congruences are generated by $k=\frac{1}{\sqrt{2}}(\partial_t+\frac{\sqrt{E^x}}{E^\varphi}\partial_x)$ and $l=\frac{1}{\sqrt{2}}(\partial_t-\frac{\sqrt{E^x}}{E^\varphi}\partial_x)$. Their expansions are given by
\be
\Theta_k&=&\frac{1}{2}\tilde{h}^{\a\b}\nabla_\a k_\b=\frac{E^x{}'}{2 \sqrt{2} \sqrt{E^x} E^\varphi}-\frac{E^x{}'}{2 \sqrt{2} E^x}=(e^{-\xi } -1)\frac{\psi '}{\sqrt{2}}\\
\Theta_l&=&\frac{1}{2}\tilde{h}^{\a\b}\nabla_\a l_\b=-\frac{E^x{}'}{2 \sqrt{2} \sqrt{E^x} E^\varphi}-\frac{E^x{}'}{2 \sqrt{2} E^x}=(-e^{-\xi } -1)\frac{\psi '}{\sqrt{2}}
\ee
where $\tilde{h}_{\a\b}$ is the induced metric on $S^2$. At the killing horizon $z\simeq \frac{2}{3}R_s$, $\Theta_k$ flips sign from positive to negative, while $\Theta_l$ keeps negative. For $z<\frac{2}{3}R_s$, $\Theta_k$ and $\Theta_l$ are of the same sign, although they can flips signs at the same time. In particular, when $z<0$, we have that $e^{-\xi}\ll 1$ is negligible in $\Theta_k$ and $\Theta_l$, so we have $\Theta_k\simeq\Theta_l\simeq -\psi'/\sqrt{2}$ in this regime (see FIG.\ref{expansionkl} from the numerical solution discussed in Section \ref{Non-singular black hole and asymptotic Nariai geometry}).

$\Theta_k,\Theta_l$ are oscillatory for $z< 20$ (see FIG.\ref{expansionkl0}) and give many transition surfaces, at which both $\Theta_k,\Theta_l$ change signs at the same time. Finally the expansions stablize at $\Theta_k=\Theta_l=0$ in the asymptotic $\mathrm{dS}_2\times S^2$ geometry, where $\psi'=0$. The oscillation of $\Theta_k,\Theta_l$ is purely a consequence of the oscillation of the $S^2$ area $E^x=e^{2\psi}$, since $\Theta_k\simeq\Theta_l\simeq -\psi'/\sqrt{2}$ in this regime.

\begin{figure}[h]
    \begin{subfigure}{0.5\textwidth}
      \includegraphics[width = 1\textwidth]{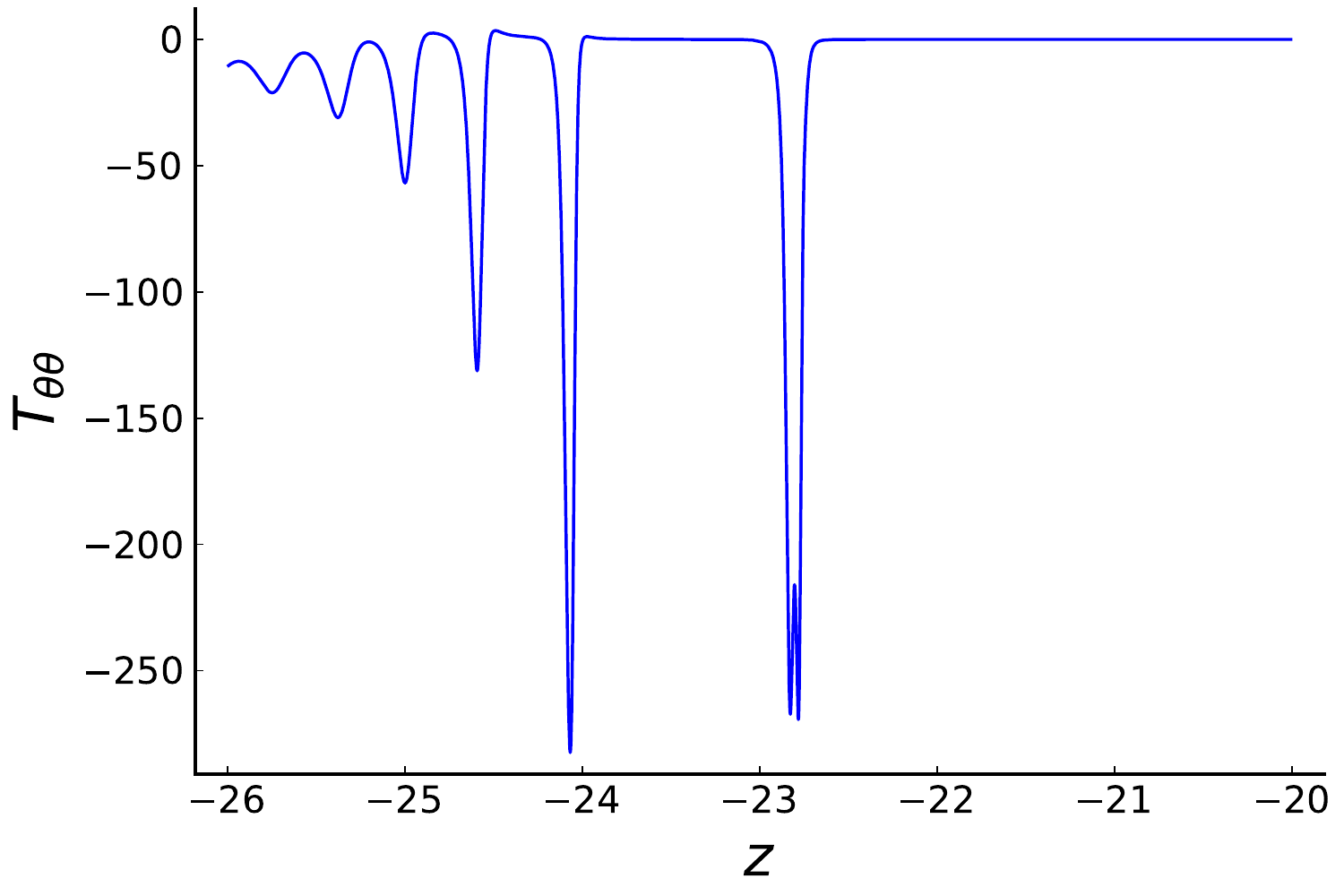} 
      \caption{}\label{pressure_sph}
    \end{subfigure}
    \begin{subfigure}{0.5\textwidth}
      \includegraphics[width = 1\textwidth]{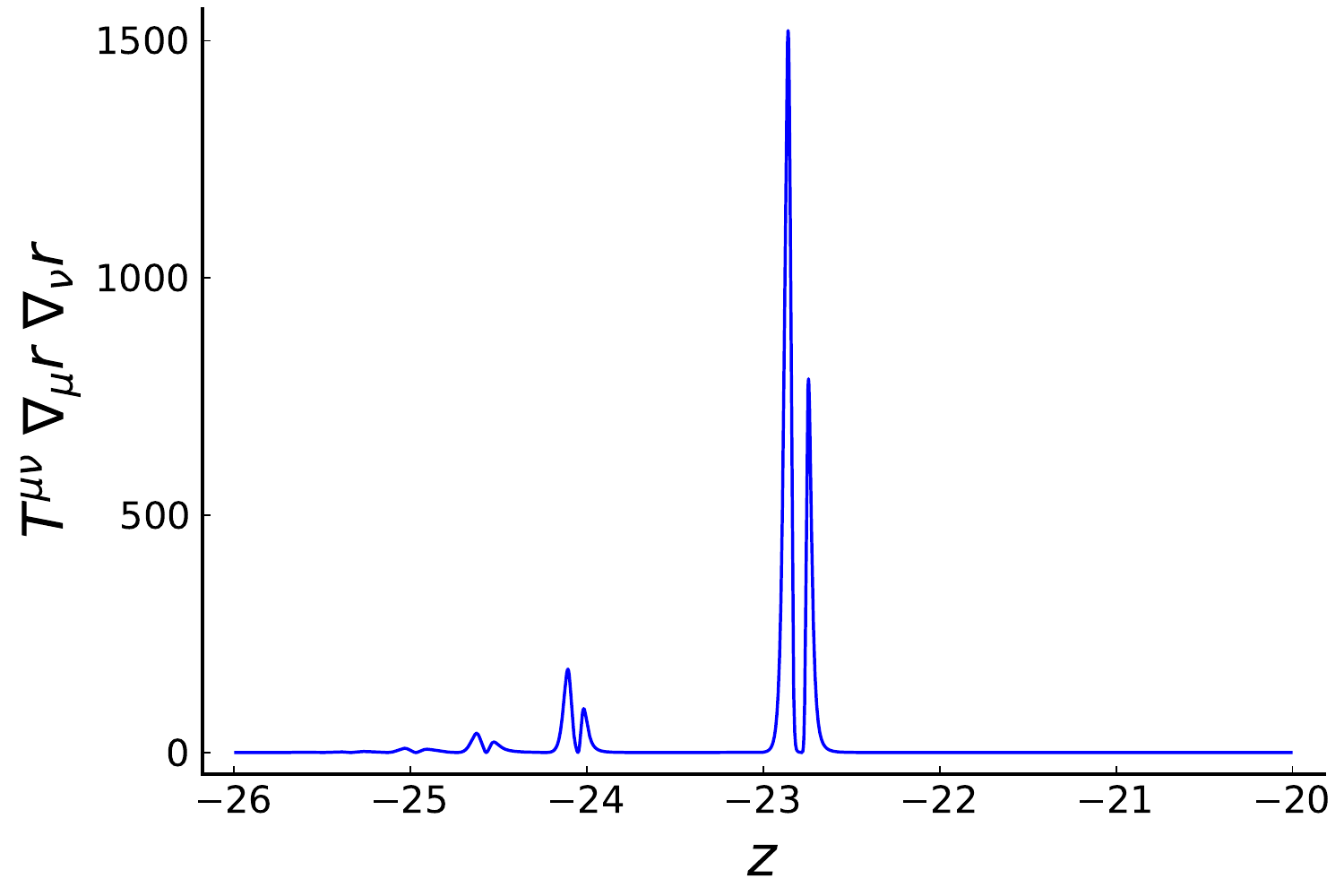} 
      \caption{}\label{Rpressure}
    \end{subfigure}
  \begin{subfigure}{0.5\textwidth}
    \includegraphics[width = 1\textwidth]{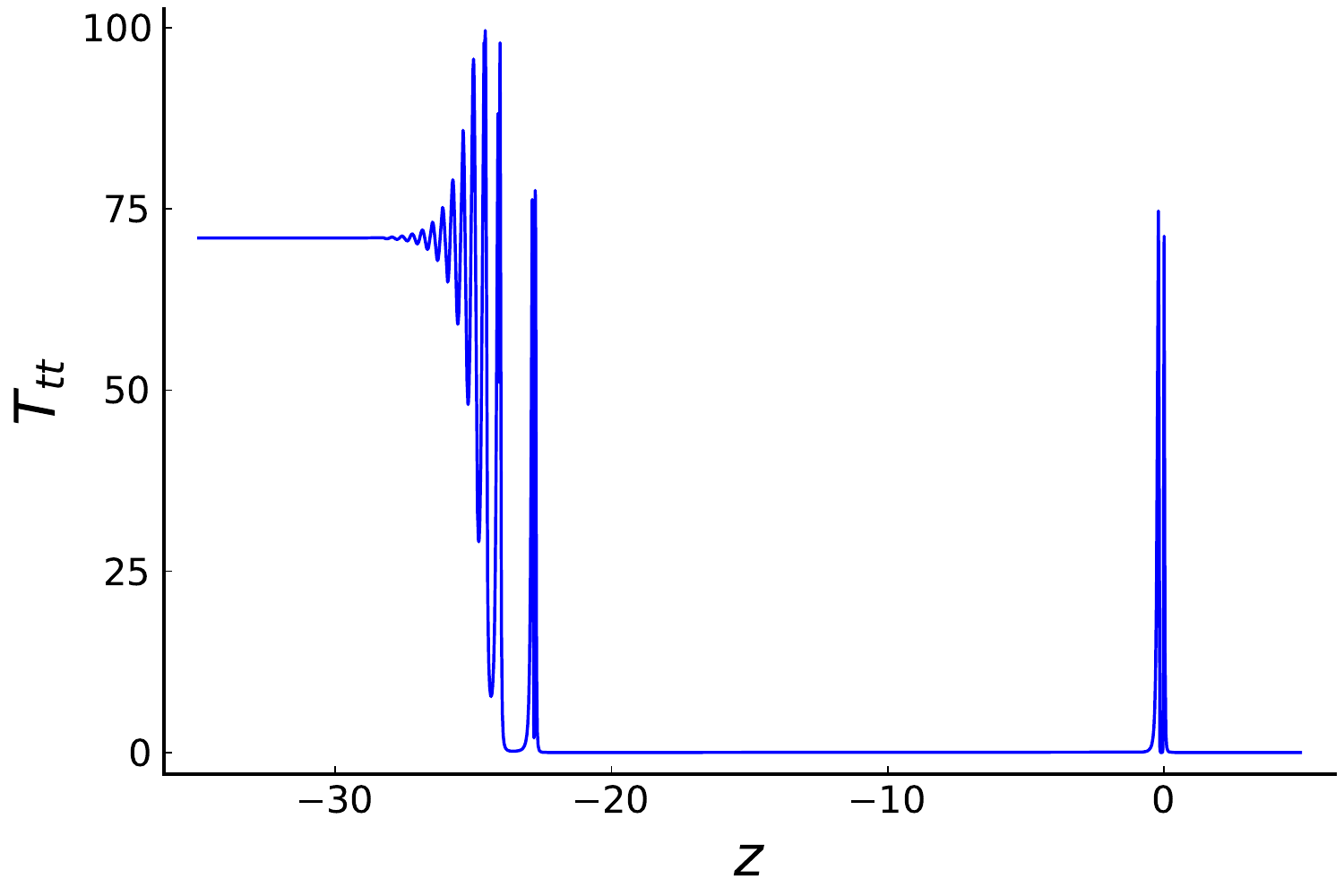} 
    \caption{}\label{T_tt}
  \end{subfigure} 
  \begin{subfigure}{0.5\textwidth}
    \includegraphics[width = 1\textwidth]{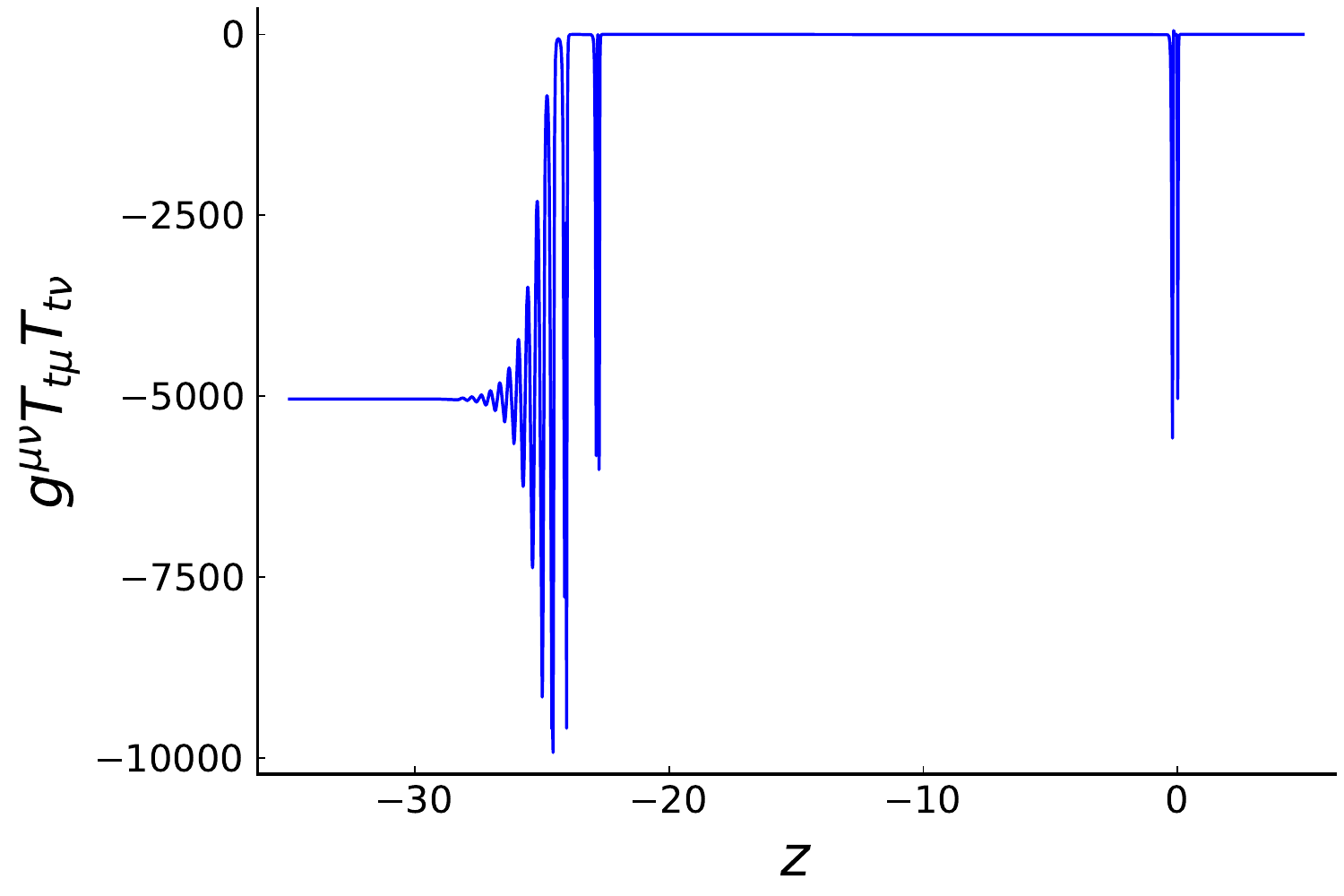} 
    \caption{The norm of the effective energy flow.}\label{Eflow}
  \end{subfigure}
\caption{(a)The plot of the effective tension on $S^2$. (b)The plot of the effective energy density. (c)The plot of the norm of the effective energy flow.}
\end{figure}

Similar to the discussin of cosmology, we extract the energy momentum tensor $T'_{\mu\nu}$ by the Einstein equation $G_{\mu\nu}= T'_{\mu\nu}$. From the viewpoint of the effective dynamics of LQG, $T'_{\mu\nu}$ is the effective stress-energy tensor counting the quantum correction to the Einstein equation, while it is also the stress-energy tensor of the mimetic field from the mimetic-gravity point of view. In the $(t,x,\theta,\varphi)$ coordinate, $T'_{\mu\nu}$ depends only on 4 independent components $T_{tt}$, $T_{tx}$, $T_{xx}$, and $T_{\theta\theta}$
\be
T'_{\mu\nu}=\begin{pmatrix}
  T_{tt} & T_{tx} & 0 & 0 \\
  T_{tx} & T_{xx} & 0 & 0 \\
  0 & 0  & T_{\theta\theta} & 0  \\
  0 & 0  & 0 & T_{\theta\theta}\sin^2(\theta)
  \end{pmatrix}
\ee
The oscillation of the $S^2$ area relates to the oscillations of $T_{\theta\theta}$ and $T^{\mu\nu}\nabla_\mu r\nabla_\nu r$ (with $r=\sqrt{E^x}$), where $T_{\theta\theta }$ is the tension of the effective quantum matter (or equivalently, the mimetic field) wrapping on $S^2$, and $T^{\mu\nu}\nabla_\mu r\nabla_\nu r$ is the pressure normal to $S^2$. See FIG.\ref{pressure_sph} for the tension $T_{\theta\theta}$ and FIG.\ref{Rpressure} for the pressure $T^{\mu\nu}\nabla_\mu r\nabla_\nu r$ (zoomed in the regime where $z<20$ and the geometry is transiting to $\mathrm{dS}_2\times S^2$) from the numerical solution in Section \ref{Non-singular black hole and asymptotic Nariai geometry}.

The effective energy density and the norm of energy flow $g^{\mu\nu}T_{t\mu}T_{t\nu}$ are plotted in FIGs.\ref{T_tt} and \ref{Eflow}. $T_{tt}$ is always positive. $g^{\mu\nu}T_{t\mu}T_{t\nu}$ is positive in a small region near $z=0$,  where the dominant energy condition is voilated, $g^{\mu\nu}T_{t\mu}T_{t\nu}$ is zero or negative elsewhere.

In order to clarify the oscillatory behavior of the geometry as approaching to the asymptotic $\mathrm{dS}_2\times S^2$, we perform the perturbations of the $\mathrm{dS}_2\times S^2$ geometry in the regime of large negative $z$:
\be
\psi(z)=\log(r_0)\lt[1+p_1(z)\rt],&&\quad \xi(z)=\lt[-\frac{z}{a_0}-a_1-\log (r_0)\rt]\lt[1+p_2(z)\rt],\\
\zeta_1(z)=\mathring{\zeta}_1\lt[1+f_1(z)\rt],&&\quad \zeta_2(z)=\mathring{\zeta}_2\lt[1+f_2(z)\rt]
\ee
where $\mathring{\zeta}_1$ and $\mathring{\zeta}_1$ are the asymptotical constant values of $\zeta_1$ and $\zeta_2$ as $z\to-\infty$ (see FIGs.\ref{plotK1} and \ref{plotK2}). The linearization of \eqref{ODEz1} - \eqref{ODEz4} and the expansion in $e^{z/a_0}$ give
\be
p_1'(z)&=&\frac{\mathring{\zeta}_1 f_1(z) \cos \left(\frac{\sqrt{\Delta } \mathring{\zeta}_1}{4}\right)+\mathring{\zeta}_2 f_2(z) \cos \left(\frac{\sqrt{\Delta } \mathring{\zeta}_2}{2}\right)}{4 \log (r_0)} +O(e^{z/a_0}),\\
p_2'(z)&=&\frac{a_0 \mathring{\zeta}_2 f_2(z) \cos \left(\frac{\sqrt{\Delta } \mathring{\zeta}_2}{2}\right)-4 p_2(z)}{4 (a_0 \text{a1}+a_0 \log (r_0)+z)}+O(e^{z/a_0}),\\
f_1'(z)&=&-\frac{ f_1(z) \sin \left(\frac{\sqrt{\Delta } \mathring{\zeta}_2}{2}\right)}{2 \sqrt{\Delta }}-\frac{2  f_1(z) \sin \left(\frac{\sqrt{\Delta } \mathring{\zeta}_1}{4}\right)}{\sqrt{\Delta }}-\frac{1}{4} \mathring{\zeta}_2 f_2(z) \cos \left(\frac{\sqrt{\Delta } \mathring{\zeta}_2}{2}\right)\nonumber\\
&&-\frac{ \mathring{\zeta}_2^2}{2\mathring{\zeta}_1} f_2(z) \cos \left(\frac{\sqrt{\Delta } \mathring{\zeta}_2}{2}\right)+O(e^{z/a_0}),\\
f_2'(z)&=&-\frac{\mathring{\zeta}_1^2 f_1(z) \cos \left(\frac{\sqrt{\Delta } \mathring{\zeta}_1}{4}\right)}{4 \mathring{\zeta}_2}-\frac{1}{2} \mathring{\zeta}_1 f_1(z) \cos \left(\frac{\sqrt{\Delta } \mathring{\zeta}_1}{4}\right)-\frac{2 f_2(z) \sin \left(\frac{\sqrt{\Delta } \mathring{\zeta}_1}{4}\right)}{\sqrt{\Delta }}-\frac{f_2(z) \sin \left(\frac{\sqrt{\Delta } \mathring{\zeta}_2}{2}\right)}{2 \sqrt{\Delta }}\nonumber\\
&&+\frac{4 p_1(z) \log (r_0)}{\mathring{\zeta}_2 r_0^2}+O(e^{z/a_0}).
\ee

We evaluate $a_0,r_0,\mathring{\zeta}_1,\mathring{\zeta}_2$ at $z=z_f$ with a large negative $z_f$ from the numerical solution (with $\Delta=10^{-2}$ and $R_s=10^5$) in Section \ref{Non-singular black hole and asymptotic Nariai geometry}. The solution neglecting $O(e^{z/a_0})$ is obtained explicitly:
\be
p_1(z)&=& (0.0738514 c_1-0.232818 c_2+0.835002 c_3) e^{3.61642 z}\nonumber\\
&&\ +e^{1.80821 z} \Big[(-0.0738514 c_1+0.232818 c_2+0.164998 c_3) \cos (17.1418 z)\nonumber\\
&&\ +(0.0866379 c_1+0.24505 c_2-0.193565 c_3) \sin (17.1418 z)\Big]\\
p_2(z)&=& \frac{e^{1.80821 z} }{(164.651\, -1. z)}\Big[(0.0972387 c_3-0.0435231 c_1) \cos (17.1418 z)\nonumber\\
&&+\ (0.00459104 c_1+0.129943 c_2-0.0102572 c_3) \sin (17.1418 z)\Big],\\
f_1(z)&=& (0.164998 c_1-0.52016 c_2+1.86555 c_3) e^{3.61642 z}\nonumber\\
&&\ +e^{1.80821 z} ((0.835002 c_1+0.52016 c_2-1.86555 c_3) \cos (17.1418 z)\nonumber\\
&&\ +(0.0880803 c_1-2.43812 c_2-0.196788 c_3) \sin (17.1418 z)),\\
f_2(z)&=& e^{1.80821 z} \Big[c_2 \cos (17.1418 z)\nonumber\\
&&\ +(0.338666 c_1+0.105485 c_2-0.756645 c_3) \sin (17.1418 z)\Big].
\ee
$c_1,c_2,c_3$ are integration constants, and there is another integration constant $c_4$ vanishing due to the boundary condition $p_1,p_2,f_1,f_2\to 0$ at $z=z_f$. The solution demonstrates the quasi-normal ocsilations of the perturbations and explains the behavior of the geometry when approaching to the $\mathrm{dS}_2\times S^2$ geometry. The frequency of the oscillation is $\o=17.1418$, while the amplitude of the oscillation is decaying exponentially as $z\to -\infty$ by the factor $e^{1.80821 z}=e^{z/a_0}$.

\section{On the consistency and uniqueness of covariant $\bar{\mu}$-scheme}\label{More on the relation with mubar scheme}

\begin{figure}[h]
  \begin{subfigure}{0.5\textwidth}
  \includegraphics[width = 1\textwidth]{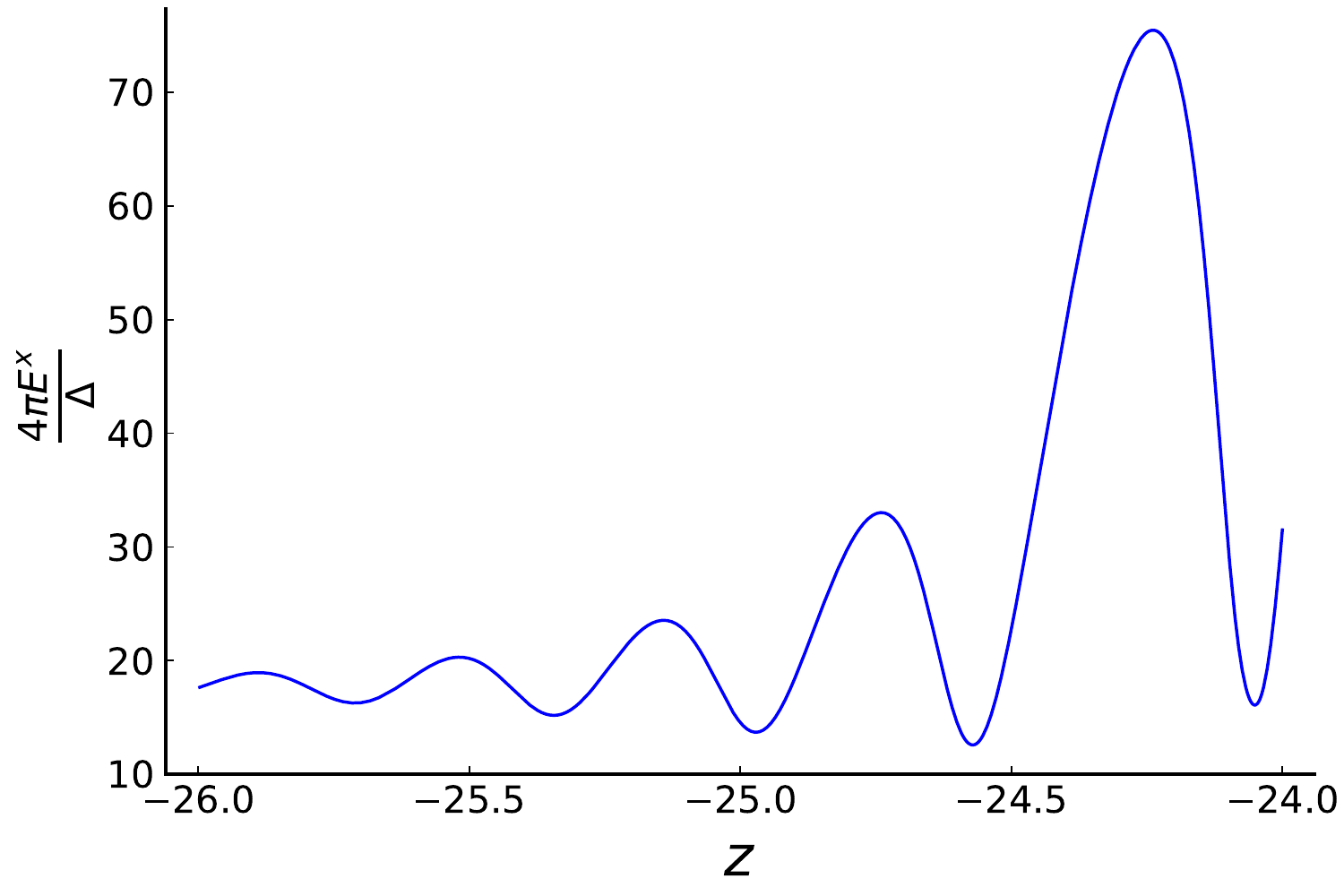} 
   \caption{}
  \end{subfigure}
  \begin{subfigure}{0.5\textwidth}
    \includegraphics[width = 1\textwidth]{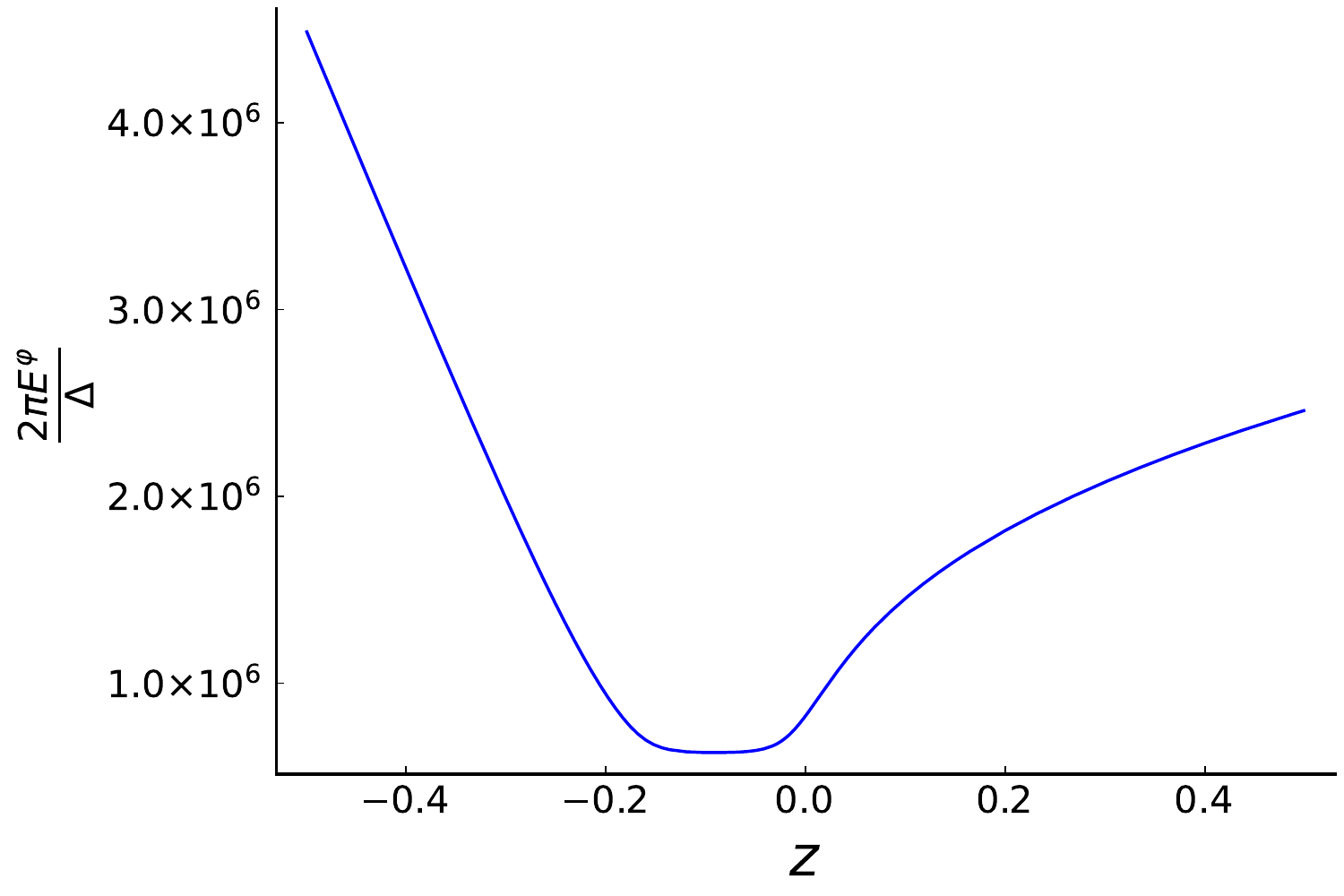} 
    \caption{}
  \end{subfigure}
  \caption{The subfigures (a) and (b) plot respectively the regimes where $E^x$ and $E^\varphi$ reach their minima in the evolution. The parameters used in the figures are the same as in FIG.\ref{plotsol}.}
\label{qualifymubar}
\end{figure}

The covariant $\bar{\mu}$-scheme Hamiltonian \eqref{ccDelta111} depends on the following $\bar{\mu}$-scheme holonomies:
\be
h_x=\exp\lt(\frac{i\b\sqrt{\Delta}\sqrt{E^x}}{ E^\varphi}2K_x\rt),\quad h_\theta= h_\varphi=\exp\lt(\frac{i\b\sqrt{\Delta}}{\sqrt{E^x}}K_\varphi\rt).\label{mubarh}
\ee
Here $\Delta$ is identified to the minimal nonzero eigenvalue in the LQG area spectrum. Recall the discussion in Section \ref{effective dynamics of spherical symmetric quantum gravity} that the $\bar{\mu}$-scheme polymerization in LQG uses the loop holonomies $h_\Delta(\Box)$ around fundamental plaquettes with the fixed area that is set to $\Delta$. There are 2 types of fundamental plaquettes $\Box(\theta,\varphi)$ and $\Box(x,\varphi)$, where $\Box(\theta,\varphi)$ is in any 2-sphere with constant $x$ and $\Box(x,\varphi)$ is in the $x$-$\varphi$ cylinder at $\theta=\pi/2$. Their physical areas are 
\be
\mathrm{Ar}(\Box(\theta,\varphi))=4\pi E^x\delta_\theta\delta_\varphi=\Delta,\qquad \mathrm{Ar}(\Box(x,\varphi))=2\pi E^\varphi\delta_x\delta_\varphi=\Delta
\ee
where $\delta_x,\delta_\theta,\delta_\varphi$ are coordinate lengths of the plaquette edges. The covariant $\bar{\mu}$-polymerization must respect these fundamental plaquettes. So $\delta_\theta\delta_\varphi<1$, i.e. $4\pi E^x/\Delta>1$, must hold in the entire evolution, such that there are always enough room on the 2-sphere to accommodate $\Box(\theta,\varphi)$ with the area $\Delta$. Indeed, this requirement is fulfilled by the black hole solution from the covariant $\bar{\mu}$-scheme dynamics. FIG.\ref{qualifymubar} shows that we have $\delta_\theta\delta_\varphi<1/10<1$ and $\delta_x\delta_\varphi\lesssim 10^{-5}<1$. Therefore the covariant $\bar{\mu}$-scheme dynamics of the non-singular black hole is self-consistent.

Recall that there is freedom in choosing the parameters $a,c,h,\a_1,\a_2$ in the covariant $\bar{\mu}$-scheme Hamiltonian, and the cosmological effective dynamics provides a restriction to the parameters due to the consistency condition \eqref{consistencycondi}. The above discussion is based on the choice \eqref{choiceparameter} satisfying the consistency condition. But there exists other choices classified below \eqref{cosm2}, which are allowed by the cosmological effective dynamics. Let us discuss the implications from these choices to the spherical symmetric effective Hamiltonian:

\begin{enumerate}

  \item $c=h$ and $\b =2a_2/h$ gives the covariant $\bar{\mu}$-scheme Hamiltonian with
  \be
  \cc_\Delta&=&\frac{\sqrt{E^x} E^\varphi}{\Delta  G}\lt[\frac{3 a^2  \sin ^2\left(\frac{4 \alpha _1 \sqrt{\Delta } \sqrt{E^x} K_x}{3 a E^\varphi}-\frac{2 \alpha _1 \sqrt{\Delta } K_{\varphi }}{3 a \sqrt{E^x}}\right)}{8 \alpha _1^2 }-\frac{3 \sin ^2\left(\frac{2 \beta  \sqrt{\Delta } \sqrt{E^x} K_x}{3 E^\varphi}+\frac{2 \beta  \sqrt{\Delta } K_{\varphi }}{3 \sqrt{E^x}}\right)}{2 \beta ^2}\rt]\nonumber\\
  &&+\, \frac{1}{4G\sqrt{E^x}}\Bigg(-\frac{2 E^{x} E^{x \prime} E^{\varphi \prime}}{E^{\varphi 2}}+\frac{4 E^{x} E^{x \prime \prime}+E^{x \prime 2}}{2 E^{\varphi}}-2 E^{\varphi}\Bigg).\label{hamchoice1}
  \ee
  The $\bar{\mu}$-scheme holonomies in this Hamiltonian contain
  \be
  h_x=\exp\lt(\frac{\rho}{2}\frac{i\b\sqrt{\Delta}\sqrt{E^x}}{ E^\varphi}2K_x\rt),\quad h_\theta= h_\varphi=\exp\lt(\rho\frac{i\b\sqrt{\Delta}}{\sqrt{E^x}}K_\varphi\rt),\quad \rho =\frac{4}{3}
  \ee
  These holonmies are along the edges with the fixed geometrical length $\rho \sqrt{\Delta}\equiv \sqrt{\Delta_0}$. Given that the cosmological effective dynamics has the $\bar{\mu}$-scheme holonomy $\exp(i\b\sqrt{\Delta}b)$, $\Delta\neq \Delta_0$ in this scheme implies that the black hole and cosmology corresponds to the $\bar{\mu}$-scheme holonomies with different lengths. Then there is ambiguity about whether $\Delta$ or $\Delta_0$ should be identified to the minimal area gap in LQG. Therefore, although this choice of parameter has no problem from the mimetic-gravity point of view, the inconsistency of $\bar{\mu}$-scheme holonomies suggests to exclude this choice from the LQG point of view,

  \item $c=-h/2$ and $\b=2\a_1/a$ gives the Hamiltonian with
  \be
  \cc_\Delta&=&\frac{\sqrt{E^x} E^\varphi}{\Delta  G}\lt[\frac{3 h^2 \sin ^2\left(\frac{8 \alpha _2 \sqrt{\Delta } \sqrt{E^x} K_x}{3 E^\varphi h}-\frac{4 \alpha _2 \sqrt{\Delta } K_\varphi}{3 \sqrt{E^x} h}\right)}{32 \alpha _2^2 }-\frac{3 \sin ^2\left(\frac{2 \beta  \sqrt{\Delta } \sqrt{E^x} K_x}{3 E^\varphi}+\frac{2 \beta  \sqrt{\Delta } K_\varphi}{3 \sqrt{E^x}}\right)}{2 \beta ^2 }\rt]\nonumber\\
  &&+\, \frac{1}{4G\sqrt{E^x}}\Bigg(-\frac{2 E^{x} E^{x \prime} E^{\varphi \prime}}{E^{\varphi 2}}+\frac{4 E^{x} E^{x \prime \prime}+E^{x \prime 2}}{2 E^{\varphi}}-2 E^{\varphi}\Bigg).
  \ee
  This Hamiltonian becomes the same as \eqref{hamchoice1} by $h\to a$ and $\a_2\to\a_1/2$. This choice of parameters has the same problem as the above, and thus it should be exclude by the consistency in LQG.
  
  \item For the choice $h=\frac{c (\a_1 c-2 \alpha_2  a)}{\alpha_2  a+\a_1 c}$ and $\b=\frac{8 \alpha_2  \a_1}{\alpha_2 a-\a_1 c}$, it is convenient to introduce $m:=\frac{\a_2}{\b c}$. The Hamiltonian has 
  \be
  \cc_\Delta&=&\frac{3\sqrt{E^x} E^\varphi}{8 \beta ^2 \Delta  G m (m+1)}\lt[(1+2m)^2 \sin ^2\left(\frac{4(m+1) \b \sqrt{\Delta } \sqrt{E^x} K_x}{3(2m+1) E^\varphi }+\frac{(4m+1)\b \sqrt{\Delta } K_\varphi}{3(2m+1) \sqrt{E^x}}\right)\rt.\nonumber\\
  &&\lt.- \sin ^2\left(\frac{4 (m+1) \beta  \sqrt{\Delta } \sqrt{E^x} K_x}{3 E^\varphi}+\frac{(1-2m) \beta  \sqrt{\Delta } K_\varphi}{3 \sqrt{E^x}}\right)\rt]\nonumber\\
  &&+\, \frac{1}{4G\sqrt{E^x}}\Bigg(-\frac{2 E^{x} E^{x \prime} E^{\varphi \prime}}{E^{\varphi 2}}+\frac{4 E^{x} E^{x \prime \prime}+E^{x \prime 2}}{2 E^{\varphi}}-2 E^{\varphi}\Bigg).\label{hamchoice3}
  \ee
  This choice includes \eqref{choiceparameter} as a special case equivallent to $m=1/2$. this specical case results in the Hamiltonian \eqref{ccDelta111} with the $\bar{\mu}$-scheme holonomies \eqref{mubarh}, whose length is the same as the $\bar{\mu}$-scheme holonomy in cosmology. So $m=1/2$ does not have the above problem of inconsistency. Requiring \eqref{hamchoice3} to only depend on the same holonomies as \eqref{mubarh} constrains 
  \be
  \lt(\frac{4 (m+1)}{3},\quad \frac{2}{3} (1-2 m),\quad \frac{4 (m+1)}{6 m+3},\quad \frac{2 (4 m+1)}{6 m+3}\rt)\in \mathbb{Z}^4.\label{integer1111}
  \ee
  If we define $n=\frac{4 (m+1)}{3}$ and $k=\frac{4 (m+1)}{6 m+3}$, \eqref{integer1111} equals $(n,\ n-2,\ k,\ 2 k)$ thus implies both $n,k\in\mathbb{Z}$. By the definition of $n,k$, they are constrained by $\frac{2}{3}(k+n)=kn$ or $\frac{1}{n}+\frac{1}{k}= \frac{3}{2}$ if $n,k\neq0$. It is only possible if $|n|\leq 2$ and $|k|\leq 2$. Indeed, we check that there are only 3 possibilities $n=0,1,2$, which give 
  \be
  m=-1,\ -1/4,\ 1/2.
  \ee
$m=-1$ is ruled out since it causes $\cc_\Delta$ to diverge. $m=-1/4$ and $m=1/2$ gives exactly the same $\cc_\Delta$ as \eqref{ccDelta111}.

\item The choice $h=\frac{c (\a_1 c+2 \alpha_2  a)}{-\alpha_2  a+\a_1 c}$ and $\b=\frac{8 \alpha_2  \a_1}{-\alpha_2 a-\a_1 c}$ gives the same Hamilonian as \eqref{hamchoice3}, so the discussion and result are the same as the above case.

\end{enumerate}
In summary, among the Hamiltonians with $\cc_\Delta$ in \eqref{ccDelta1} derived from the mimetic gravity Lagrangian, $\cc_\Delta$ in \eqref{ccDelta111} stands out uniquely by requiring (1) the consistency condition \eqref{consistencycondi} in cosmology, and (2) the consistency between the lengths of the $\bar{\mu}$-scheme holonomies in black hole and cosmology.

\section{Mimetic-CGHS model and light-cone effective dynamics}\label{Mimetic-CGHS model and light-cone effective dynamics}

\subsection{Covariant $\bar{\mu}$-scheme in the light-cone gauge}

The covariant $\bar{\mu}$-scheme is important conceptually because it guarantees the general covariance at the level of the effective theory. The covariant $\bar{\mu}$-scheme is also technically powerful: All earlier studies of LQG effective dynamics are based on 3+1 decomposition and canonical formulation. But here the effective dynamics is generally covariant, and it can adapt to any coordinate system. In particular, the effective dynamics can be formulated in the light-cone coordinates. This formulation is useful in the black hole model with null-shell collapse, as we are going discuss below.

In this section, we focus on the 2d mimetic-CGHS model, whose Lagrangian is $S_2^{(\eta=1)}$ in \eqref{dilatongravity}. We would like to formulation the dynamics in the light-cone gauge: we introduce the null coordinate $(u,v)$ in 2d. All dynamical fields are functions of $u,v$. We impose the gauge fixing condition to the equations of motion \eqref{mimeticin2d} - \eqref{varihij} (and set $\eta=1$),
\be
h_{uu}=h_{vv}=0,\qquad h_{uv}=-\frac{1}{2}e^{2\omega(u,v)}.
\ee
Firstly, the mimetic constraint gives the relation between $\partial_u\phi$ and $\partial_v\phi$
\be
\phi ^{(1,0)}(u,v)= \frac{e^{2 \omega (u,v)}}{4 \phi ^{(0,1)}(u,v)}.
\ee
This relation can be used to reduces all $u$-derivative of $\phi$ to the $v$-derivative. Furthermore, Eqs.\eqref{varipsi2d} and \eqref{varihij} gives 2 dynamical equations
\be
2 \psi ^{(0,1)}(u,v) \psi ^{(1,0)}(u,v)+2 \psi ^{(1,1)}(u,v)+\omega ^{(1,1)}(u,v)+\frac{1}{2} e^{2 \omega (u,v)}+\cw_1[\phi,\psi,\o,\l]&=&0,\label{cghs1}\\
8 \psi ^{(0,1)}(u,v) \psi ^{(1,0)}(u,v)+4 \psi ^{(1,1)}(u,v)+2 e^{2 \omega (u,v)}+\cw_2[\phi,\psi,\o,\l]&=&0,\label{cghs2}
\ee
and 2 constraint equations 
\be
4 \psi ^{(0,2)}(u,v)-8 \psi ^{(0,1)}(u,v) \omega ^{(0,1)}(u,v)+\cw_3[\phi,\psi,\o,\l]&=&0,\label{cghs3}\\
4 \psi ^{(2,0)}(u,v)-8 \psi ^{(1,0)}(u,v) \omega ^{(1,0)}(u,v)+\cw_4[\phi,\psi,\o,\l]&=&0.\label{cghs4}
\ee
Here $\cw_I$ ($I=1,\cdots,4$) are corrections coming from the coupling to the mimetic scalar. The expressions of $\cw_I$ are given in Appendix \ref{WI in mimetic-CGHS}. $\l\to 0$ and $L_\phi\to0$ leads to $\cw_I\to0$ and reduces \eqref{cghs1} - \eqref{cghs4} to the equations of motion \cite{Strominger:1994tn} in the standard CGHS model in the light-cone gauge. 

The details of $\cw_I$ depends on the mimetic potential $\tilde{L}(X,Y)$. We insert the expression of $\tilde{L}$ in \eqref{tildeLUV} - \eqref{tildeLUV2} and fix the free parameters to
\be
a= c=\a_1= 1, \quad h= -1.
\ee
We also introduce $\zeta_1(u,v)$ and $\zeta_2(u,v)$ satisfying
\be
 Y-X=-\frac{\sin \left(\a_2 \sqrt{\Delta }\, \zeta _2(u,v)\right)}{2 \a_2 \sqrt{\Delta }},\qquad  X+Y=-\frac{\sin \left( \sqrt{\Delta } \,\zeta _1(u,v)\right)}{2\sqrt{\Delta }}
\ee
Inserting the expression of $X,Y$ in \eqref{XYab} gives
\be
\omega ^{(0,1)}(u,v)&=&\frac{\phi ^{(0,1)}(u,v)^2 \sin \left(\a_2 \sqrt{\Delta } \zeta_2(u,v)\right)}{4 \a_2 \sqrt{\Delta }}+\frac{\phi ^{(0,2)}(u,v)}{2 \phi ^{(0,1)}(u,v)},\label{cghs5}\\
\psi ^{(0,1)}(u,v)&=&-\frac{\phi ^{(0,1)}(u,v) \sin \left(\sqrt{\Delta } \zeta _1(u,v)\right)}{2 \sqrt{\Delta }}-\frac{\phi ^{(0,1)}(u,v) \sin \left(\a_2 \sqrt{\Delta } \zeta _2(u,v)\right)}{2 \a_2 \sqrt{\Delta }}\nonumber\\
&&-\,4 \psi ^{(1,0)}(u,v) e^{-2 \omega (u,v)} \phi ^{(0,1)}(u,v)^2.\label{cghs6}
\ee
These reduces some 2nd order derivatives of $\o,\psi$ to the 1st order derivatives of $\zeta_1,\zeta_2$, and they may be understood as the analog of the Legandre transformation adapted to the light-cone gauge. 

The full set of equations of motion contains 6 equations that are \eqref{cghs5}, \eqref{cghs6}, and \eqref{cghs1} - \eqref{cghs4} with \eqref{cghs5} and \eqref{cghs6} inserted. The explicit expressions of the equations are given in Appendix \ref{The mimetic-CGHS equations in terms of}. The dynamics determined by these equations are referred to as the effective dynamics, because we would like to understand the mimetic gravity as the effective description of the fundamental quantum gravity theory. 

The set of equations of motion admits the following vacuum solution, which endows $\sm_2$ a flat geometry,
\be
\zeta_1= 0,\quad \zeta_2= 0,\quad \omega = 0,\quad \psi= \frac{u-v}{2},\quad \phi=\frac{u+v}{2},\quad \lambda = 0.\label{vacsol}
\ee
Here the mimetic scalar $\phi$ again plays the role as the global time function $t=({u+v})/{2}$.

\subsection{Conformal matter and null shell in CGHS model}

We study the gravitational collapse in the mimetic-CGHS model by coupling to $S_2^{(\eta=1)}$ a set of $N$ conformal scalar fields $\{f^{(i)}(u,v)\}_{i=1}^N$ in 2d. The total action is
\be
S_{\rm tot}=S_2^{(\eta=1)}+S_M,\qquad  S_M=-\frac{1}{4 \pi} \sum_{i=1}^N \int \rmd^2 x \sqrt{-h}\left(\nabla f^{(i)}\right)^2
\ee
Any 2d metric can be turned into the flat metric locally by conformal transformation. In the conformal gauge, the equation of motion for $f_i$ is $\partial_u\partial_v f^{(i)}=0$ independent of the 2d metric. The general solution $f^{(i)}(u,v)$ is a superposition of the left-moving and right-moving modes $f^{(i)}_+(u)$ and $f^{(i)}_-(v)$
\be
f^{(i)}(u,v)=f^{(i)}_+(u)+f^{(i)}_-(v).
\ee

It is useful to introduce the Kruskal-like exponential coordinate for discussing the CGHS black hole: 
\be
x^+=e^u,\qquad x^-=-e^{-v}.
\ee
The derivatives are denoted by $\partial_\pm =\partial/\partial x^{\pm}$. 

We turn off the right-moving mode of the conformal scalar: $f^{(i)}_-=0$, and we take $\{f^{(i)}_+\}$ to be the shock-waves traveling along the $x^-$-direction with the total magnitude $A$. The stress-energy tensor gives
\be
\sum_{i=1}^N\partial_+f^{(i)}\partial_+f^{(i)}=2A\delta(x^+-x^+_0),\qquad A>0,\quad x_0^+>0
\ee
It gives a null shell that is the source of the gravitational collapse and results in the formation of black hole. The spacetime region $x^+< x^+_0$ is assumed to be the vacuum, while the black hole forms in the region $x^+ > x^+_0$.

\begin{figure}[h]
  \begin{center}
  \includegraphics[width = 0.6\textwidth]{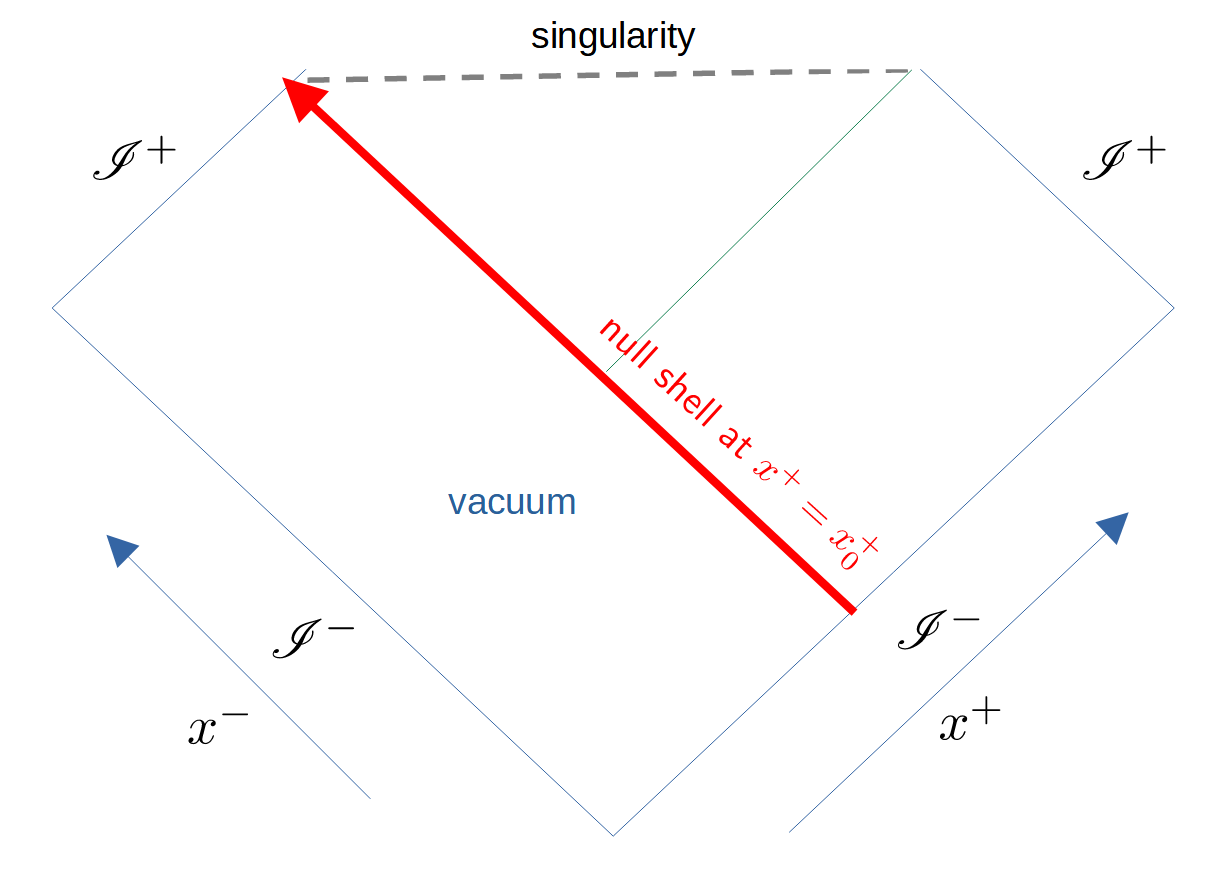} 
  \end{center}
  \caption{The conformal diagram of the classical CGHS black hole with the null shell.}
  \label{cghs_spacetime}
\end{figure}

Let us firstly turn off the mimetic-coupling, i.e. $\cw_I=0$, and review briefly the black hole in standard CGHS model. The solution in the standard CGHS in presence of the null shell gives
\be
e^{2 \psi}=-e^{-2\o}{x^+x^-}=-A\left(x^{+}-x_0^{+}\right) \Theta\left(x^{+}-x_0^{+}\right)- x^{+} x^{-} .
\ee
This solution reduces to the vacuum \eqref{vacsol} when $x^+< x^+_0$, and reduces to the classical CGHS black hole solution when $x^+> x^+_0$:
\be
e^{2 \psi}=-e^{-2\o}{x^+x^-}={M}- x^{+} \tilde{x}^{-},\quad M=Ax_0^+,\quad \tilde{x}^-=x^-+A.\label{CGHSbh}
\ee
$M$ is the mass of the black hole. The singularity is at $x^+\tilde{x}^-=M$, where both $\psi$ and $\o$ diverge. The apparent horizon is at $\partial_+\psi(x^-_H) =0$ and $\partial_- e^{2\psi}(x^-_H) <0$, where ${x}_H^-=-A$. The 2d Ricci scalar is 
\be
R=\frac{4 M}{M - x^{+} \tilde{x}^{-}},
\ee
which diverges at the singularity. The conformal diagram of the 2d spacetime is given in FIG.\ref{cghs_spacetime}.

In the standard CGHS model, when the quantum effect and the back-reaction of the Hawking radiation are taken into account, $\psi$ and $\o$ becomes not divergent, but their derivatives are still divergent at the singularity \cite{Russo:1992ht}. In particular, one finds that just above the null shell, 
\be
\partial_{+} \psi\left(x_0^{+}, x^{-}\right)=-\frac{1}{2 x_0^{+}}\left(\frac{M }{\sqrt{-x_0^{+} x^{-}}\sqrt{ -x_0^{+} x^{-}-N  / 12}}-1\right),\label{partialpluspsi}
\ee
which diverges at $x^-=-\frac{N}{12x_0^+}$. The scalar curvature $R$ also diverges at the same location.

\subsection{The effective dynamics along the null shell}

We turn on the coupling to the mimetic field and consider the light-cone effective dynamics in presence of the null shell. Here as the first step, we only focus on the effective dynamics at $x^+=x^+_0$ just above the null shell, while the effective dynamics in the full spacetime is postponed to the future research.

The vacuum spacetime in $x^+<x^+_0$ is given by the solution \eqref{vacsol}. As the junction condition to match the spacetime regions $x^+<x^+_0$ and $x^+>x^+_0$, the fields $\psi$, $\o$, and $\phi$ are assumed to be continuous across the null shell, i.e. along the shell
\be
\psi=\frac{u_0-v}{2},\quad \o=0,\quad \phi = \frac{u_0+v}{2},\qquad u_0=\ln x^+_0.\label{junctioncondi}
\ee
Their $v$-derivatives (or $x^-$-derivative) are also continuous across the shell, but their $u$-derivatives (or $x^+$-derivative) are not assumed to be continuous. The Lagrangian multiplier $\l$ is not assumed to be continuous across the shell, since $\l$ relates to the derivatives of $\psi$, $\o$, and $\phi$ by the equations of motion.

Inserting \eqref{junctioncondi} in the mimetic-CGHS equations of motion and setting $u=u_0$, we reduces the equations along the null shell. These equations are reduced from PDEs to ODEs with only $v$-derivatives, when we view $\partial_u\psi,\ \partial^2_u\psi\ \partial_u\o$ as the additional independent fields. Firstly, the mimetic constraint, \eqref{cghs5}, and \eqref{cghs6} gives
\be
\partial_u\phi(u_0,v)=\frac{1}{2},\qquad \zeta_2(u_0,v)= 0,\qquad \partial_u\psi(u_0,v) = \frac{2  \sqrt{\Delta }-\sin \left(  \sqrt{\Delta }\, \zeta_1(u_0,v)\right)}{4  \sqrt{\Delta }}.\label{partialupsi}
\ee
Interestingly, the third equation implies hat $\partial_+\psi=(x_0^+)^{-1}\partial_u\psi$ is bounded and can never diverge on the null shell, in contrast to \eqref{partialpluspsi} in the standard CGHS model. Moreover, one of the constraint equations \eqref{cghs3} is used to solve $\l$
\be
\lambda (u_0,v)= 8 \psi ^{(1,1)}(u_0,v) \sin ^2\left(\frac{1}{2}  \sqrt{\Delta } \zeta _1(u_0,v)\right) \sec \left(  \sqrt{\Delta } \zeta _1(u_0,v)\right).\label{sollambda} 
\ee
Inserting \eqref{partialupsi} and \eqref{sollambda} in the rest three equations derived from \eqref{cghs1}, \eqref{cghs2}, and \eqref{cghs4}, we obtain
\be
\omega ^{(1,1)}\left(u_0,v\right)&=& -\frac{1}{128 \Delta  \left(\cos \left(\sqrt{\Delta } \zeta _1\left(u_0,v\right)\right)-2\right)}\Bigg[16 \sqrt{\Delta } \sin \left(2 \sqrt{\Delta } \zeta _1\left(u_0,v\right)\right)\label{omega11}\\
&&+7 \cos \left(\sqrt{\Delta } \zeta _1\left(u_0,v\right)\right)-4 \cos \left(2 \sqrt{\Delta } \zeta _1\left(u_0,v\right)\right)+\cos \left(3 \sqrt{\Delta } \zeta _1\left(u_0,v\right)\right)-4\Bigg]\nonumber\\
\zeta _1^{(0,1)}\left(u_0,v\right)&=&\frac{\sin \left(\sqrt{\Delta } \zeta _1\left(u_0,v\right)\right)}{\sqrt{\Delta }}-\frac{\cos \left(\sqrt{\Delta } \zeta _1\left(u_0,v\right)\right)}{4 \Delta }+\frac{\cos \left(2 \sqrt{\Delta } \zeta _1\left(u_0,v\right)\right)+3}{16 \Delta },\label{eomf0}\\
\psi ^{(2,0)}\left(u_0,v\right)&=& -\frac{\omega ^{(1,0)}\left(u_0,v\right) \sin \left(\sqrt{\Delta } \zeta _1\left(u_0,v\right)\right)}{2 \sqrt{\Delta }}+2 \omega ^{(1,1)}\left(u_0,v\right) \cos \left(\sqrt{\Delta } \zeta _1\left(u_0,v\right)\right)\nonumber\\
&&+\omega ^{(1,0)}\left(u_0,v\right)-2 \omega ^{(1,1)}\left(u_0,v\right)\label{psi20}
\ee
Here is the scheme to solve these equations: The key equation is the ODE \eqref{eomf0}. Once we obtain the solution $\zeta_1(u_0,v)$ to \eqref{eomf0}, inserting the solution in \eqref{partialupsi} and \eqref{omega11} gives $\partial_u\psi(u_0,v)$ and $\omega ^{(1,1)}\left(u_0,v\right)$. The 2d scalar curvature is obtained by $R=8 e^{-2 \omega } \omega ^{(1,1)}$. Then $\partial_u\o(u_0,v)$ is obtained by integrating $\omega ^{(1,1)}\left(u_0,v\right)$. Finally $\psi ^{(2,0)}\left(u_0,v\right)$ is obtained by inserting the results in \eqref{psi20}.

The first step is to solve \eqref{eomf0}. We transform \eqref{eomf0} to the $x^\pm$-coordinates,
\be
-x^-\partial_-\zeta _1\left(x^+_0,x^-\right)&=&\frac{\sin \left(\sqrt{\Delta } \zeta _1\left(x^+_0,x^-\right)\right)}{\sqrt{\Delta }}-\frac{\cos \left(\sqrt{\Delta } \zeta _1\left(x^+_0,x^-\right)\right)}{4 \Delta }\nonumber\\
&&+\frac{\cos \left(2 \sqrt{\Delta } \zeta _1\left(x^+_0,x^-\right)\right)+3}{16 \Delta },\label{eomf}
\ee
It is important to transform to $x^\pm$-coordinates, because it turns out that the solution will extend from $x^-<0$ to $x^->0$. To be concrete, we set $x_0=1$ in the following discussion.

\begin{figure}[h]
  \begin{center}
  \includegraphics[width = 0.6\textwidth]{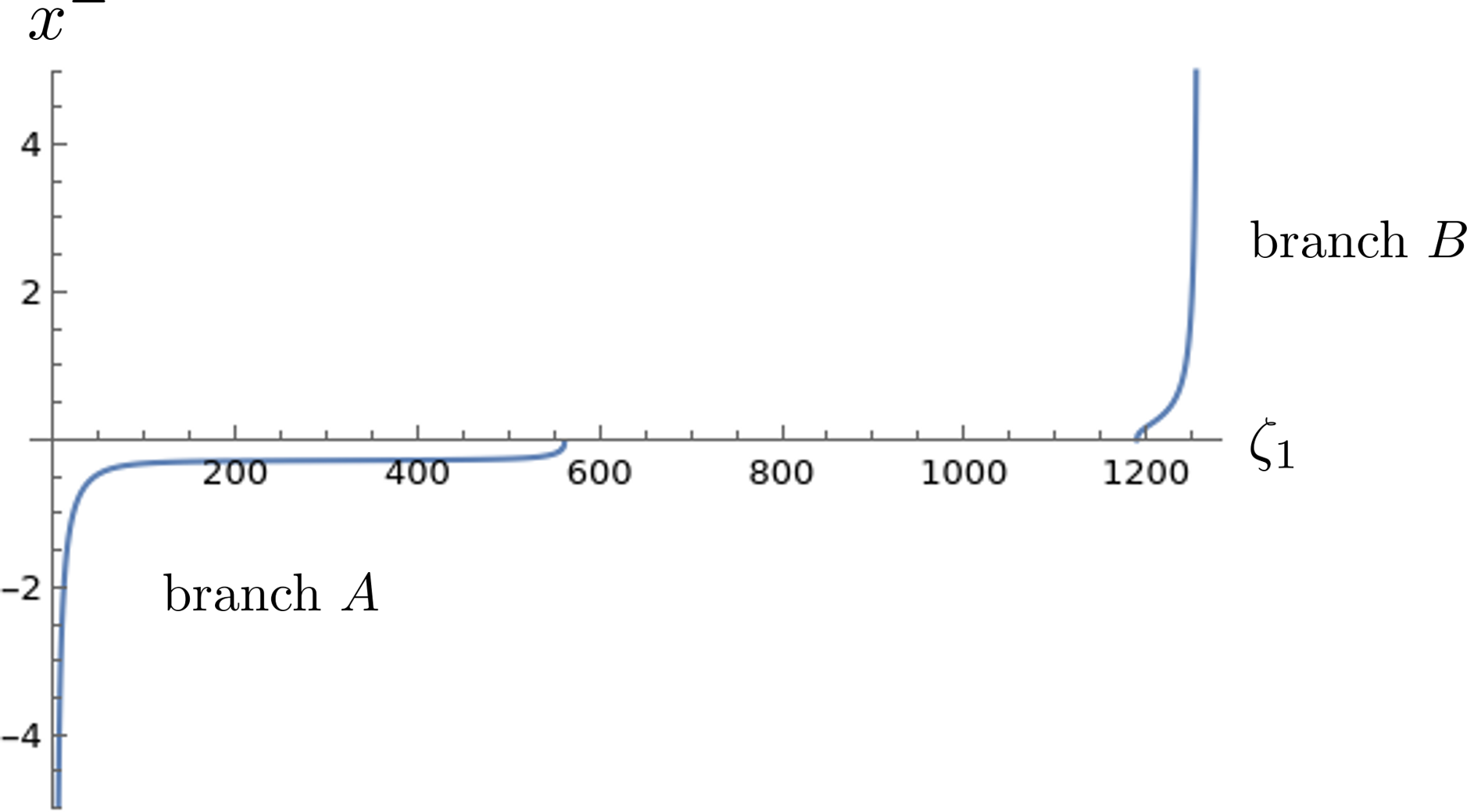} 
  \end{center}
  \caption{The plot of the solution $x^-(\zeta_1)$ in \eqref{xmi} with the value of $C$ in \eqref{solc1}.}
  \label{xf}
\end{figure}

Eq.\eqref{eomf} has the following symmetry:
\be
\text{Translation:}&&\quad\zeta_1\to\zeta_1+\frac{2\pi}{\sqrt{\Delta}}\mathbb{Z},\\
\text{Reflection:}&&\quad x^-\to-x^-,\quad \zeta_1(x^-)\to \zeta_1(-x^-)
\ee
The symmetries will be used for generating solutions. It is convenient to use $\partial_- \zeta_1(x^-)=1/\partial_{\zeta_1} x^-(\zeta_1)$ to write \eqref{eomf} as 
\be
-\frac{\partial_{\zeta_1} x^-}{x^-}=\lt[\frac{\sin \left(\sqrt{\Delta } \zeta _1\right)}{\sqrt{\Delta }}-\frac{\cos \left(\sqrt{\Delta } \zeta _1\right)}{4 \Delta }+\frac{\cos \left(2 \sqrt{\Delta } \zeta _1\right)+3}{16 \Delta }\rt]^{-1},\label{eomf1}
\ee
where the refection symmetry is even more clear. Eq.\eqref{eomf1} can be integrated and gives the solution $x^-$ as the function of $\zeta_1$.
The solution is given by
\be
x_-(\zeta_1)&=&\frac{1}{2}e^{-\frac{4C}{\sqrt{\Delta}}}\cot\lt( \frac{\sqrt{\Delta} \zeta_1}{4}\rt) \prod_{i=1}^6 \lt(p_i + \tan\lt( \frac{\sqrt{\Delta} \zeta_1}{4}\rt) \rt)^{\frac{p_i-\sqrt{\Delta}(1+p_i^2)^2}{\sqrt{\Delta}(1-2 p_i^2 -3 p_i^4)+3 p_i}}
\ee
where $p_i$ are roots of the polynomial $\sqrt{\Delta}(x^6+x^4-x^2-1) -2 x^3$. 
To be explicit, we insert the numerical value $\Delta =10^{-4}$, the solution is expressed as
\be
x_-(\zeta_1)&=&\frac{\cot\left(0.0025 \zeta _1\right)}{2}\Big[e^{-0.04 C} \left(25 \tan ^2\left(0.0025 \zeta _1\right)\rt.\nonumber\\
&&\lt.-140.504 \tan \left(0.0025 \zeta _1\right)-25\right){}^{0.360335} \left((70.252\, -131.552 i) \tan \left(0.0025 \zeta _1\right)\rt.\nonumber\\
&&\lt. +25 \tan ^2\left(0.0025 \zeta _1\right)-25\right){}^{0.319832\, +0.0216294 i} \left((70.252\, +131.552 i) \tan \left(0.0025 \zeta _1\right)\rt.\nonumber\\
&&\lt. +25 \tan ^2\left(0.0025 \zeta _1\right)-25\right){}^{0.319832\, -0.0216294 i}\Big],\label{xmi}
\ee
The integration constant $C$ is determined by the following boundary condition: Near $\mathscr{I}^-$ where $x^-\to-\infty$, the solution should reduce asymptotically to the standard CGHS solution. By Eq.\eqref{CGHSbh} or the asymptotic behavior of \eqref{partialpluspsi}, we have that as $x^-\to-\infty$,
\be
\partial_u\psi(x^+_0,x^-)\sim \frac{1}{2}\lt(\frac{M}{x^-}+1\rt)
\ee
where we have set $x_0^+=1$. Then Eq.\eqref{partialupsi} implies asymptotically
\be
\zeta_1(x^+_0,x^-)\sim -\frac{2 M}{x^-}.
\ee
In practice, we impose the boundary condition $\zeta_1(x^+_0,x^-)= -{2 M}/{x^-}$ at $x^-=x^-_{\rm ini}<0$ with $|x^-_{\rm ini}|$ large but finite. The numerical values of $M$ and $x^-_{\rm ini}$ are set to be $M=10$ and $x^-_{\rm ini}=-10^6$, and they determine the integration constant $C$ to be
\be
C= -0.276316-0.0628319 i.\label{solc1}
\ee
The solution $x^-(\zeta_1)$ is plotted in FIG.\ref{xf} in the range of a single period $(0,4\pi/\sqrt{\Delta}]$. The refection symmetry and a part of the translational symmetry are broken by this solution. The remaining translational symmetry is $\zeta_1\to\zeta_1+4\pi/\sqrt{\Delta}$. The range in $\zeta_1$, whose value of $x^-$ is not plotted in FIG.\ref{xf}, gives the complex $x^-$ with nonzero imaginary part, and this range is disregarded because we are only interested in the real solution. There are 2 branches of real solutions pictured in FIG.\ref{xf}, where the left and right branches are called the branch $A$ and $B$ respectively.

Focusing on the single period, if we rotate FIG.\ref{xf} by 90 degrees and view it as plotting $\zeta_1(x^-)$, the function $\zeta_1(x^-)$ is discontinuous, so it does not satisfy \eqref{eomf} at $x^-=0$. Indeed, the solution \eqref{xmi} gives 
\be 
\zeta_1'(x^-=0)=\frac{1}{\partial_{\zeta_1} x^-(\zeta_1(x^-=0))}=0
\ee
for all $C$, which is valid for all roots $\zeta_1(x^-=0)$. This implies that \eqref{xmi} is not invertible at $x^-=0$. The branch $A$ is determined by the boundary condition. The solution $\zeta_1(x^-)$ of \eqref{eomf} is a differentiable continuation of the branch $A$ in FIG.\ref{xf} from $x^-<0$ to $x^->0$. There are 2 immediate choices by using the symmetries:

\begin{description}
  \item[Solution A:] We ignore the branch $B$. The continuation is the refection of the branch $A$ with respect to $x^-=0$. 
  
  \item[Solution B:] The continuation is obtained by translating the branch $B$ by $2\pi/\sqrt{\Delta}$ and connecting to the branch $A$ at $x^-=0$.
\end{description}

\begin{figure}[h]
  \begin{subfigure}{0.5\textwidth}
    \includegraphics[width = 1\textwidth]{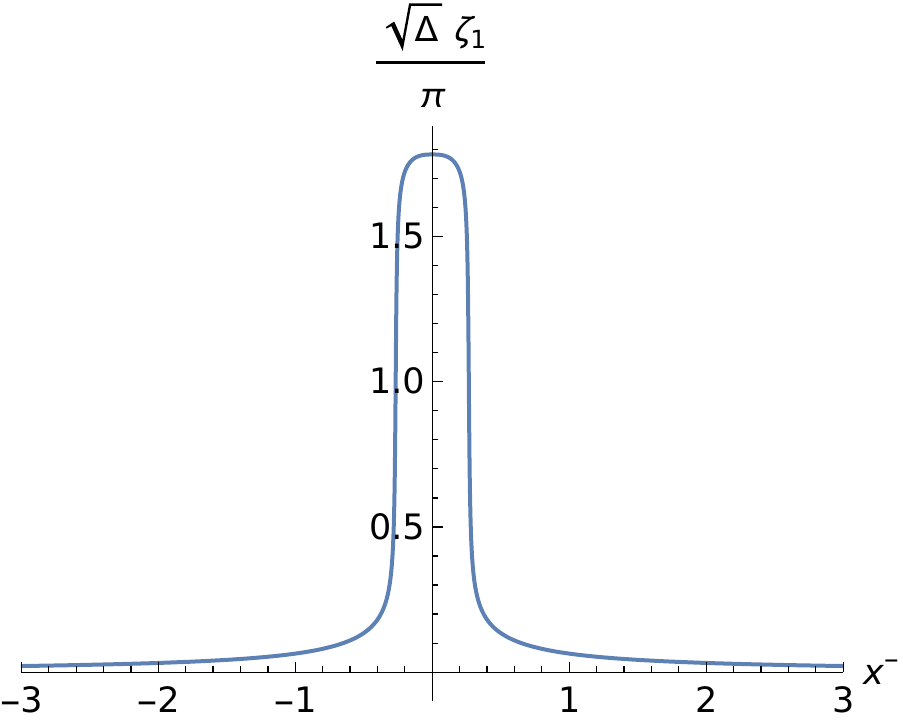} 
    \caption{Solution A}
  \end{subfigure}
  \begin{subfigure}{0.5\textwidth}
    \includegraphics[width = 1\textwidth]{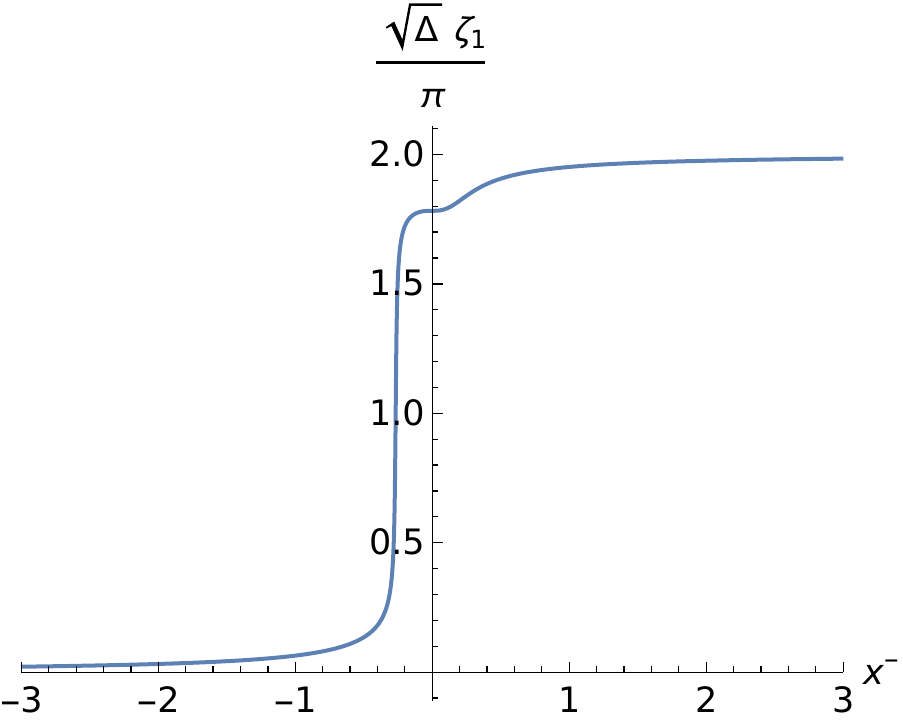} 
    \caption{Solution B}
  \end{subfigure}
  \caption{Two solutions of \eqref{eomf} obtained by two different continuations.}.
  \label{solchoices}
\end{figure}

Both solutions satisfy \eqref{eomf} on entire $x^-\in(-\infty,\infty)$, and are plotted in FIG.\ref{solchoices}. In each solution, $\zeta_1,\ \partial_-\zeta_1,\ \partial^2_-\zeta_1$ are continuous at $x^-=0$ where two branches of solutions are connected. Indeed both $\partial_-\zeta_1 $ and $ \partial^2_-\zeta_1$ vanishes at $x^-=0$, as can be checked from \eqref{xmi} with arbitrary $C$. However, $ \partial^3_-\zeta_1$ diverges at $x^-=0$, so the both solutions are 2nd order differentiable but not 3rd order differentiable.

\begin{figure}[h]
  \begin{subfigure}{0.5\textwidth}
    \includegraphics[width = 1\textwidth]{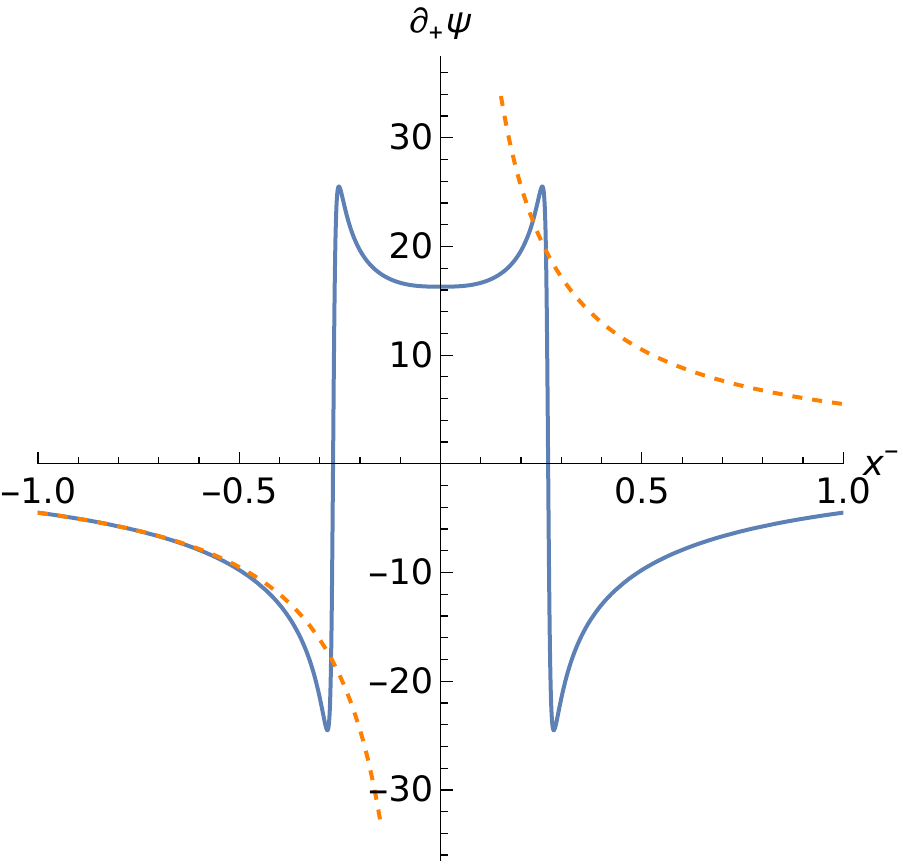} 
    \caption{Solution A}
  \end{subfigure}
  \begin{subfigure}{0.5\textwidth}
    \includegraphics[width = 1\textwidth]{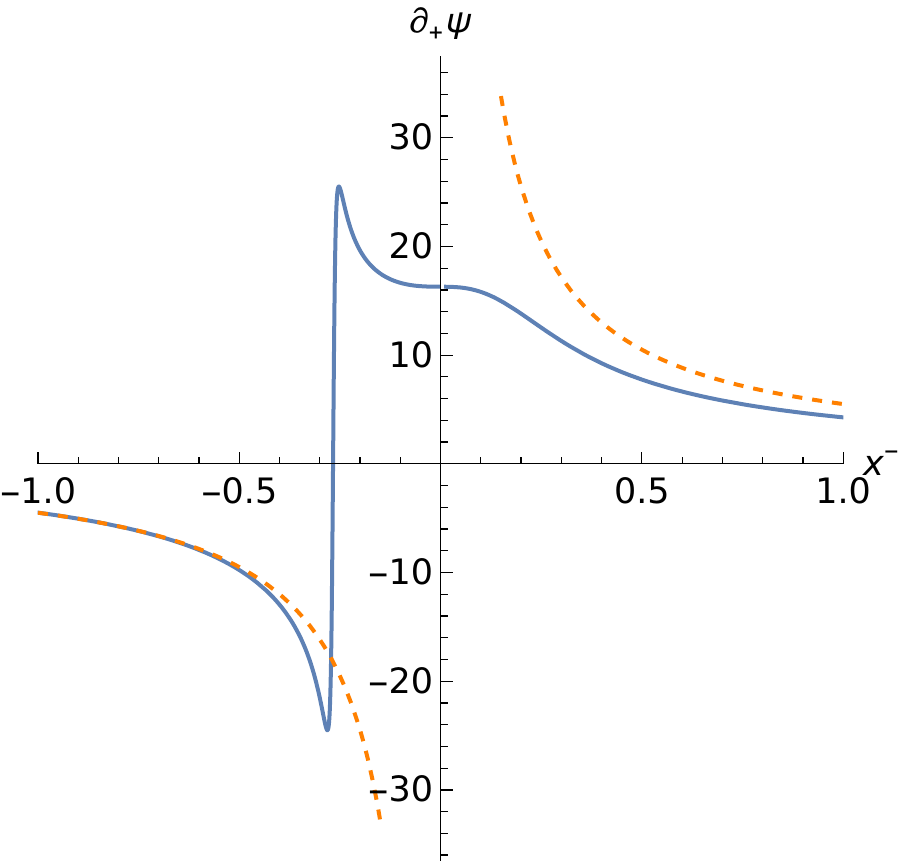} 
    \caption{Solution B}
  \end{subfigure}

  \caption{$\partial_+\psi(x_0^+,x_-)$ resulting from the solution A and B, comparing to the classical solution (orange dashed curves).}.
  \label{dpluspsiAB}
\end{figure}

$\partial_+\psi(x_0^+,x_-)$ is obtained by inserting the solutions in \eqref{partialupsi}, and is plotted in FIG.\ref{dpluspsiAB} for both solutions. In contrast to \eqref{CGHSbh} or \eqref{partialpluspsi} in the standard CGHS model, $\partial_+\psi(x_0^+,x_-)$ is finite in the entire range of $x^-\in (-\infty,\infty)$ along the null shell, in both cases. $\partial_+\psi(x_0^+,x_-)$ reduces to the classical solution $\partial_+\psi(x_0^+=1,x_-)=\frac{1}{2}(1+M/x^-)$ for large negative $x^-$. In addition, $\partial_+\psi(x_0^+,x_-)$ from the solution B reduces to \eqref{partialpluspsi} (or the classical solution) for large positive $x^-$. Although the classical solution is considered as unphysical for positive $x^-$, the result in \cite{Ashtekar:2010qz} suggests that once Hawking radiation and the backreaction are taken into account, the positive $x^-$ regime in \eqref{partialpluspsi} should be physical and connect to the future null infinity $\mathscr{I}^+$. The solution B should be consistent with the scenario of \cite{Ashtekar:2010qz} due to the same asymptotic behavior for large positive $x^-$ along the null shell.

\begin{figure}[h]
  \begin{subfigure}{0.5\textwidth}
    \includegraphics[width = 1\textwidth]{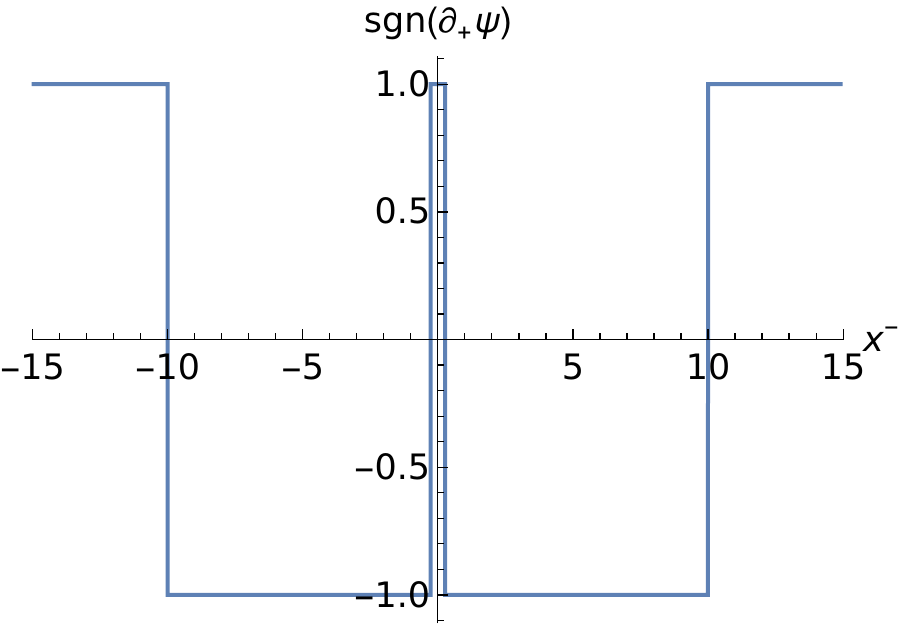} 
    \caption{Solution A}
  \end{subfigure}
  \begin{subfigure}{0.5\textwidth}
    \includegraphics[width = 1\textwidth]{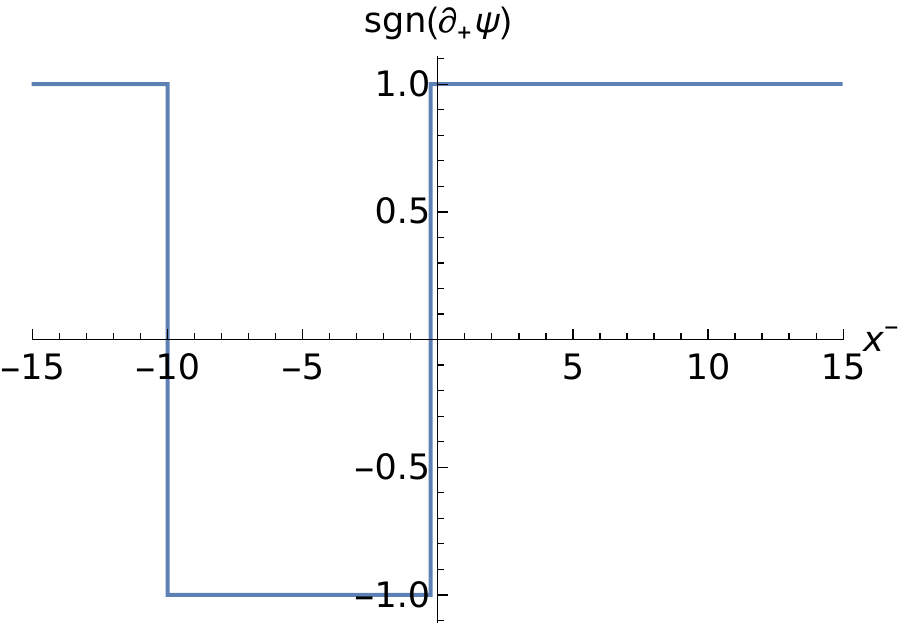} 
    \caption{Solution B}
  \end{subfigure}
  \caption{The apparent horizon is located at $\partial_+\psi=0$. The correction to the classical horizon $x_H=M$ is small in both cases: $|x^-_H-M|=9.17\times 10^{-4}$.}.
  \label{horizonAB}
\end{figure}

\begin{figure}[h]
  \begin{subfigure}{0.5\textwidth}
    \includegraphics[width = 1\textwidth]{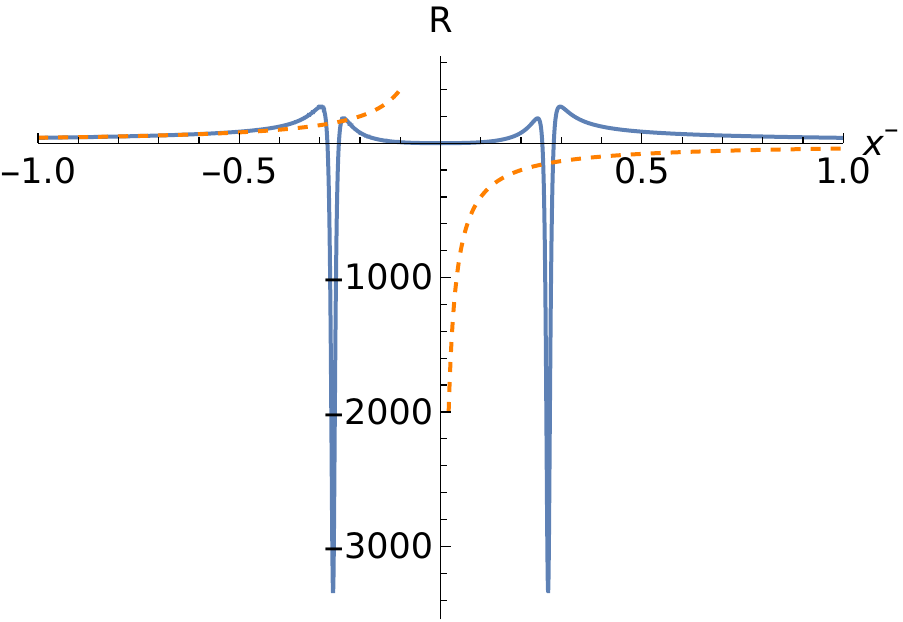} 
    \caption{Solution A}
  \end{subfigure}
  \begin{subfigure}{0.5\textwidth}
    \includegraphics[width = 1\textwidth]{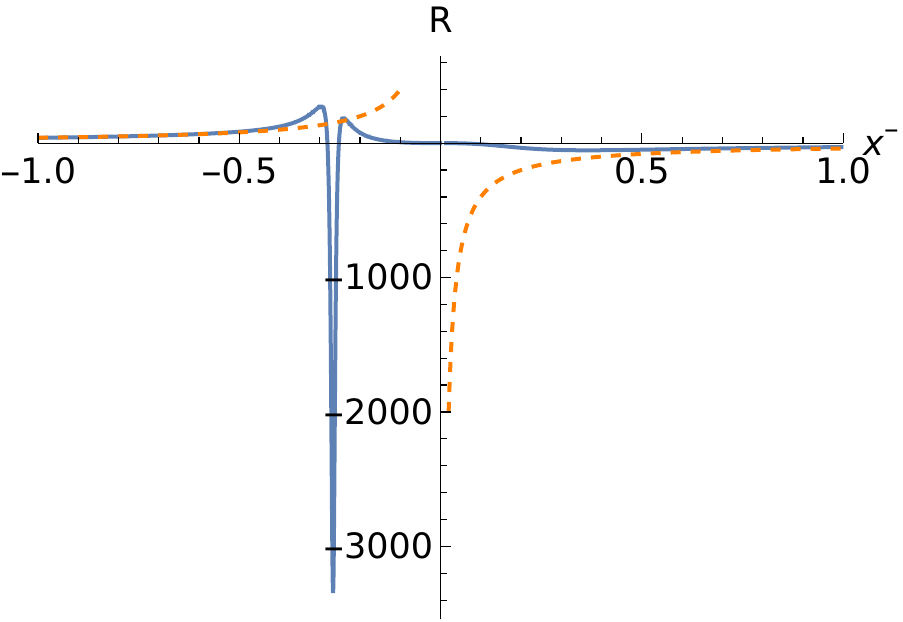} 
    \caption{Solution B}
  \end{subfigure}
  \caption{The scalar curvature $R$ resulting from the solutions A and B, comparing to the classical solution (orange dashed curves). Both $x^-\to \pm\infty$ are asymptotically flat regimes in both cases. }
  \label{scalarRAB}
\end{figure}

\begin{figure}[h]
  \begin{center}
  \includegraphics[width = 0.6\textwidth]{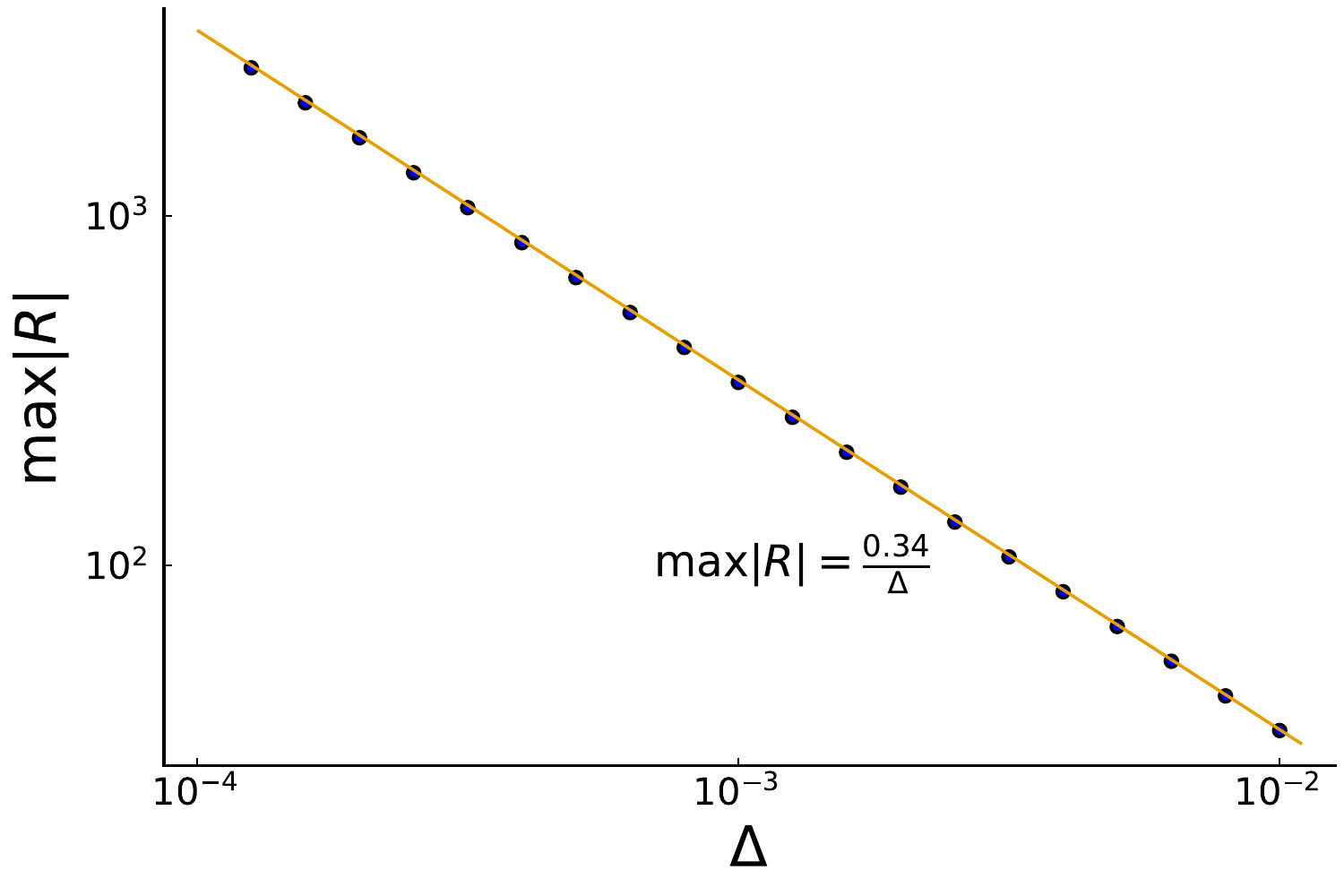} 
   \caption{The solutions of \eqref{eomf} are computed for different values of $\Delta$. This log-log plot demonstrates that $\mathrm{max}|R|$ is inverse proportional to $\Delta$. }
  \label{R_Delta_cghs}
   \end{center}
\end{figure}

The 2d scalar curvature $R=8 e^{-2 \omega } \omega ^{(1,1)}$ is obtained by inserting the solution in \eqref{omega11}. $R$ in both cases are plotted in FIG.\ref{scalarRAB}. It again shows the solution B is prefereed since it reduces to the standard CGHS situation for both $x-\to \pm\infty$. $R$ is asymptotically vanishing as $x^-\to \pm\infty$ independent of the choice of solutions. It indicates that the null shell always starts and ends at the asymptotically flat regimes. At least in a neighborhood of the null shell, the future null infinity $\mathscr{I}^+$ (the left $\mathscr{I}^+$ in FIG.\ref{cghs_spacetime}) can extend from the vacuum to the other side of the null shell. Importantly, The curvature is finite on the entire range of the null shell. The maximal $|R|$ depends on $\Delta$ and scales as 
\be
\mathrm{max}|R|\propto \frac{1}{\Delta},
\ee
as demonstrated in FIG.\ref{R_Delta_cghs}. The classical singularity is resolved with the Planckian curvature, as $\Delta\sim \ell_P^2$.

The solutions A and B does not exhaust all possible differentiable continuation of the branch $A$ in FIG.\ref{xf}. Indeed, the branch $A$ can be connected to the refection of the branch $A$ or the translated branch $B$ with a different value of the integration constant $C$, since \eqref{xmi} implies both $\partial_-\zeta_1 $ and $ \partial^2_-\zeta_1$ vanishes at $x^-=0$ with an arbitrary $C$. The result is still a solution of \eqref{eomf} on the entire range of $x^-$. The solution in $x^->0$ can corresponds to a different mass $M$. It means that the remnant mass of the black hole cannot be predicted by the effective dynamics only along the null shell. The different continuations of the solution correspond to the different possible scenarios in the 2d black hole dynamics. The above solution B is the unique one to have $\partial_+\psi$ and $R$ reduce to the standard CGHS situation for both $x^-\to \pm\infty$. A unique scenario might also be fixed by solving the full PDEs of the light-cone effective dynamics, which is beyond the scope of this paper. However, as the universal feature of all scenarios, both $\partial_+\psi$ and the 2d curvature are finite along the entire null shell, the spacetime extends to the regime that has been forbidden by the singularity, and the null shell ends at a new asymptotically flat regime. 

\section{Conclusion and outlook}

In this work, we propose the covariant $\bar{\mu}$-scheme effective dynamics of the spherical symmetric LQG. This effective dynamics can be derived from the covariant mimetic gravity Lagrangian in 4d, with the prescribed higher derivative couplings. The effective theory contains the LQC effective dynamics as a subsection. The theory gives the non-singular black hole solution, which resolves the singularity in the Schwarzschild spacetime. The non-singular black hole spacetime has the complete $\mathscr{I}^+$. In the interior of the black hole, the spacetime evolves to $\mathrm{dS}_2\times S^2$ geometry as the asymptotic final state. The covariant mimetic gravity Lagrangian allows us to formulate the effective dynamics beyond the 3+1 canonical formulation. In particular, it is useful to formulate the effective dynamics in the light-cone gauge. The application to the CGHS model resolves the singularity along the null shell.

As the applications of the covariant $\bar{\mu}$-scheme effective dynamics to the spherical symmetric LQG, the solutions of black hole and cosmology has more symmetries than only the sperical symmetry. The additional symmetries are used for simplifying the PDEs to ODEs. However, more interesting situations with richer dynamical propertes often need to relax the additional symmetry and solve the full PDEs. One interesting dynamical situations is the gravitational collapse with massive or null matter (see \cite{Giesel:2021dug,Husain:2022gwp,Lewandowski:2022zce} for some recent progress on the effective dynamics of gravitational collapse). The null shell collapse is considered here for the CGHS model, but understanding the full dynamics on the 2d spacetime requires to solve the full PDEs. Another interesting situation is to include the backreaction of the Hawking radiation, which should result in the dynamical black hole solution. The black hole solution in this work does not have the white-hole type marginal anti-trap surface, but it might be the consequence from the additional killing symmetry $\partial_t+\partial_x$. Treating the full PDEs should give the more interesting dynamical black hole solutions and provide different scenarios of the black hole final states.

\section*{Acknowledgments}

M.H. acknowledges Abhay Ashtekar for many enlightening discussions and his encouragement. In particularly, this work receives valuable contributions from Abhay Ashtekar on the covariance, the foliation and geometry of the non-singular black hole, the null expansion and the physical interpretation. The authors also acknowledges Eugenio Bianchi, Kristina Giesel, Lingzhen Guo, and Stefan Weigl for helpful discussions on aspects of Hamiltonian dynamics.  M.H. receives support from the National Science Foundation through grants PHY-1912278 and PHY-2207763, and the sponsorship provided by the Alexander von Humboldt Foundation during his visit at FAU Erlangen-N\"urnberg. In addition, M.H. acknowledges IQG at FAU Erlangen-N\"urnberg, IGC at Penn State University, Perimeter Institute for Theoretical Institute, and University of Western Ontario for the hospitality during his visits.


\appendix

\section{Conformal diagram and maximal extension}\label{conformal diagram}

The 2d metric is given by
\be
h_{ij}\rmd x^i \rmd x^j= -\rmd t^2 + \frac{E^{\varphi}(z)^2}{E^x(z)} \rmd x^2,\qquad z=x-t. 
\ee
We introduce the new coordinate $\tau$ given by
\be 
\tau = t- \int_{z_0}^z dz' \frac{E^{\varphi}(z')^2 }{ E^{\varphi}(z')^2 - E^x(z')}
\ee 
The 2d metric is expressed as below in the $(\t,z)$-coordiante:
\be 
h_{ij}\rmd x^i \rmd x^j = - \frac{E^x -(E^{\varphi})^2}{E^x} \rmd \tau^2 + \frac{(E^{\varphi})^2 }{E^x -(E^{\varphi})^2} \rmd z^2
\ee 
The coordinate transformation is singular at the killing horizon where $E^x=(E^\varphi )^2$. 

We define two null coordinates
\be 
u = \int_{z_0}^z dz' \frac{E^{\varphi}(z') \sqrt{E^{x}(z') } }{ E^{\varphi}(z')^2 - E^x(z')} + \tau, \qquad 
v = \int_{z_0}^z dz' \frac{E^{\varphi}(z') \sqrt{E^{x}(z') } }{ E^{\varphi}(z')^2 - E^x(z')} - \tau
\ee 
and their rescaling
\be 
U = e^{A_0 - \frac{u}{2 B_0}}, \qquad V = \sgn((E^{\varphi})^2 - E^x) e^{A_0 - \frac{v}{2 B_0}}
\ee 
where $B_0= \left|\frac{E^{\varphi} \sqrt{E^{x}} }{ 2 E^{\varphi}E^{\varphi}{}'- E^x{}'} \right| \Big|_{(E^{\varphi})^2 - E^x=0}$. For sufficiently large black hole, the value of $B_0$ is well approximated by the Schwarzchild one, which is $B_0 = R_s := \lt.\sqrt{E^x} \rt|_{(E^{\varphi})^2 - E^x=0}$. We chose $A_0$ such that when $z \to \infty$ we have $UV =1$. When $z_0$ is sufficiently large, $A_0$ is approximately given by the Schwarzchild one which reads
\be 
A_0 = \lt( \frac{E^x}{2 (E^{\varphi})^2} + \frac{1}{2} \log \lt|  \frac{ (E^{\varphi})^2 - E^x}{ (E^{\varphi})^2}  \rt| \rt)\bigg|_{z = z_0}
\ee 
$UV =0$ indicates the location of the horizon. Using $U$ and $V$ we define
\be 
T = \frac{1}{2} (U+V), \qquad X = \frac{1}{2} (U-V)
\ee 
Here $z$ is a function of $UV$ thus a function of $T^2 - X^2$, and $\tau$ is a function of $T/X$. As a result, we can define the extension $(T,X) \to (-T,-X)$. 
With the Schwarzschild geometry, we recover the Kruskal-Szekeres coordinates
\be 
T = e^{\frac{r}{2 R_s}} \cosh\lt(\frac{\tau}{2 R_s} \rt)\sqrt{1- \frac{r}{R_s}}\, , \qquad X =  e^{\frac{r}{2 R_s}} \sinh\lt(\frac{\tau}{2 R_s} \rt)\sqrt{1- \frac{r}{R_s}}\,
\ee 
for $z$ inside the horizon and 
\be 
T =   e^{\frac{r}{2 R_s}} \sinh\lt(\frac{\tau}{2 R_s} \rt)\sqrt{ \frac{r}{R_s} -1 }, \qquad X = e^{\frac{r}{2 R_s}} \cosh\lt(\frac{\tau}{2 R_s} \rt)\sqrt{ \frac{r}{R_s} -1}
\ee 
for $z$ outside the horizon.

\begin{figure}[h]
    \centering
    \includegraphics[width=0.45\linewidth]{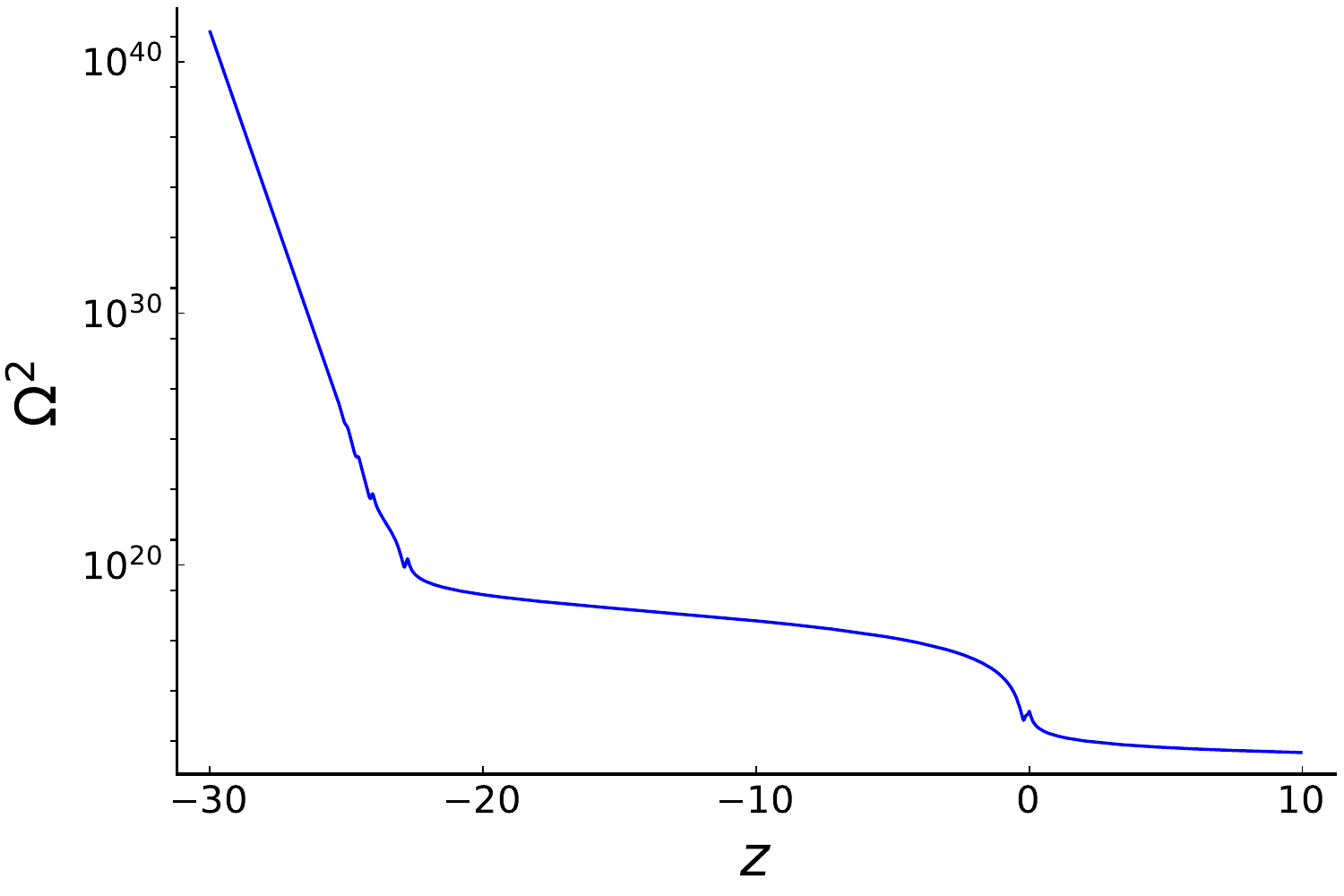}
    \includegraphics[width=0.45\linewidth]{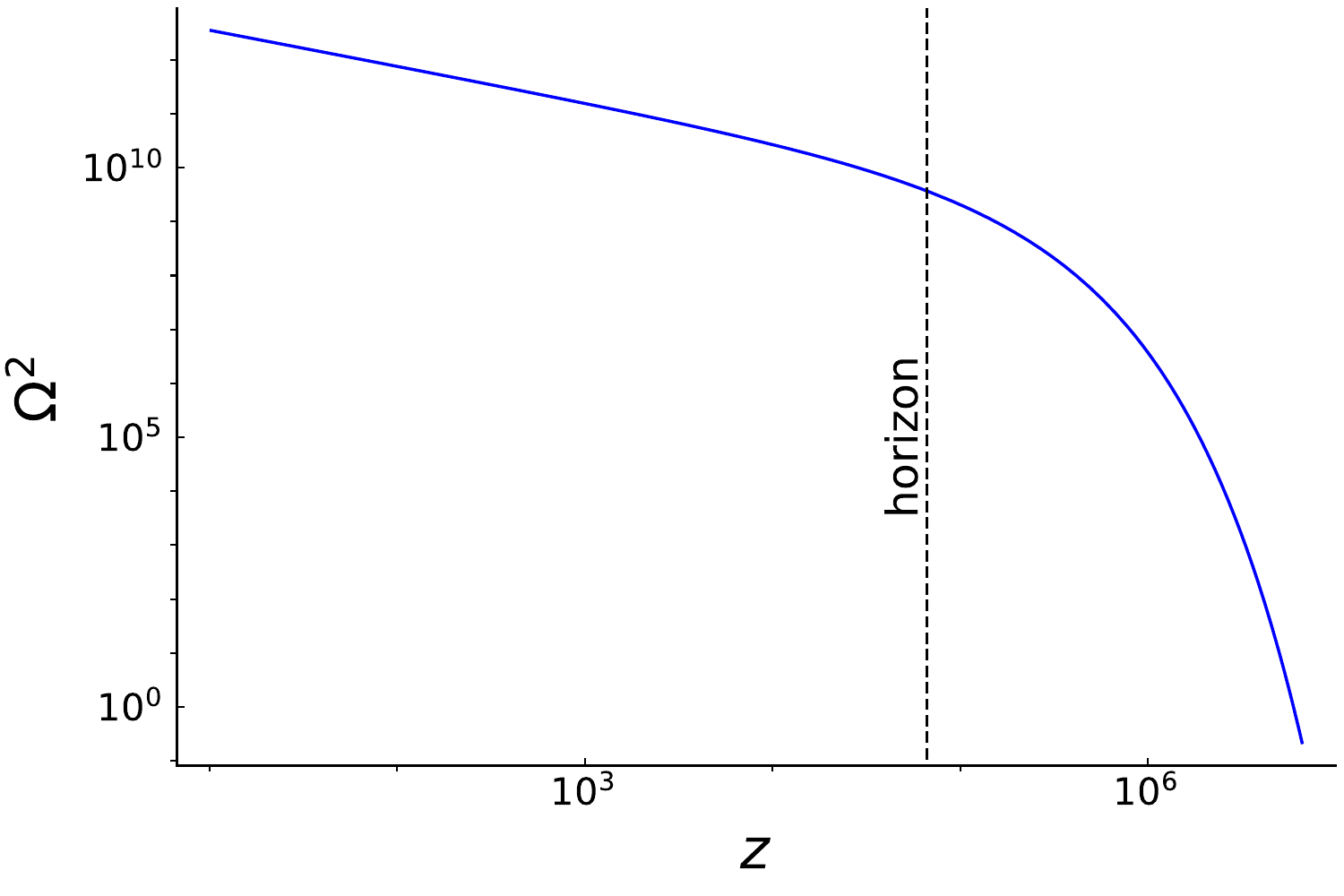}
    \caption{Log-log plot of the conformal factor $\O^2$ as a function of $z$.}
    \label{fig:omega}
\end{figure}

The 2d metric is conformally flat
\be 
h_{ij}\rmd x^i \rmd x^j = \Omega^2 (-dT^2 + dX^2)
\ee 
with the conformal factor as a function of $z$ only 
\be 
\Omega^2 =\sgn((E^{\varphi})^2 - E^x) e^{-2A_0 + \frac{u+v}{2 R_s}  } \frac{E^x}{R_s[(E^{\varphi})^2 - E^x]}
\ee 
$\Omega^2$ of the non-singular black hole solution discussed in Section \ref{Application II} is plotted in FIG.\ref{fig:omega}. The right panel in FIG.\ref{fig:omega} shows that $\O^2$ is continuous at the horizon, while the left panel shows the exponential growth of $\O^2$ as $z\to -\infty$.

\begin{figure}[h]
  \centering
  \includegraphics[width=0.6\linewidth]{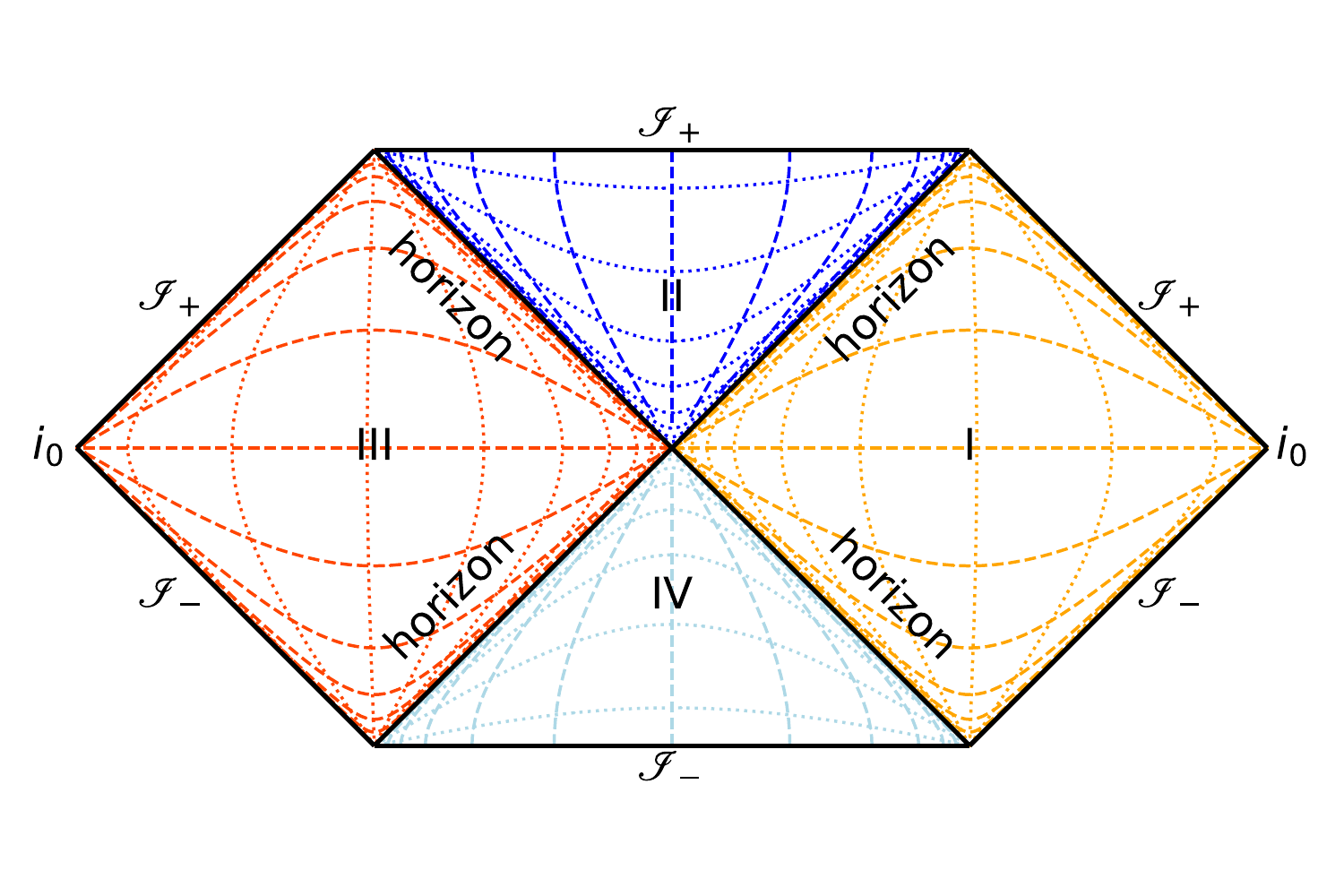}
  \caption{The conformal diagram of the 2d spacetime from the non-singular black hole solution. The dashed lines are the $\t,z$ coordinate lines, where constant $z$ lines are timelike outside (spacelike inside) the killing horizon.}
  \label{fig:penrose}
\end{figure}

We make the conformal compatification of the 2d spacetime by introducing
\be 
\tilde{U} = \arctan(U), &&\quad \tilde{V} = \arctan(V) \\
2\tilde{T} = \tilde{U} + \tilde{V}, &&\quad 2\tilde{X} = \tilde{U} - \tilde{V}
\ee 
where $\tilde{U},\tilde{V}\in[-\pi/2,\pi/2]$. The metric is given by
\be 
h_{ij}\rmd x^i \rmd x^j = \tilde{\Omega}^2 (-d\tilde{T}^2 + d\tilde{X}^2)
\ee 
with 
\be 
\tilde{\Omega}^2 = \frac{\Omega^2}{\cos(\tilde{U})^2\cos(\tilde{V})^2}  
\ee 
where the factor $[{\cos(\tilde{U})\cos(\tilde{V})}]^{-1} $ comes from conformal compactification of flat spacetime. The conformal diagram is shown in FIG.\ref{fig:penrose}

\section{$\cw_I$ in the mimetic-CGHS equations \eqref{cghs1} - \eqref{cghs4}}\label{WI in mimetic-CGHS}

We denote by $L_{a,b}\equiv \tilde{L}^{(a,b)}(X,Y)$.
\be
\cw_1&=&\frac{1}{32} \left(8 \left(-\left(L_{0,2}+2 L_{1,1}+L_{2,0}\right) \psi ^{(1,1)}(u,v)-2 \left(L_{1,1}+L_{2,0}\right) \omega ^{(1,1)}(u,v)+\tilde{L} e^{2 \omega (u,v)}\right)\nonumber\rt.\\
&&\lt. +16 \left(L_{0,2}+2 L_{1,1}+L_{2,0}\right) e^{-2 \omega (u,v)} \phi ^{(0,1)}(u,v)^2 \left(2 \psi ^{(1,0)}(u,v) \omega ^{(1,0)}(u,v)-\psi ^{(2,0)}(u,v)\right)\rt.\nonumber\\
&&\lt. +\frac{e^{2 \omega (u,v)}}{\phi ^{(0,1)}(u,v)^3} \left(\phi ^{(0,2)}(u,v) \left(\left(L_{0,2}+2 L_{1,1}+L_{2,0}\right) \psi ^{(0,1)}(u,v)+4 \left(L_{1,1}+L_{2,0}\right) \omega ^{(0,1)}(u,v)\right)\rt.\rt.\nonumber\\
&& \lt.\lt. +2 \left(L_{1,1}+L_{2,0}\right) \phi ^{(0,3)}(u,v)\right)+\frac{e^{2 \omega (u,v)} }{\phi ^{(0,1)}(u,v)^2}\left(-\left(L_{0,2}+2 L_{1,1}+L_{2,0}\right) \psi ^{(0,2)}(u,v)\rt.\rt.\nonumber\\
&&\lt.\lt.-4 \left(L_{1,1}+L_{2,0}\right) \omega ^{(0,2)}(u,v)+4 \left(L_{0,1}+L_{1,0}\right) \phi ^{(0,2)}(u,v)\right)-16 \left(L_{1,1}+L_{2,0}\right) \phi ^{(0,2)}(u,v) \phi ^{(1,1)}(u,v)\rt.\nonumber\\
&&\lt. +\frac{4 }{\phi ^{(0,1)}(u,v)}\left(L_{0,1} \left(-e^{2 \omega (u,v)}\right) \left(\psi ^{(0,1)}(u,v)+2 \omega ^{(0,1)}(u,v)\right)\rt.\rt.\nonumber\\
&& \lt.\lt.-L_{1,0} e^{2 \omega (u,v)} \left(\psi ^{(0,1)}(u,v)+2 \omega ^{(0,1)}(u,v)\right)+\phi ^{(1,1)}(u,v) \left(\left(L_{0,2}+2 L_{1,1}+L_{2,0}\right) \psi ^{(0,1)}(u,v)\rt.\rt.\rt.\nonumber\\
&&\lt.\lt.\lt. +4 \left(L_{1,1}+L_{2,0}\right) \omega ^{(0,1)}(u,v)\right)+2 \left(L_{1,1}+L_{2,0}\right) \phi ^{(1,2)}(u,v)\right)-16 \left(L_{0,1}+L_{1,0}\right) \psi ^{(1,0)}(u,v) \phi ^{(0,1)}(u,v)\rt.\nonumber\\
&&\lt.-\frac{4 \left(L_{1,1}+L_{2,0}\right) e^{2 \omega (u,v)} \phi ^{(0,2)}(u,v)^2}{\phi ^{(0,1)}(u,v)^4}\right)\nonumber\\
\cw_2&=&\frac{e^{2 \omega (u,v)}}{2 \phi ^{(0,1)}(u,v)^2} \left(\phi ^{(0,1)}(u,v) \left(-\left(L_{0,1}+L_{1,0}\right) \psi ^{(0,1)}(u,v)-4 L_{1,0} \omega ^{(0,1)}(u,v)\rt.\rt.\nonumber\\
&&\lt.\lt. +2 \phi ^{(0,1)}(u,v) (\lambda (u,v)+\tilde{L})\right)+2 L_{1,0} \phi ^{(0,2)}(u,v)\right)-2 \left(L_{0,1}+L_{1,0}\right) \psi ^{(1,0)}(u,v) \phi ^{(0,1)}(u,v)\nonumber \\
\cw_3&=&-2 \left(L_{1,1}+L_{2,0}-2\right) \psi ^{(0,2)}(u,v)-8 L_{2,0} \omega ^{(0,2)}(u,v)\nonumber\\
&&+8 \left(L_{1,1}+L_{2,0}\right) e^{-2 \omega (u,v)} \phi ^{(0,1)}(u,v)^2 \left(2 \psi ^{(1,0)}(u,v) \omega ^{(0,1)}(u,v)-\psi ^{(1,1)}(u,v)\right)\nonumber\\
&&+\frac{2 \phi ^{(0,2)}(u,v) \left(\left(L_{1,1}+L_{2,0}\right) \psi ^{(0,1)}(u,v) +4 L_{2,0} \omega ^{(0,1)}(u,v)\right)+4 L_{2,0} \phi ^{(0,3)}(u,v)}{\phi ^{(0,1)}(u,v)}\nonumber\\
&&+4 \phi ^{(0,1)}(u,v) \left(\left(L_{0,1}-L_{1,0}\right) \psi ^{(0,1)}(u,v)-2 \left(L_{1,1}+L_{2,0}\right) \psi ^{(1,0)}(u,v) e^{-2 \omega (u,v)} \phi ^{(0,2)}(u,v)\right)\nonumber\\
&&-\frac{8 L_{2,0} \phi ^{(0,2)}(u,v)^2}{\phi ^{(0,1)}(u,v)^2}-4 \lambda (u,v) \phi ^{(0,1)}(u,v)^2-4 \psi ^{(0,2)}(u,v)\nonumber\\
\cw_4&=&-\frac{8 }{4 \phi ^{(0,1)}(u,v)^4}\lt[\left(L_{1,1}+L_{2,0}-2\right) \phi ^{(0,1)}(u,v)^4 \left(\psi ^{(2,0)}(u,v)-2 \psi ^{(1,0)}(u,v) \omega ^{(1,0)}(u,v)\right)\rt.\nonumber\\
&&\lt. +2 e^{2 \omega (u,v)} \phi ^{(0,1)}(u,v)^2 \left(\left(L_{1,1}+L_{2,0}\right) \left(2 \psi ^{(1,0)}(u,v) \omega ^{(0,1)}(u,v)+\psi ^{(1,1)}(u,v)\right)+4 L_{2,0} \omega ^{(1,1)}(u,v)\right)\rt.\nonumber\\
&&\lt. -2 e^{2 \omega (u,v)} \phi ^{(0,1)}(u,v) \left(\left(L_{1,1}+L_{2,0}\right) \psi ^{(1,0)}(u,v) \phi ^{(0,2)}(u,v)+\left(L_{1,1}+L_{2,0}\right) \psi ^{(0,1)}(u,v) \phi ^{(1,1)}(u,v)\rt.\rt.\nonumber\\
&&\lt.\lt. +2 L_{2,0} \left(2 \omega ^{(0,1)}(u,v) \phi ^{(1,1)}(u,v)+\phi ^{(1,2)}(u,v)\right)\right)-4 \left(L_{0,1}-L_{1,0}\right) \psi ^{(1,0)}(u,v) e^{2 \omega (u,v)} \phi ^{(0,1)}(u,v)^3\rt.\nonumber\\
&&\lt. +8 L_{2,0} e^{2 \omega (u,v)} \phi ^{(0,2)}(u,v) \phi ^{(1,1)}(u,v)+\lambda (u,v) e^{4 \omega (u,v)} \phi ^{(0,1)}(u,v)^2\rt]+8 \psi ^{(1,0)}(u,v) \omega ^{(1,0)}(u,v)-4 \psi ^{(2,0)}(u,v)\nonumber
\ee

\section{The mimetic-CGHS equations in terms of $\psi,\o,\zeta_1,\zeta_2,\phi, \l$}\label{The mimetic-CGHS equations in terms of}

Inserting \eqref{cghs5} and \eqref{cghs6}, Eq.\eqref{cghs1} reduces to
\be
0&=&\frac{1}{\Delta }\Big[64 \Delta  \psi ^{(1,1)} \sec \left(\sqrt{\Delta } \zeta _1\right)+128 \Delta  \omega ^{(1,1)} \sec \left(\sqrt{\Delta } \zeta _1\right)+128 \Delta  \psi ^{(0,1)} \psi ^{(1,0)}+64 \Delta  \psi ^{(1,1)}\nonumber\\
&&-4 \sqrt{\Delta } \zeta _1 e^{2 \omega } \sin \left(\sqrt{\Delta } \zeta _1\right)+8 \sqrt{\Delta } \zeta _2 e^{2 \omega } \sin \left(\frac{\sqrt{\Delta } \zeta _2}{2}\right)+16 e^{2 \omega } \cos \left(\frac{\sqrt{\Delta } \zeta _2}{2}\right)-4 e^{2 \omega } \cos \left(\sqrt{\Delta } \zeta _1\right)\nonumber\\
&&-e^{2 \omega } \cos \left(2 \sqrt{\Delta } \zeta _1\right)+4 e^{2 \omega } \cos \left(\sqrt{\Delta } \zeta _2\right)-15 e^{2 \omega }\Big]+\frac{8 e^{2 \omega } \left(\psi ^{(0,1)}+2 \omega ^{(0,1)}\right) \left(\sin \left(\sqrt{\Delta } \zeta _1\right)-\sqrt{\Delta } \zeta _1\right)}{\sqrt{\Delta } \phi ^{(0,1)}}\nonumber\\
&&-256 e^{-2 \omega } \left(\phi ^{(0,1)}\right)^2 \left(2 \psi ^{(1,0)} \omega ^{(1,0)}-\psi ^{(2,0)}\right) \sin ^2\left(\frac{\sqrt{\Delta } \zeta _1}{2}\right) \sec \left(\sqrt{\Delta } \zeta _1\right)\nonumber\\
&&-\frac{8 e^{2 \omega }}{\sqrt{\Delta } \left(\phi ^{(0,1)}\right)^2} \Big[2 \sqrt{\Delta } \psi ^{(0,1)} \omega ^{(0,1)} \left(\sec \left(\sqrt{\Delta } \zeta _1\right)-1\right)-\sqrt{\Delta } \psi ^{(0,2)} \sec \left(\sqrt{\Delta } \zeta _1\right)\nonumber\\
&&+16 \sqrt{\Delta } \left(\omega ^{(0,1)}\right)^2 \left(\sec \left(\sqrt{\Delta } \zeta _1\right)-1\right)-\sqrt{\Delta } \zeta _1 \phi ^{(0,2)}+\phi ^{(0,2)} \sin \left(\sqrt{\Delta } \zeta _1\right)+\sqrt{\Delta } \psi ^{(0,2)}\Big]\nonumber\\
&&+\frac{32 \psi ^{(1,0)} \phi ^{(0,1)} \left(\sin \left(\sqrt{\Delta } \zeta _1\right)-\sqrt{\Delta } \zeta _1\right)}{\sqrt{\Delta }}+\frac{128 e^{2 \omega } \omega ^{(0,1)} \phi ^{(0,2)} \left(\sec \left(\sqrt{\Delta } \zeta _1\right)-1\right)}{\left(\phi ^{(0,1)}\right)^3}\nonumber\\
&&-\frac{32 e^{2 \omega } \left(\phi ^{(0,2)}\right)^2 \left(\sec \left(\sqrt{\Delta } \zeta _1\right)-1\right)}{\left(\phi ^{(0,1)}\right)^4},
\ee
Eq.\eqref{cghs2} reduces to
\be
0&=&\left(\phi ^{(0,1)}\right)^2 \Bigg[256 \Delta  e^{2 \psi } \psi ^{(0,1)} \psi ^{(1,0)}+128 \Delta  e^{2 \psi } \psi ^{(1,1)}-4 \sqrt{\Delta } \zeta _1 e^{2 (\psi +\omega )} \sin \left(\sqrt{\Delta } \zeta _1\right)\nonumber\\
&& +8 \sqrt{\Delta } \zeta _2 e^{2 (\psi +\omega )} \sin \left(\frac{\sqrt{\Delta } \zeta _2}{2}\right)+16 e^{2 (\psi +\omega )} \cos \left(\frac{\sqrt{\Delta } \zeta _2}{2}\right)-4 e^{2 (\psi +\omega )} \cos \left(\sqrt{\Delta } \zeta _1\right)\nonumber\\
&&-e^{2 (\psi +\omega )} \cos \left(2 \sqrt{\Delta } \zeta _1\right)+4 e^{2 (\psi +\omega )} \cos \left(\sqrt{\Delta } \zeta _2\right)+32 \Delta  \lambda  e^{2 (\psi +\omega )}+32 \Delta  e^{2 \omega }-15 e^{2 (\psi +\omega )}\Bigg]\nonumber\\
&&-8 \sqrt{\Delta } \phi ^{(0,1)} e^{2 (\psi +\omega )} \Bigg[\psi ^{(0,1)} \left(\sqrt{\Delta } \zeta _1-\sin \left(\sqrt{\Delta } \zeta _1\right)\right)\nonumber\\
&&+4 \omega ^{(0,1)} \left(\sqrt{\Delta } \zeta _1+\sqrt{\Delta } \zeta _2-2 \sin \left(\frac{\sqrt{\Delta } \zeta _2}{2}\right)-\sin \left(\sqrt{\Delta } \zeta _1\right)\right)\Bigg]\nonumber\\
&&-32 \sqrt{\Delta } e^{2 \psi } \psi ^{(1,0)} \left(\phi ^{(0,1)}\right)^3 \left(\sqrt{\Delta } \zeta _1-\sin \left(\sqrt{\Delta } \zeta _1\right)\right)+16 \sqrt{\Delta } \phi ^{(0,2)} e^{2 (\psi +\omega )} \left(\sqrt{\Delta } \zeta _1+\sqrt{\Delta } \zeta _2\rt.\nonumber\\
&&\lt.-2 \sin \left(\frac{\sqrt{\Delta } \zeta _2}{2}\right)-\sin \left(\sqrt{\Delta } \zeta _1\right)\right),
\ee
Eq.\eqref{cghs3} reduces to
\be
0&=&2 \left(\psi ^{(0,2)} \sec \left(\sqrt{\Delta } \zeta _1\right)+4 \omega ^{(0,2)} \left(\sec \left(\sqrt{\Delta } \zeta _1\right)-\sec \left(\frac{\sqrt{\Delta } \zeta _2}{2}\right)\right)-2 \psi ^{(0,1)} \omega ^{(0,1)}+\left(\psi ^{(0,1)}\right)^2\right)\nonumber\\
&&-\frac{2 \left(\psi ^{(0,1)} \phi ^{(0,2)} \left(\sec \left(\sqrt{\Delta } \zeta _1\right)-1\right)+2 \left(2 \omega ^{(0,1)} \phi ^{(0,2)}+\phi ^{(0,3)}\right) \left(\sec \left(\sqrt{\Delta } \zeta _1\right)-\sec \left(\frac{\sqrt{\Delta } \zeta _2}{2}\right)\right)\right)}{\phi ^{(0,1)}}\nonumber\\
&&-16 e^{-2 \omega } \left(\phi ^{(0,1)}\right)^2 \left(2 \psi ^{(1,0)} \omega ^{(0,1)}-\psi ^{(1,1)}\right) \sin ^2\left(\frac{\sqrt{\Delta } \zeta _1}{2}\right) \sec \left(\sqrt{\Delta } \zeta _1\right)\nonumber\\
&&+\phi ^{(0,1)} \Bigg[\psi ^{(0,1)} \left(\frac{-2 \sqrt{\Delta } \zeta _2+4 \sin \left(\frac{\sqrt{\Delta } \zeta _2}{2}\right)+\sin \left(\sqrt{\Delta } \zeta _1\right)}{\sqrt{\Delta }}-\zeta _1\right)\nonumber\\
&&+8 e^{-2 \omega } \psi ^{(1,0)} \phi ^{(0,2)} \left(\sec \left(\sqrt{\Delta } \zeta _1\right)-1\right)\Bigg]+\frac{8 \left(\phi ^{(0,2)}\right)^2 \left(\sec \left(\sqrt{\Delta } \zeta _1\right)-\sec \left(\frac{\sqrt{\Delta } \zeta _2}{2}\right)\right)}{\left(\phi ^{(0,1)}\right)^2}\nonumber\\
&&-2 \lambda  \left(\phi ^{(0,1)}\right)^2,
\ee
Eq.\eqref{cghs4} reduces to
\be
0&=&16 \sec \left(\sqrt{\Delta } \zeta _1\right) \left(\left(\psi ^{(1,0)}\right)^2 \cos \left(\sqrt{\Delta } \zeta _1\right)-2 \psi ^{(1,0)} \omega ^{(1,0)}+\psi ^{(2,0)}\right)\nonumber\\
&&+\frac{e^{4 \omega }}{\left(\phi ^{(0,1)}\right)^5} \Bigg[\psi ^{(0,1)} \phi ^{(0,2)} \left(\sec \left(\sqrt{\Delta } \zeta _1\right)-1\right)\nonumber\\
&&+2 \left(10 \omega ^{(0,1)} \phi ^{(0,2)}+\phi ^{(0,3)}\right) \left(\sec \left(\sqrt{\Delta } \zeta _1\right)-\sec \left(\frac{\sqrt{\Delta } \zeta _2}{2}\right)\right)\Bigg]\nonumber\\
&&+\frac{4 e^{2 \omega }}{\left(\phi ^{(0,1)}\right)^2} \Bigg[2 \psi ^{(1,0)} \omega ^{(0,1)} \left(\sec \left(\sqrt{\Delta } \zeta _1\right)-1\right)+\psi ^{(1,1)} \left(\sec \left(\sqrt{\Delta } \zeta _1\right)-1\right)\nonumber\\
&&+4 \omega ^{(1,1)} \left(\sec \left(\sqrt{\Delta } \zeta _1\right)-\sec \left(\frac{\sqrt{\Delta } \zeta _2}{2}\right)\right)\Bigg]-\frac{2 e^{4 \omega }}{\left(\phi ^{(0,1)}\right)^4} \Bigg[\psi ^{(0,1)} \omega ^{(0,1)} \left(\sec \left(\sqrt{\Delta } \zeta _1\right)-1\right)\nonumber\\
&&+2 \left(4 \left(\omega ^{(0,1)}\right)^2+\omega ^{(0,2)}\right) \left(\sec \left(\sqrt{\Delta } \zeta _1\right)-\sec \left(\frac{\sqrt{\Delta } \zeta _2}{2}\right)\right)\Bigg]+\frac{2 e^{2 \omega } \psi ^{(1,0)}}{\sqrt{\Delta } \phi ^{(0,1)}} \Bigg[-\sqrt{\Delta } \zeta _1-2 \sqrt{\Delta } \zeta _2\nonumber\\
&&+4 \sin \left(\frac{\sqrt{\Delta } \zeta _2}{2}\right)+\sin \left(\sqrt{\Delta } \zeta _1\right)\Bigg]-\frac{4 e^{2 \omega } \psi ^{(1,0)} \phi ^{(0,2)}}{\left(\phi ^{(0,1)}\right)^3} \left(\sec \left(\sqrt{\Delta } \zeta _1\right)-1\right)\nonumber\\
&&-\frac{8 e^{4 \omega }}{\left(\phi ^{(0,1)}\right)^6} \left(\phi ^{(0,2)}\right)^2 \left(\sec \left(\sqrt{\Delta } \zeta _1\right)-\sec \left(\frac{\sqrt{\Delta } \zeta _2}{2}\right)\right)-\frac{\lambda  e^{4 \omega }}{\left(\phi ^{(0,1)}\right)^2}.
\ee
The above equations are supplemented by 
\be
\omega ^{(0,1)}(u,v)&=&\frac{\phi ^{(0,1)}(u,v) \sin \left(\sqrt{\Delta } \zeta_2(u,v)\right)}{4 \sqrt{\Delta }}+\frac{\phi ^{(0,2)}(u,v)}{2 \phi ^{(0,1)}(u,v)},\\
\psi ^{(0,1)}(u,v)&=&-\frac{\phi ^{(0,1)}(u,v) \sin \left(\sqrt{\Delta } \zeta _1(u,v)\right)}{2 \sqrt{\Delta }}-\frac{\phi ^{(0,1)}(u,v) \sin \left(\sqrt{\Delta } \zeta _2(u,v)\right)}{2 \sqrt{\Delta }}\nonumber\\
&&-\,4 \psi ^{(1,0)}(u,v) e^{-2 \omega (u,v)} \phi ^{(0,1)}(u,v)^2.
\ee

\bibliographystyle{jhep}
\bibliography{reference}

\end{document}